\documentclass[usegraphicx,usenatbib]{mn2e}

\usepackage{times}
\usepackage{amsmath}
\usepackage{amssymb}
\usepackage{epsfig}
\usepackage{float}
\usepackage{lmodern}

\begin{document}

\title[The AGN fuelling reservoir in MCG--6-30-15]{Tracing the origin of the AGN fuelling reservoir in MCG--6-30-15}
\author[S. I. Raimundo et al. ]{S. I. Raimundo$^{1}$\thanks{E-mail: s.raimundo@dark-cosmology.dk}, R. I. Davies$^{2}$, R. E. A Canning$^{3,4}$, A. Celotti$^{5,6,7}$, A. C. Fabian$^{8}$, 
\newauthor P. Gandhi$^{9}$  
\\
$^{1}$Dark Cosmology Centre, Niels Bohr Institute, University of Copenhagen, Denmark\\
$^{2}$Max-Planck-Institut f\"ur extraterrestrische Physik, Postfach 1312, D-85741, Garching, Germany\\
$^{3}$Kavli Institute for Particle Astrophysics and Cosmology (KIPAC), Stanford University, 452 Lomita Mall, Stanford, CA 94305-4085, USA\\
$^{4}$Department of Physics, Stanford University, 452 Lomita Mall, Stanford, CA 94305-4085, USA\\
$^{5}$SISSA - Scuola Internazionale Superiore di Studi Avanzati, Via Bonomea 265, I-34135 Trieste, Italy\\
$^{6}$INFN - Sezione di Trieste, via Valerio 2, I-34127 Trieste, Italy\\
$^{7}$INAF - Osservatorio Astronomico di Brera, via Bianchi 46, I-23807 Merate, Italy\\ 
$^{8}$Institute of Astronomy, University of Cambridge, Madingley Road, Cambridge CB3 0HA, UK\\
$^{9}$Department of Physics and Astronomy, University of Southampton, Highfield, Southampton SO17 1BJ, UK
}

\maketitle
\begin{abstract}
The active galaxy MCG--6-30-15 has a 400 pc diameter stellar kinematically distinct core, counter-rotating with respect to the main body of the galaxy. Our previous high spatial resolution (0${''}$.1) H-band observations of this galaxy mapped the stellar kinematics and [Fe II] 1.64 $\micron$ gas dynamics though mainly restricted to the spatial region of the counter-rotating core.
In this work we probe the stellar kinematics on a larger field-of-view and determine the ionised and molecular gas dynamics to study the formation of the counter-rotating core and the implications for AGN fuelling. We present integral field spectroscopy observations with SINFONI in the H and K-bands in the central 1.2 kpc and with VIMOS HR-blue in the central 4 kpc of the galaxy.
Ionised gas outflows of v$_{\rm out} \sim$ 100 km/s are traced by the [Ca VIII] 2.32 $\micron$ coronal line and extend out to at least a radius of $r \sim$ 140 pc. The molecular gas, traced by the H$_{2}$ 2.12 $\micron$ emission is also counter rotating with respect to the main body of the galaxy, indicating that the formation of the distinct core was associated with inflow of external gas into the centre of MCG--6-30-15. The molecular gas traces the available gas reservoir for AGN fuelling and is detected as close as r $\sim$ 50 - 100 pc. External gas accretion is able to significantly replenish the fuelling reservoir suggesting that the event that formed the counter-rotating core was also the main mechanism providing gas for AGN fuelling. 
 
\vspace{0.2cm}
\end{abstract} 


\footnotetext{Based on observations collected at the European Organisation for Astronomical Research in the Southern Hemisphere, Chile, during program 093.B-0734 and on observations made with ESO Telescopes at the La Silla Paranal Observatory under programme 073.B-0617.}
\section{Introduction}
The observed correlations between properties of supermassive black holes and their host galaxies (\citealt{magorrian98}, \citealt{ferrarese&merritt00}, \citealt{gebhardt00}, \citealt{marconi&hunt03}, \citealt{haring&rix04}) suggest a co-evolution in which the Active Galactic Nuclei (AGN) phase may play a central role in regulating gas accretion and star formation in the galaxy. Understanding how the host galaxy fuels the AGN and how the AGN in turn affect gas accretion is currently one of the most important questions in AGN and galaxy evolution studies.

To understand the physics of AGN fuelling it is essential to study not only the black hole accretion physics but also the environment of the host galaxies, in particular their central regions. Due to the higher spatial resolution achievable, several signatures of fuelling have been observed in the low redshift Universe, providing the best laboratory to study AGN fuelling. The molecular gas survey NUGA for example, has found a wide variety of fuelling tracers on the form of dynamical perturbations in the range 0.1 $-$ 1 kpc in lower luminosity AGN (Seyferts and Low Ionisation Nuclear Emission-Line Regions $-$ LINERs) (e.g. \citealt{garcia-burillo03}, \citealt{burillo&combes12}). The ionised gas in the central kiloparsec also seems to be more disturbed in Seyfert host galaxies than in inactive galaxies \citep{dumas07}. For even smaller physical scales, \cite{hicks13} and \cite{davies14} compared 5 pairs of matched active and inactive galaxies at a resolution of $\sim$50 pc using infrared integral field spectroscopy. They found that at the scales of $<$ 200 - 500 pc, where the dynamical time approaches the AGN activity timescale, there are systematic differences between properties of stars and gas for active and inactive galaxies, associated with the black hole fuelling process. The active galaxies in their sample show a nuclear structure in stars and gas, composed of a relatively young stellar population and of a significant gas reservoir,
that could contribute to fuel the black hole. This nuclear structure is not observed in the quiescent galaxies. The presence of circum-nuclear discs of molecular gas seems to be required for nuclear activity, with gas inflows and outflows being detected only in active galaxies. 
Gas inflow/outflow structures are commonly detected in the inner regions of AGN host galaxies (e.g. \citealt{storchi-bergmann07}, \citealt{davies09}, \citealt{muller-sanchez11}, \citealt{davies14}), and usually most of the inflow and outflow mass is in the form of molecular gas. At higher spatial resolution, the Atacama Large Millimeter/submillimeter Array (ALMA) has also provided a new view of the small spatial scales around the AGN, with tracers of molecular gas inflow and outflow being detected in most of the AGN observed (e.g. \citealt{combes13}, \citealt{garcia-burillo14_short}, \citealt{smajic14}, \citealt{combes14}).
Considering the recent work in this field, it has become clear that the central hundreds of parsecs of galaxies are essential to understand the AGN fuelling process.

The S0 galaxy MCG--6-30-15 is a narrow line Seyfert 1 galaxy, with an AGN luminosity of L$_{\rm X}(2-10 \thinspace\rm keV) \sim 4 \times 10^{42}$ erg\thinspace s$^{-1}$ (\citealt{winter09}; \citealt{vasudevan09}), and a total bolometric luminosity of $\sim 3 \times 10^{43}$ erg\thinspace s$^{-1}$ \citep{lira15}. This galaxy has been particularly targeted by X-ray studies, since it was the first source for which the broadened Fe K$_{\alpha}$ emission line was observed, showing that this line was being emitted from the very inner regions of the accretion disc (\citealt{tanaka95}). Our study of the inner 500 pc of MCG--6-30-15 in \cite{raimundo13}, was the first integral field spectroscopy study on this galaxy. 
Due to the unprecedented spatial resolution reached (0.${''}$1 $\sim$ 16 pc in the H-band), we discovered a previously unknown stellar kinematically distinct core in the centre of MCG--6-30-15 counter-rotating with respect to the main body of the galaxy. We first estimated a size of R $\sim$ 125 pc for the distinct core due to the limited field-of-view of our initial observations, we will show in this work that the distinct core has a size of R $\sim$ 200 pc. 
The age of the stars in the counter-rotating core was estimated to be $<$100 Myr \cite{raimundo13} which is in line with the findings of \cite{bonatto00}, who determined that  $>$ 35 per cent of the stellar population is the result of bursts of star formation distributed in age between 2.5 Myr and 75 Myr.

Stellar kinematically distinct cores (KDCs) such as the one in MCG--6-30-15, are characterised by the presence of a stellar core with kinematics and angular momentum distinct from the main body of the galaxy. KDCs can be defined by an abrupt change of more than 30 degrees in the local kinematic orientation of the galaxy (e.g. \citealt{mcdermid06}; \citealt{krajnovic11}), with counter-rotating cores having angles of 180 degrees. The ATLAS$^{\rm 3D}$ project found that KDCs are present in a small fraction ($\sim$ 7 per cent) of early-type galaxies \citep{krajnovic11}, and seem to be more common in slow rotators than in fast rotators \citep{emsellem11_short}. In spiral galaxies, stellar counter-rotating discs are present in $<$ 8 per cent of the galaxies and are rarer than counter-rotating gas discs ($<$ 12 per cent) - \cite{pizzella04}. In S0 galaxies such as MCG--6-30-15, it is estimated that $<10$ per cent host counter-rotating stars (\citealt{kuijken96}) and $\sim$ 25 - 40 per cent host a counter-rotating gas disc (\citealt{kuijken96,kannapan&fabricant01,pizzella04,bureau&chung06}), or $\sim$ 70 per cent if considering isolated galaxies \citep{katkov14} which is consistent with an external origin for the counter-rotating gas \citep{bertola92}. \cite{pizzella04} argue that the formation of counter-rotating gas discs is favoured in S0 galaxies due to their low gas fraction. In spiral galaxies which are more gas rich, the acquired external gas can be swept away due to the interactions with the pre-existing gas which would make the counter-rotating disc observable only when its mass is higher than the one of the pre-existing gas. The difference between the fraction of S0 galaxies with observed counter-rotating gas and observed counter-rotating stars may be due to the inherent difficulty of detecting small numbers of counter-rotating stars (e.g. \citealt{bureau&chung06}). 

The presence of large kinematic misalignments of counter-rotating gas, in particular in S0 galaxies, has been taken as strong evidence of the external origin of the gas (e.g. \citealt{bertola92}, \citealt{davis16}). If the gas originates from stellar mass loss in the galactic disc, then stars and gas are expected to co-rotate. Internal physical processes such as weak bars or spirals can cause mild kinematic misalignments between gas and stars (up to an angular difference $\Delta$ PA $\lesssim$ 55 degrees) \citep{dumas07}. However, large kinematic misalignments or counter-rotating gas ($\Delta$ PA = 180 degrees) would require a significant amount of energy to change the angular momentum of the co-rotating gas, which is unlikely to be provided by internal processes, especially for large gas masses (e.g. \citealt{dumas07}, \citealt{davies14}).
Simulations of internal dynamical perturbations, for example of nuclear spirals generated by asymmetries in the galactic potential \citep{maciejewski04} or gas flow in barred spiral galaxies \citep{regan99} do not seem to generate counter-rotating gas structures. 

The main theories for the formation of KDCs involve the inflow of material (stars/gas) with a distinct angular momentum into the centre of the galaxy. However there is more than one mechanism than can drive this material inwards to produce the KDCs and can result in two different KDC populations. In optical wavelengths, \cite{mcdermid06} found that there is one population of KDCs which are typically large (diameter $\geq$ 1 kpc) with old stellar populations (age $\geq$ 10 Gyr) found in slow-rotating early-type galaxies. In this case the stellar population within the KDC and in the main body of the galaxy show little or no difference. This can be explained if the distinct core was formed a long time ago and the differences in origin have already been diluted or if the KDC is a superposition of stellar orbits and caused by the geometry of the galaxy. It has been suggested (e.g. \citealt{hernquist&barnes91}) and shown from simulations (\citealt{bois11}, \citealt{tsatsi15}), that these large KDCs can be formed by high mass ratio mergers (1:1 or 2:1) between disc galaxies with prograde or retrograde orbits. A non-merger scenario has also been suggested from simulations, where counter-rotating stellar discs are formed early in the history of the galaxy from the infall of gas with misaligned spin through large scale filaments \citep{algorry14}. The second population of KDCs found by \cite{mcdermid06} consists of smaller structures (diameter $<$ 1 kpc), found mostly in fast-rotators, typically close to counter-rotating in relation to the outer parts of the galaxy and showing younger stellar populations. These KDCs were likely formed by recent accretion of gas which then formed stars in-situ. The fact that the distinct core has an angle PA$_{\rm kin} \sim 180^{\circ}$ in relation to the outer parts of the galaxy may indicate that in fact dissipation processes are important in the formation of these smaller KDCs.

MCG--6-30-15 is one of the galaxies with higher AGN luminosity among the galaxies with known stellar distinct cores, and therefore we are interested in understanding the impact that the formation of the KDC may have on AGN fuelling. \cite{davies14} studied a sample of S0 galaxies and found that the ones with AGN typically have gas detected on large scales ($>$ 1 kpc), while the inactive ones do not. They also find a fraction of sources with counter-rotating gas consistent with what is expected for external accretion. These observations suggest that in S0 galaxies, AGN are typically fuelled by external accretion. They have also shown that in two of the galaxies they observed (IC 5267 and NGC 3368) the external counter-rotating gas can reach the inner tens of parsecs of the galaxy.

In our previous work \citep{raimundo13}, we studied the stellar kinematics out to $r < 250$ pc in MCG--6-30-15. Since our data was limited to the H-band and to a 3${''}$ x 3${''}$ field-of-view, we only had one tracer for the ionised gas on the form of [Fe II] 1.64 $\micron$ emission, mainly restricted to the spatial region of the counter-rotating core. In the present work our goal is foremost to probe the dynamics of the molecular gas, for which there are no previous observations. 
In this work we will investigate if the formation of the counter-rotating core is associated with inflow of gas into the central regions of the galaxy and if this gas can create or replenish the molecular gas reservoir for AGN fuelling. We will first describe the SINFONI and VIMOS observations and the data reduction and analysis. We will then present the results on the stellar, ionised and molecular gas properties in the galaxy. Using the stellar and gas dynamics we set new constraints on the black hole mass. Finally we will discuss, based on these observations, the origin of the stellar counter-rotating core and its effect on AGN fuelling.

We adopt the standard cosmological parameters of H$_0 = 70$ km s$^{-1}$ Mpc$^{-1}$, $\Omega_{\rm m} = 0.27$ and $\Omega_{\rm \Lambda} = 0.73$. The scale for our measurements is approximately 0.158 kpc/'' at a redshift $z = 0.0077$ (\citealt{fisher95} value quoted at the NASA/IPAC Extragalactic database). The distance to MCG--6-30-15 used in this paper is 33.2 Mpc \citep{wright06}.\\
\section{Data Analysis}
\label{sec:analysis}
\subsection{SINFONI data reduction}
\begin{figure*}
\centering
\epsfig{file=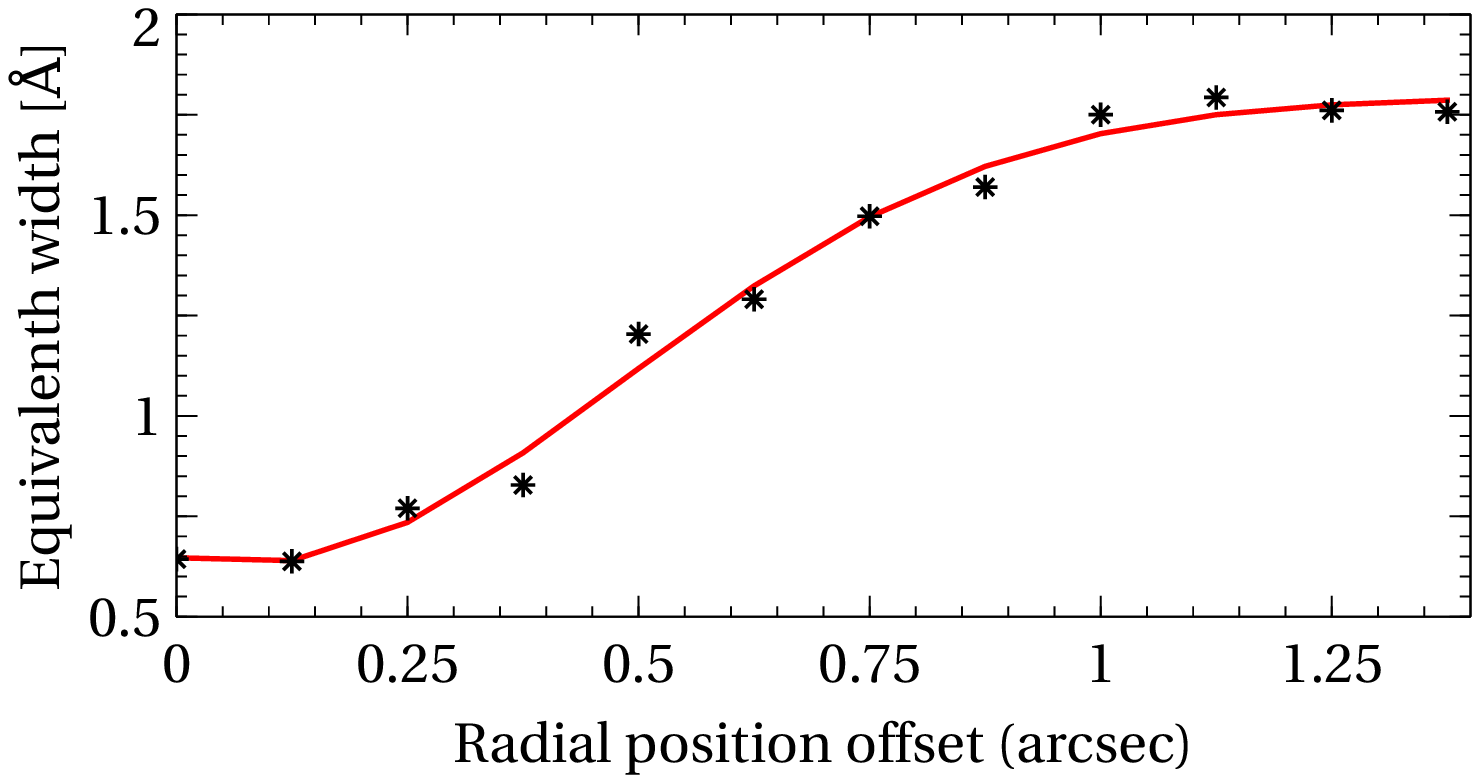,width=0.9\columnwidth,clip=}\hspace{-0.1cm}
\epsfig{file=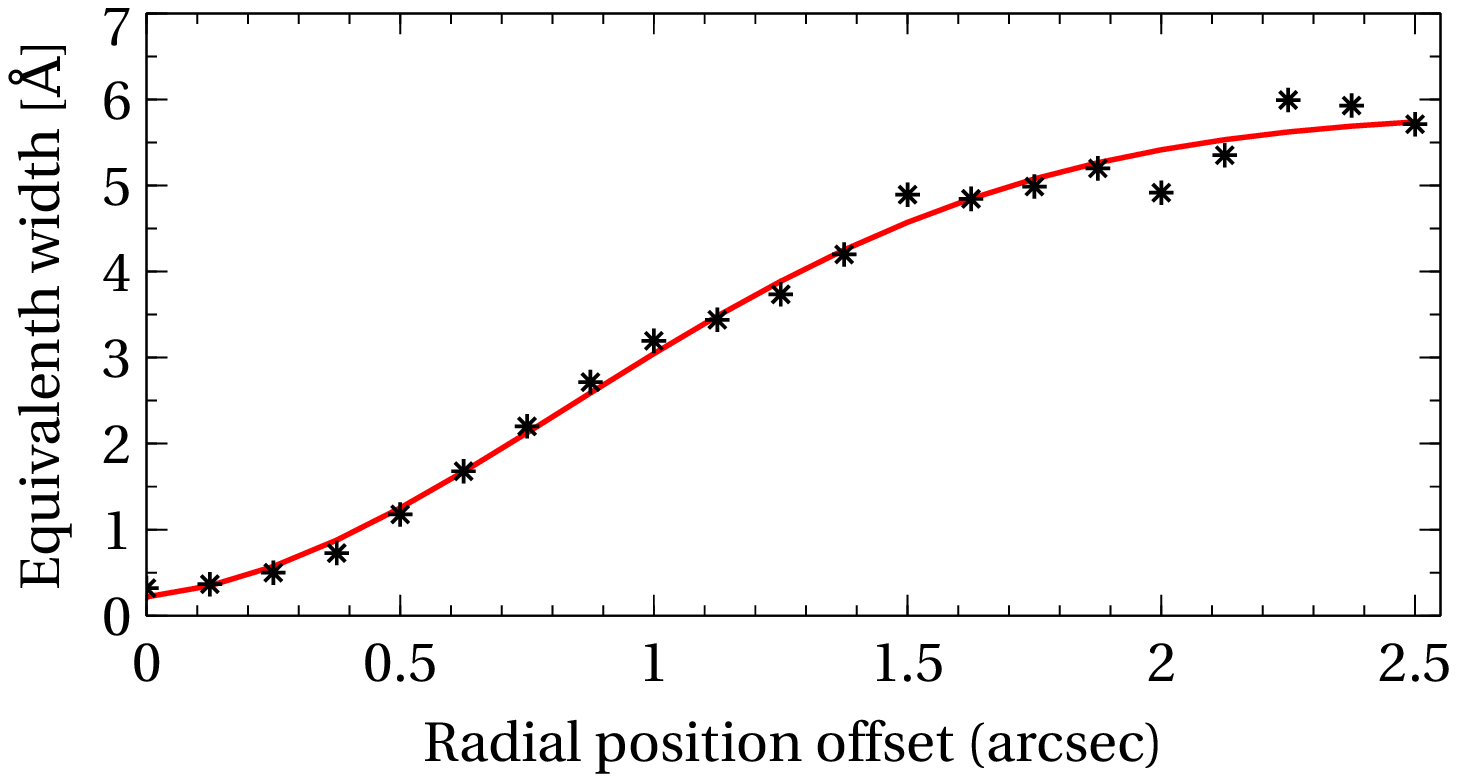,width=0.9\columnwidth,clip=}
\caption {Left: Equivalent width of the CO (4-1) 1.578 $\micron$ line in the H-band as a function of radius. Right: Equivalent width of the $^{12}$CO (3-1) 2.3227 $\micron$ line as a function of radius. The centre indicates the position of the black hole and is measured from the peak of the broad hydrogen brackett emission. The equivalent widths are measured in annuli of 1 pixel (0${''}$.125) width. The red lines in each of the plots are the best fit gaussian curves to the data points. They are used in the subsequent decomposition of AGN and stellar continuum for the H and K-bands.}
\label{eq_width}
\end{figure*}
\begin{figure*}
\centering
\hspace{-1.5cm}\epsfig{file=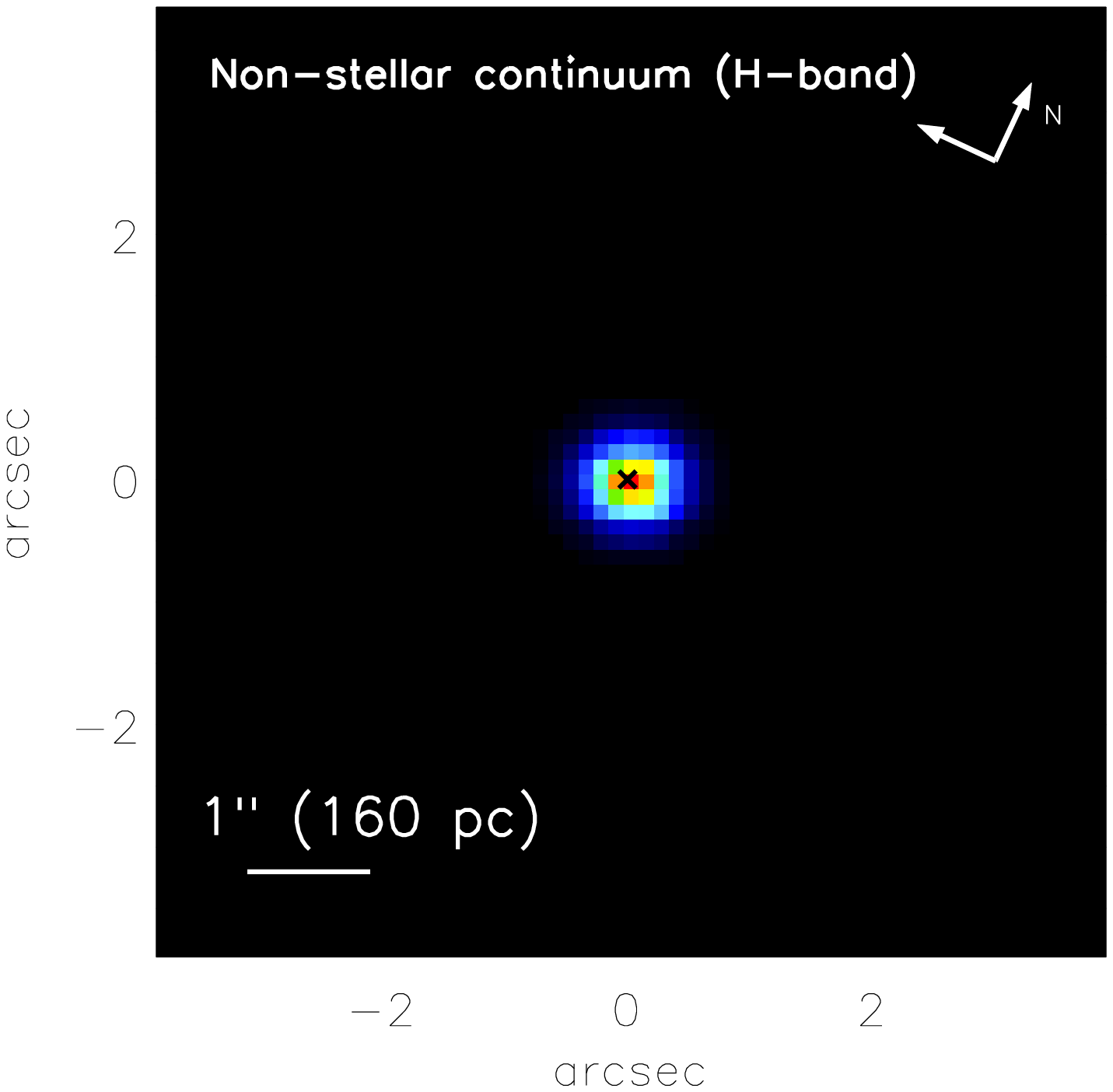,width=0.34\linewidth,clip=}\hspace{-1.8cm}
\epsfig{file=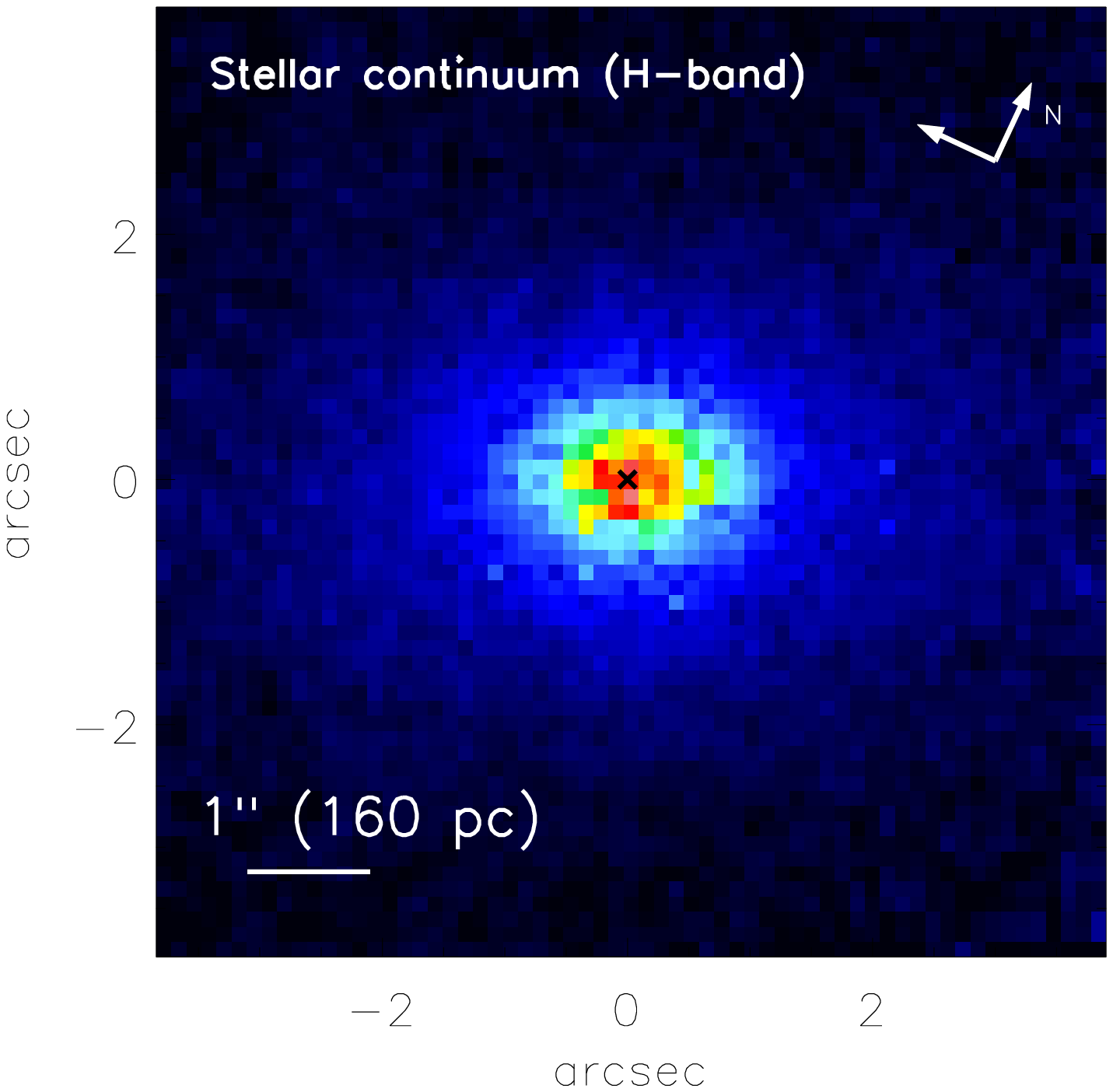,width=0.34\linewidth,clip=}\hspace{-1.8cm}
\epsfig{file=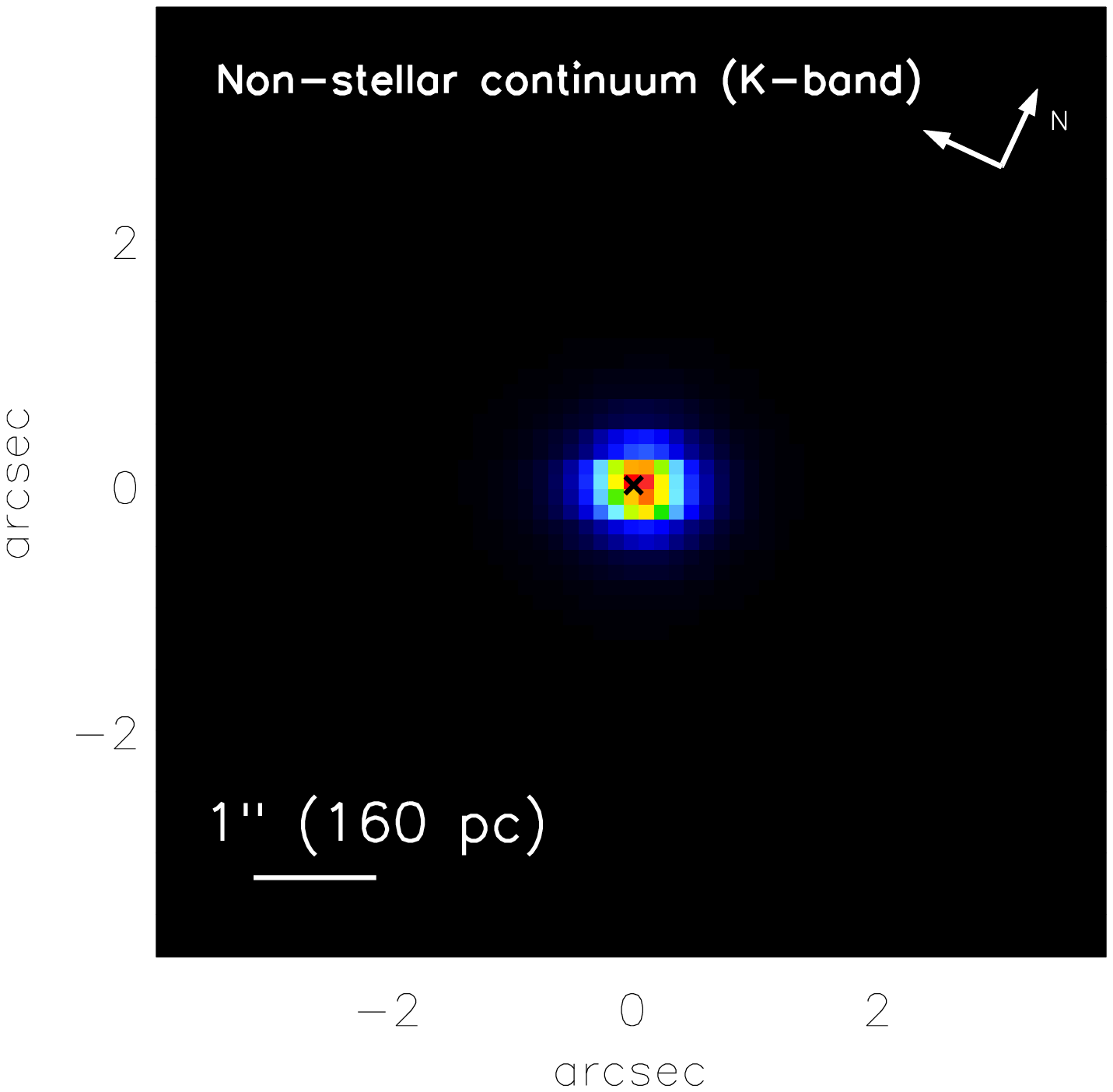,width=0.34\linewidth,clip=}\hspace{-1.8cm}
\epsfig{file=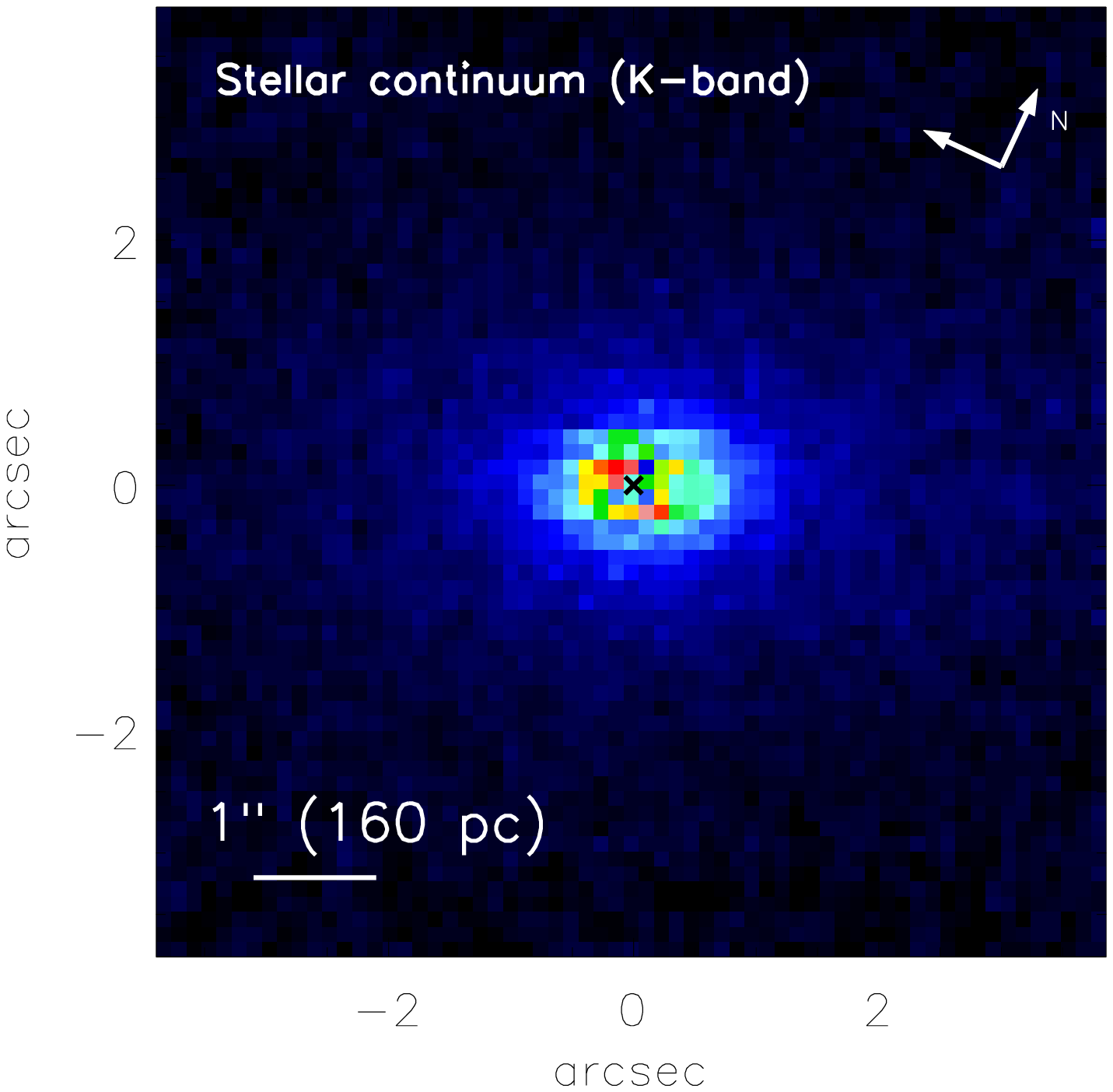,width=0.34\linewidth,clip=}\hspace{-1.8cm}
\caption {Non-stellar and stellar continuum maps after the continuum decomposition based on the stellar absorption lines equivalent widths. The two plots on the left show the H-band continuum and the two plots on the right show the K-band continuum. The position of the AGN is indicated by the black cross.}
\label{stellar_agn_cont}
\end{figure*}
The SINFONI/VLT H-band data with R$\sim$3000, were taken on the night of 2 June 2014 with a pixel scale of 0${''}$.25$\times$0${''}$.125, sampling a field-of-view of 8${''}\times$ 8${''}$. Each exposure was of 300 seconds, in an object (O) and sky (S) sequence of OSO. The total on-source exposure time was 55 min.
The SINFONI/VLT K-band data with R$\sim$4000, were taken on four nights (4, 12 June and 2, 3 July 2014) with a pixel scale 0${''}$.25$\times$0${''}$.125 and also sampling a field-of-view of 8${''}\times8{''}$. Each exposure was of 300 seconds, in an object (O) and sky (S) sequence of OSO. 
The total on-source exposure time was 2h30 min. The data reduction was done with the ESO pipeline for SINFONI {\sc esorex} version 3.10.2 and complemented with independent {\sc IDL} routines. The two main extra routines used were the implementation of the sky emission subtraction described in \cite{davies07} and a bad pixel removal routine for 3D data cubes \citep{davies10} - a 3D version of LA cosmic \citep{vandokkum01}. For the analysis, the cubes were shifted considering the offsets recorded in their headers, combined and re-sampled to a spatial scale of 0${''}$.125$\times$0${''}$.125. The observations were taken with a rotation angle of 25 degrees to obtain a better spatial sampling along the major axis of the galaxy. In the plots in this paper the x-axis will be parallel to the major axis of the galaxy. The compass in each of the plots indicates the relative position in the sky. A more detailed description of the typical data reduction steps for the H-band can be found in \cite{raimundo13}.

The point spread function (PSF) is determined from the intensity of the broad component of the atomic hydrogen Brackett recombination lines. The broad line region where these lines are produced is unresolved in our observations and therefore the line width of the broad hydrogen emission provides a method to estimate the PSF. Several transitions of this Brackett series are observed both in the H and K-bands. In the H-band there are five strong lines which we use to measure the PSF and also later in this section to remove the AGN contribution from the spectra. In the K-band we use the strongest line of the (7-4) $\lambda$ = 2.166 $\micron$ transition (Br$_{\gamma}$). We fit these lines with a double Gaussian (one narrow Gaussian to model the core and one broader Gaussian to model the extended emission), to empirically model the PSF. The full width half maximum of each of the gaussians is, for the H-band (FWHM = 3.6 pixels $\sim0{''}.4$) and (FWHM = 6.7 pixels $\sim0{''}.8$) and for the K-band (FWHM = 3.9 pixels $\sim0{''}.5$) and (FWHM = 8.6 pixels $\sim1{''}$). As a second check, we also use the unresolved non-stellar continuum to determine the PSF. The FWHM of the gaussians for the non-stellar continuum is, as expected, similar to what was determined with the broad Brackett emission: H-band (FWHM = 3.3 pixels $\sim0{''}.4$) and (FWHM = 5.4 pixels $\sim0{''}.7$); K-band (FWHM = 3.9 pixels $\sim0{''}.5$) and (FWHM = 8.7 pixels $\sim1{''}.1$).

The instrumental broadening is calculated by selecting and measuring the spectral broadening of an unblended sky emission line. The broadening varies as a function of wavelength and spatial position in our data cube, therefore we select sky lines close to the wavelength of interest for our measurements. For the K-band we do this calculation for sky lines close to the wavelength of the [Si VI] 1.963 $\micron$, Br$_{\gamma}$ 2.166 $\micron$, 1-0 S(1) H$_{2}$ 2.1218 $\micron$ and [Ca VIII] 2.3211 $\micron$ emission lines for every spaxel in the sky emission cube. The instrumental broadening at the wavelength of each line is determined by doing a field-of-view spatial average of the best fit gaussian $\sigma$. We obtain values that vary between $\sigma_{\rm inst}\sim$ 42-48 km/s in the K-band. The spectral resolution at the wavelength of the 1-0 S(1) H$_{2}$ emission is HWHM $\sim$ 3.3 \AA. For the H-band we obtain values of $\sigma_{\rm inst}\sim$ 44-54 km/s which at the wavelength of the [Fe II] 1.644 $\micron$ emission corresponds to HWHM $\sim$ 3.2 \AA\thinspace and a resolution of R$\sim$2500.
\begin{figure}
\centering
\includegraphics[width=0.8\columnwidth, height=8cm]{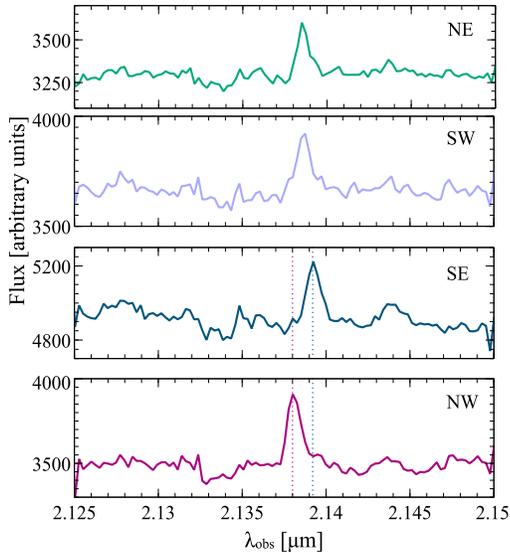}
\caption {H$_{2}$ 2.1218 $\micron$ emission from four regions of $5 \times 5$ pixels ($0{''}.625 \times 0{''}.625$) in the field-of-view, from top to bottom: North-East, South-West, South-East and North-West from the nucleus. Relative to the images in Fig. \ref{stellar_agn_cont} these regions are located at the left, right, top and bottom of the AGN position. The wavelength range in the plot shows the H$_{2}$ gas emission line. The emission at the SE and NW of the nucleus shows a clear redshift and blueshift. The vertical dashed lines indicate the position of the peak of the emission in each of these two regions. The x-axis is the observed wavelength.}
\label{H2}
\end{figure}
As mentioned above, the broad hydrogen recombination lines of the Brackett series are present in both the H and K-bands. In the H-band several of these lines overlap with stellar absorption lines of interest and therefore it was necessary to remove them before determining the stellar kinematics. The procedure was similar to the one described in \cite{raimundo13}, where the broad emission lines were fitted simultaneously, while fixing their rest-frame wavelengths to a common redshift and relative intensities to the ones predicted theoretically using as an approximation \cite{hummer&storey87} for case B recombination. In the K-band only two broad emission lines were observed and they did not overlap with the wavelength range of the CO stellar absorption lines or the emission lines. For the K-band it was not necessary to remove the AGN broad hydrogen emission prior to the analysis of the stellar properties. 

The flux calibration is done using the 2MASS measured fluxes in the H-band and Ks-band for a 5${''}$ diameter aperture. To determine the 2MASS fluxes we use the zero magnitude calibrations in \cite{cohen03}. We then compare the 2MASS fluxes with the median number counts of the integrated spectra in our data cubes, for the same aperture. In our data cubes the median number counts are determined within the waveband of [1.504-1.709] $\micron$ for the H-band and [1.989-2.316] $\micron$ for the K-band. These wavebands correspond to the wavelengths at which the transmission in the 2MASS filters is higher than 0.2. By using different apertures in 2MASS we determine that our calibration is consistent to $\sim$20 per cent.
\subsection{VIMOS data reduction}
To investigate the emission in the optical wavelengths we analysed archival IFU data of MCG--6-30-15 from 2004 using VIMOS/VLT. The data were taken with the old HR-blue grism (replaced by a new set of grisms in 2012) with a wavelength range 4200 \AA\ - 6150 \AA\ and a pixel scale of 0${''}$.67/pixel in a field-of-view of 27${''} \times 27{''}$. We reduced the data in each of the four quadrants using the ESO VIMOS pipeline and the standard recipes for doing the bias subtraction and flat field correction, bad pixel removal, wavelength calibration and relative flux calibration with \textit{vmbias}, \textit{vmifucalib}, \textit{vmifustandard} and \textit{vmifuscience}. In quadrant three we observed that there were zig-zagging patterns on its reconstructed image after \textit{vmifuscience} which is caused by a fibre misidentification during \textit{vmifucalib}. This has been observed in other datasets (e.g. \citealt{rodriguez-zaurin11}). To solve this problem we did not use a fibre identification file but ran the recipe with a `first guess' fibre identification method, specifically the arc lamp lines are identified using models of the spectral distortions as first-guesses. Doing such a fibre identification improved the result but did not completely solve the problem. We then decided to change the maximum percentage of rejected positions in the fibre spectra tracing to a conservative value, i.e. when the fraction of rejected pixel positions was more than 17 per cent, the fibre was flagged as `dead'. As we only have one exposure for each quadrant this results in some fibres being excluded (having no associated data) in the final data cube which is not ideal but preferable to having misidentified fibres. The final reconstructed image of the field-of-view and the final cube were obtained by combining the reduced data on the four quadrants using \textit{vmifucombine} and \textit{vmifucombinecube} respectively.
\begin{figure}
\centering
\epsfig{file=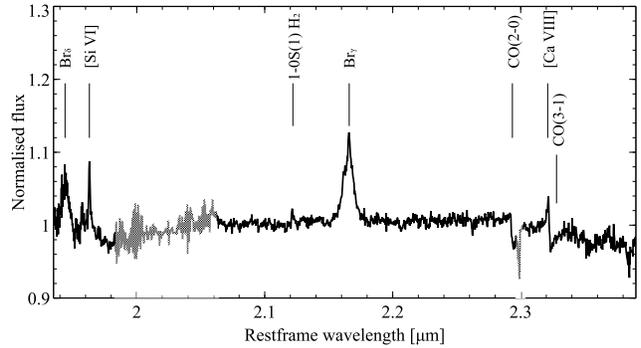,width=1.0\linewidth,clip=}
\caption{Spectrum of the nuclear region integrated in a 0${''}$.625 radius aperture. The inner r $<$ 0${''}$.25 are excluded. The regions in grey indicate the presence of residuals from the telluric absorption correction.}
\label{central_spec_K}
\end{figure}
Final fibre-to-fibre transmission corrections were done using dedicated \textsc{idl} routines and the fibre-to-fibre integrated flux in the sky line of [O I] $\lambda$ 5577 \AA. The same sky line was also used to verify the wavelength calibration and to determine the instrumental broadening which is $\sigma$ = 22 km/s. The scope of our analysis is to determine the relative line intensity map and velocity profile for the [O III] emission and therefore no absolute flux corrections were done. Sky lines are observed in the spectra, however as they are not confused with the object emission they are not subtracted in the data cube. 

Several emission lines are observed in the spectra: the strongest emission is [O III] 5007 \AA\ but H$_{\beta}$ 4861 \AA\ with a nuclear broad component and a more spatially extended narrow component is also detected. Stellar absorption lines albeit with a lower signal-to-noise ratio are also observed. The broad H$_{\beta}$ component originates from the AGN and is therefore unresolved in our data. We use its flux distribution to obtain an estimate for the PSF and the spatial resolution. Fitting the broad H$_{\beta}$ flux spatial distribution with a circular gaussian we find FWHM = 1.7 pix which for our scale of 0.67${''}$/pix corresponds to FWHM = 1.${''}$1.
\begin{figure*}
\centering
\includegraphics[width=1.7\columnwidth]{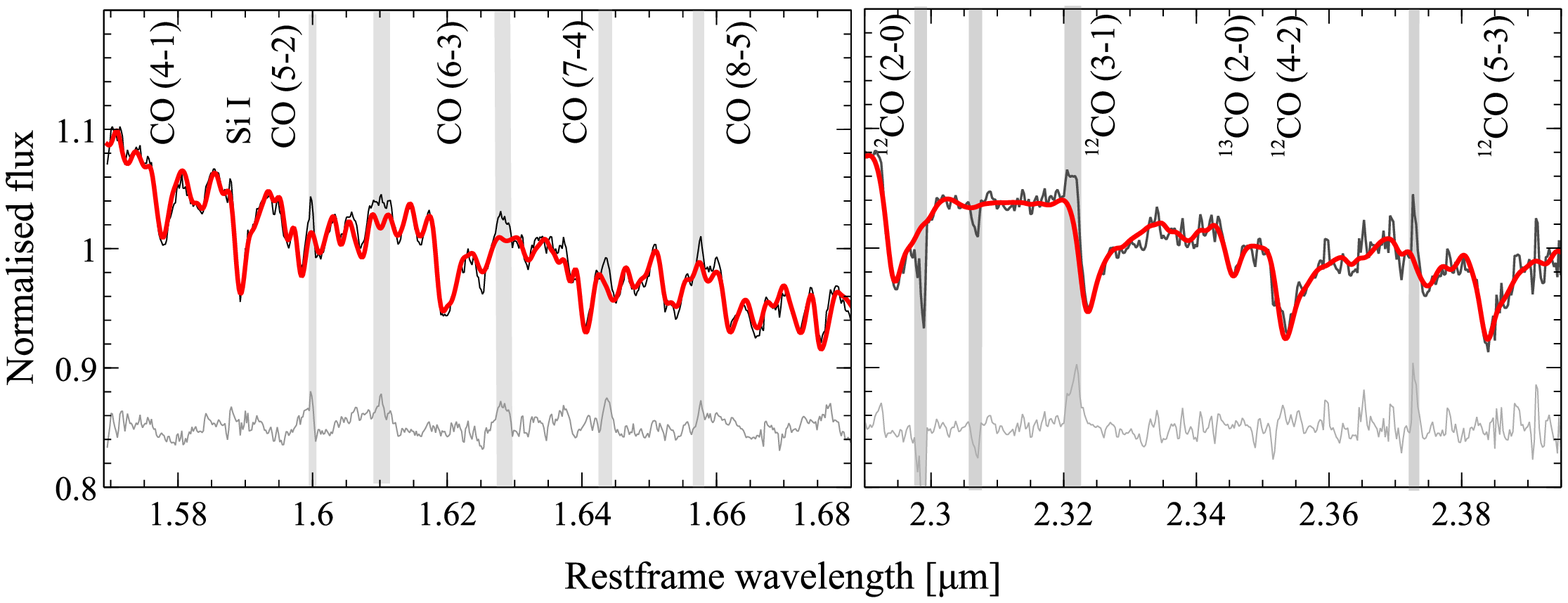}
\caption {Data (black line), best fit model (red line) and residuals (bottom grey spectrum shifted by a constant factor to be visible in the plot) for the galactic spectrum fit with pPXF. Left: H-band integrated emission in a r $ < 2{''}$ region, excluding the inner r $< 0{''}.5$. Right: K-band integrated emission in a r $< 2{''}.5$ region excluding the central r $< 0{''}.25$. The vertical shaded bands indicate the wavebands excluded from the fit. The x-axis is the rest-frame wavelength after correcting for the systemic velocity of the galaxy.}
\label{central_H}
\end{figure*} 
\subsection{AGN and stellar continuum}
\label{sec:continuum}
The observed near infrared continuum is a combination of stellar and non-stellar contributions. To evaluate the contribution of each of these components to the total continuum in the H-band we use the equivalent width of the CO (4-1) 1.578 $\micron$ stellar absorption line, while in the K-band we use the $^{12}$CO (3-1) 2.323 $\micron$ line (Fig.~\ref{eq_width}). Close to the AGN the stellar absorption line will be diluted and we expect the equivalent width to decrease as the distance from the AGN decreases. In the regions further away from the centre of the galaxy (r $\gtrsim$ 1${''}$.3), where the AGN contribution is negligible, the equivalent width will be due to stellar processes only and determines the intrinsic line equivalent width ($W_{\rm intr}$) (\citealt{davies07}, \citealt{burtscher15}). The plots in Fig.~\ref{eq_width} give us the observed equivalent width as a function of radius ($W_{\rm obs}$). The fraction of the continuum that is due to non-stellar processes will then be:
\begin{eqnarray}
f_{\rm AGN}(r)=1-\frac{W_{\rm obs}}{W_{\rm intr}}(r),
\end{eqnarray}
and we can determine the AGN contribution to the continuum at each radial position.
The total continuum at each position of the field-of-view is fitted using a second degree polynomial. The AGN continuum is calculated by multiplying the total continuum by $f_{\rm AGN}(r)$. The stellar continuum is determined from the depth of the stellar absorption lines present in our spectra. The continuum decomposition for the H and K-bands is shown in Fig.~\ref{stellar_agn_cont}. As expected the non-stellar continuum is associated with the AGN and is limited to a small region close to the nucleus (r $\lesssim$ 0.5 - 0.8 arcsec). The stellar continuum shows a more extended distribution.
\section{Results and Discussion}
\label{sec:results}
\begin{figure*}
\centering
\epsfig{file=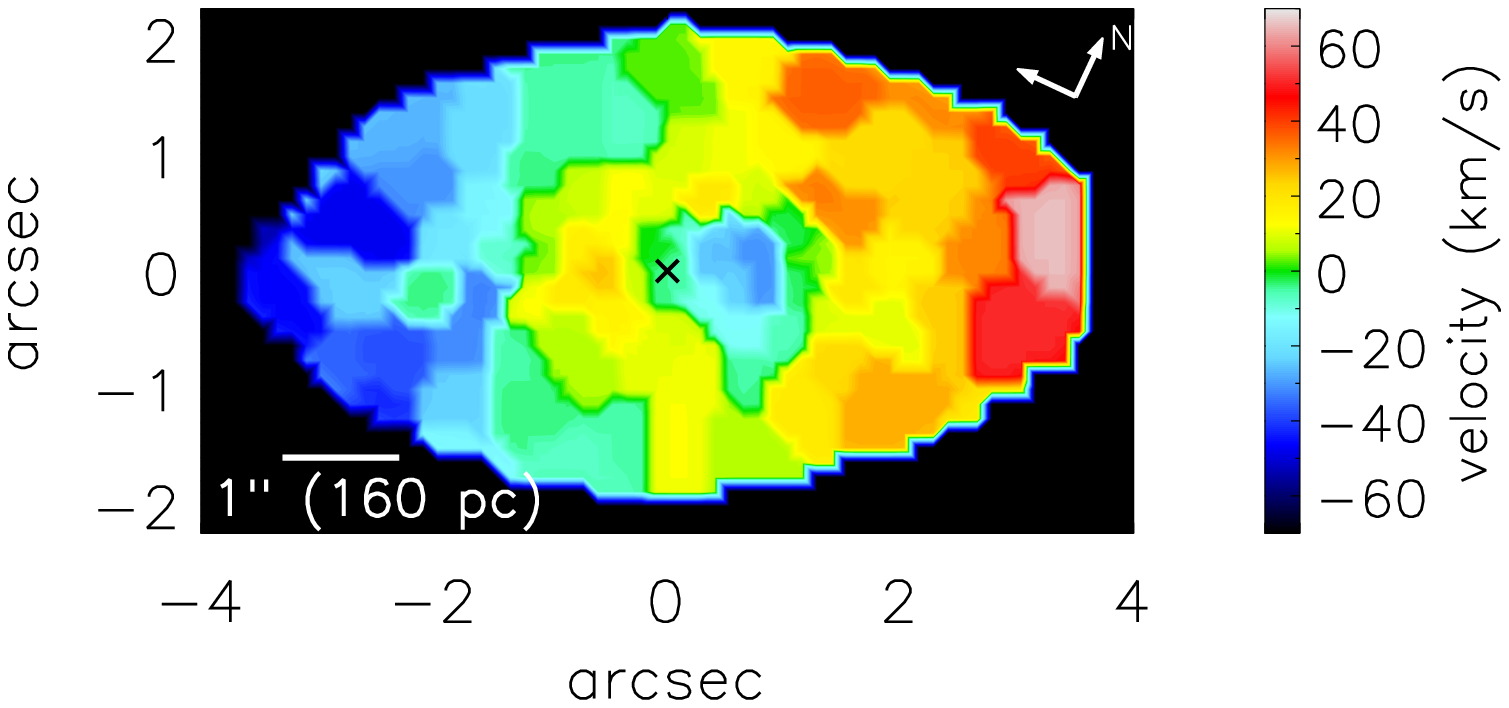,width=0.51\linewidth,clip=}
\hspace{-0.7cm}
\epsfig{file=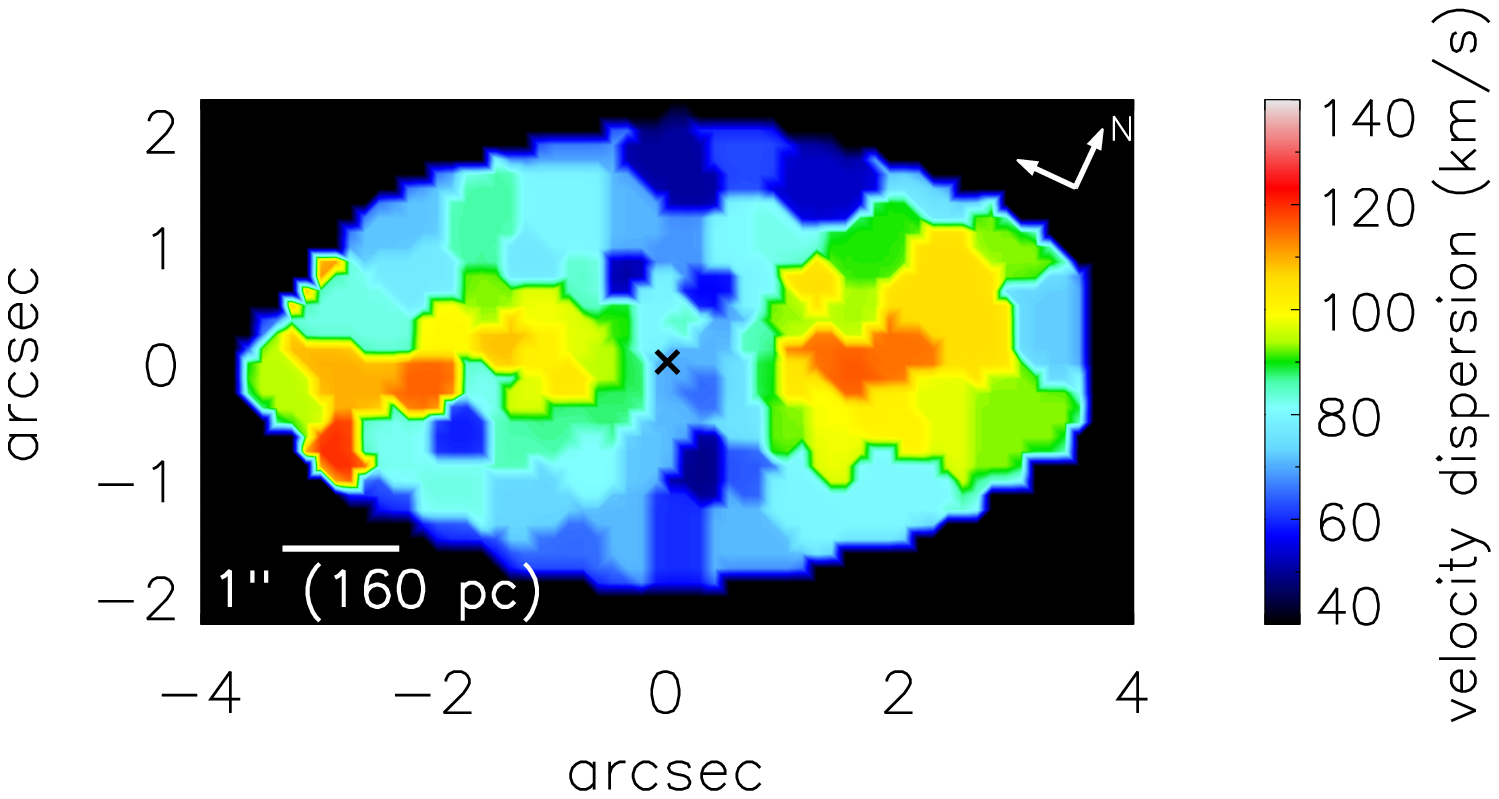,width=0.51\linewidth,clip=} \\
\vspace{-1.5cm}
\epsfig{file=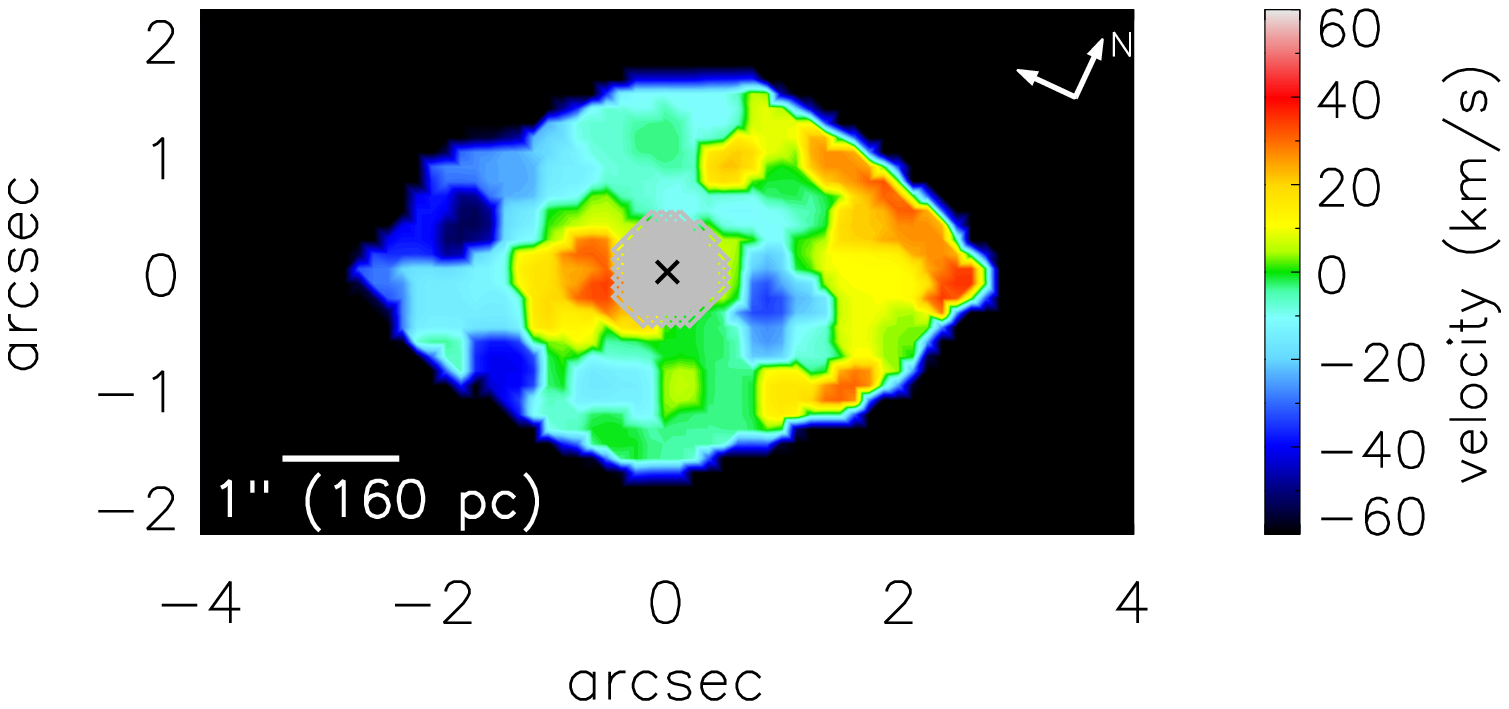,width=0.51\linewidth,clip=}
\hspace{-0.7cm}
\epsfig{file=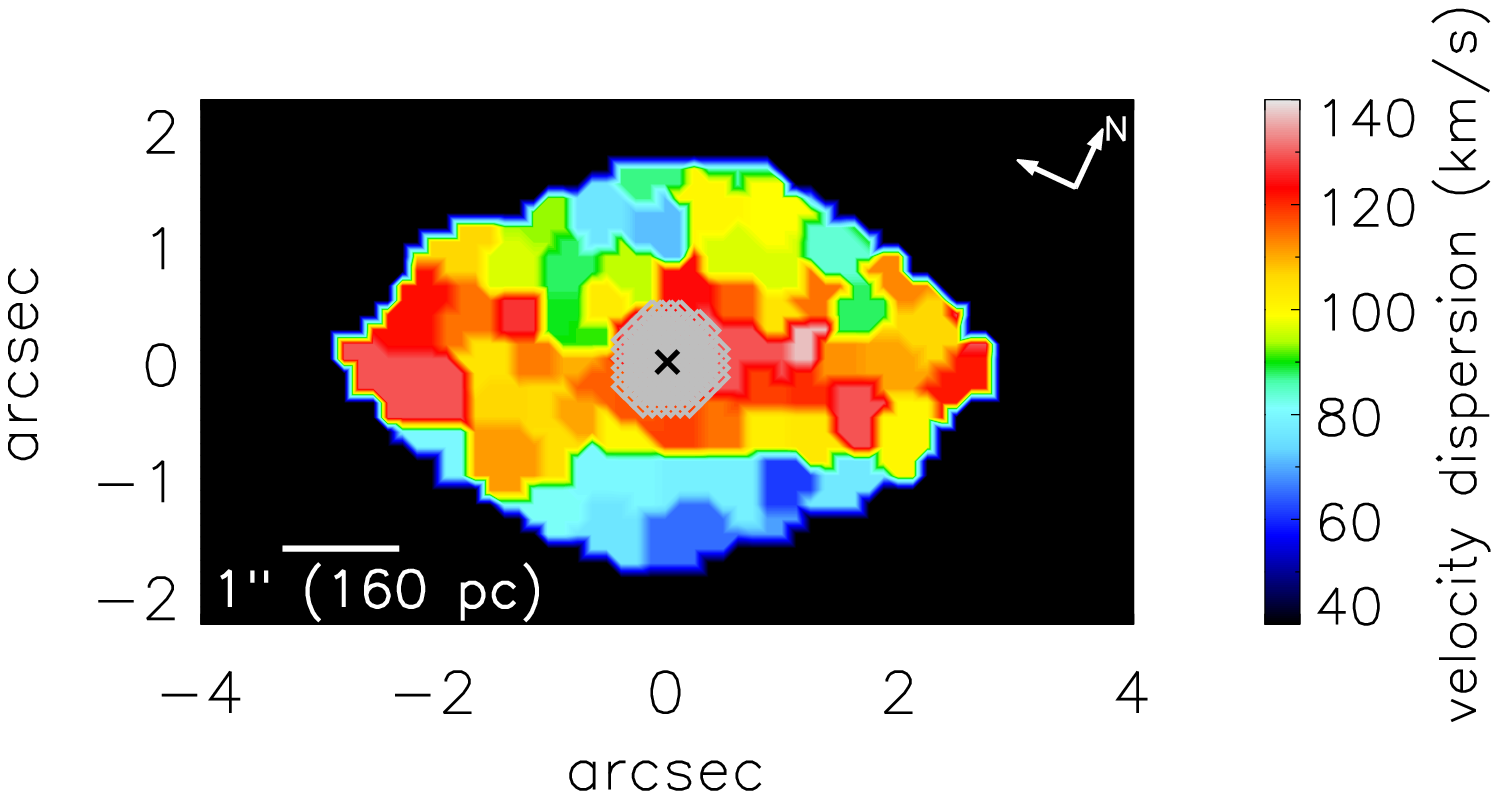,width=0.51\linewidth,clip=}
\caption {Stellar line of sight velocity and velocity dispersion maps for the H-band (top) and K-band (bottom). Left: Line of sight velocity field corrected for the galaxy's systemic velocity of 2410 km/s. Right: Velocity dispersion corrected for the instrumental broadening. The map was binned to a minimum S/N of 5. The central regions were masked out due to the low S/N ratio of the stellar absorption features and the outer regions were masked out due to a combination of low flux and low S/N.}
\label{vel_field}
\end{figure*}
In the H-band, after subtracting the broad hydrogen emission lines, various stellar absorption line features such as the CO absorption bands, Si I and Mg I are observed in the spectra. The only emission line observed, apart from the hydrogen recombination lines is the [Fe II] 1.64 $\micron$ emission line.
In the K-band we detect several stellar absorption features, including the CO absorption bands which are the strongest absorption lines in our spectra. The rovibrational H$_{2}$ 2.12 $\micron$ emission is also detected and a tracer of the molecular gas distribution (Fig.~\ref{H2}). In addition there are three other emission lines from ionised gas that are observed ([Si VI] 1.963 $\micron$, Br$_{\gamma}$ 2.166 $\micron$ and [Ca VIII] 2.3211 $\micron$) and will be analysed in the following sections. The nuclear spectrum of MCG--6-30-15 in the K-band showing the Br$_{\gamma}$ and coronal emission lines can be seen in Fig.~\ref{central_spec_K}. H$_{2}$ is not significantly detected in the nuclear region.
\subsection{Stellar kinematics}
The stellar kinematics are obtained by fitting the observed galaxy spectra using the Penalised Pixel-Fitting (pPXF) method (\citealt{cappellari04}; \citealt{vandermarel&franx93}). The goal of this method, and of our analysis, is to recover the line-of-sight velocity distribution from the stellar absorption features of our observed spectra. This method assumes a parameterisation for the stellar line-of-sight velocity distribution on the form of a Gauss-Hermite series expansion, with the first two moments being the velocity and the velocity dispersion (V, $\sigma$). It then uses a set of stellar templates and a polynomial component to account for the AGN continuum, to find the best-weighted template combination and line-of-sight velocity distribution to fit the input galaxy spectra. Due to the limited S/N of our data we only fit the first two moments of the velocity distribution. 

In the H-band we use the stellar templates of \cite{le11} at a resolution of $R\sim5000$. This library contains only ten G, K and M giant stars, but as described in \cite{raimundo13}, K and M stars provide a good match to our stellar absorption features. In the K-band we use the stellar templates from \cite{winge09} in the near-IR wavelength range of 2.24 - 2.43 $\micron$, at a resolution of R = 5900 obtained with GNIRS. We convolve these spectra with a gaussian of $\sigma$ = 2.3 \AA\thinspace to match them to the lower resolution of our data.

The stellar kinematics from both the H and K bands are consistent with what was found in \cite{raimundo13}. In \cite{raimundo13} we only had H-band data in a $3{''} \times 3{''}$ field-of-view, which allowed us to discover the counter-rotating core but not to study the main body of the galaxy extensively. With the larger field-of-view of our new H and K-band observations ($8{''} \times 8{''}$) we are able to probe larger distances from the nucleus. Fig.~\ref{central_H} shows the fit to the integrated emission in the H and K bands. The black line shows the integrated galaxy spectrum and the red line shows the best fit model from pPXF. The residuals are shown in light grey at the bottom of the plot. Regions with emission lines and telluric subtraction residuals were masked out of the fit and are identified by the vertical shaded bars. 
\begin{figure}
\centering
\includegraphics[width=1.0\columnwidth]{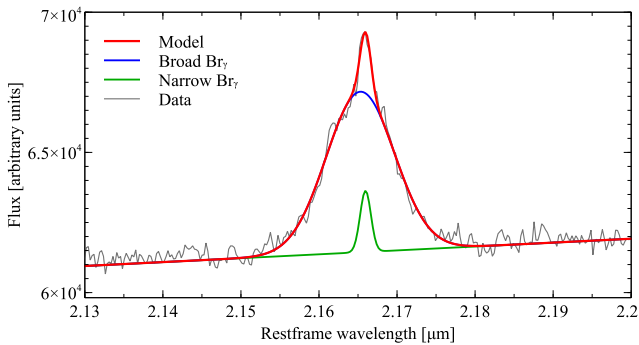}
\caption{Fit to the hydrogen Brackett emission line in the K-band which consists of two components: a broad one associated with the AGN and a narrow one which may a contribution from star formation. The emission in the plot corresponds to an integrated region of 5 x 5 pixels (0.${''}$625 x 0.${''}$625) centred at the spatial position of the AGN. The x-axis is the rest-frame wavelength.}
\label{brackett}
\end{figure}
\begin{figure*}
\centering
\epsfig{file=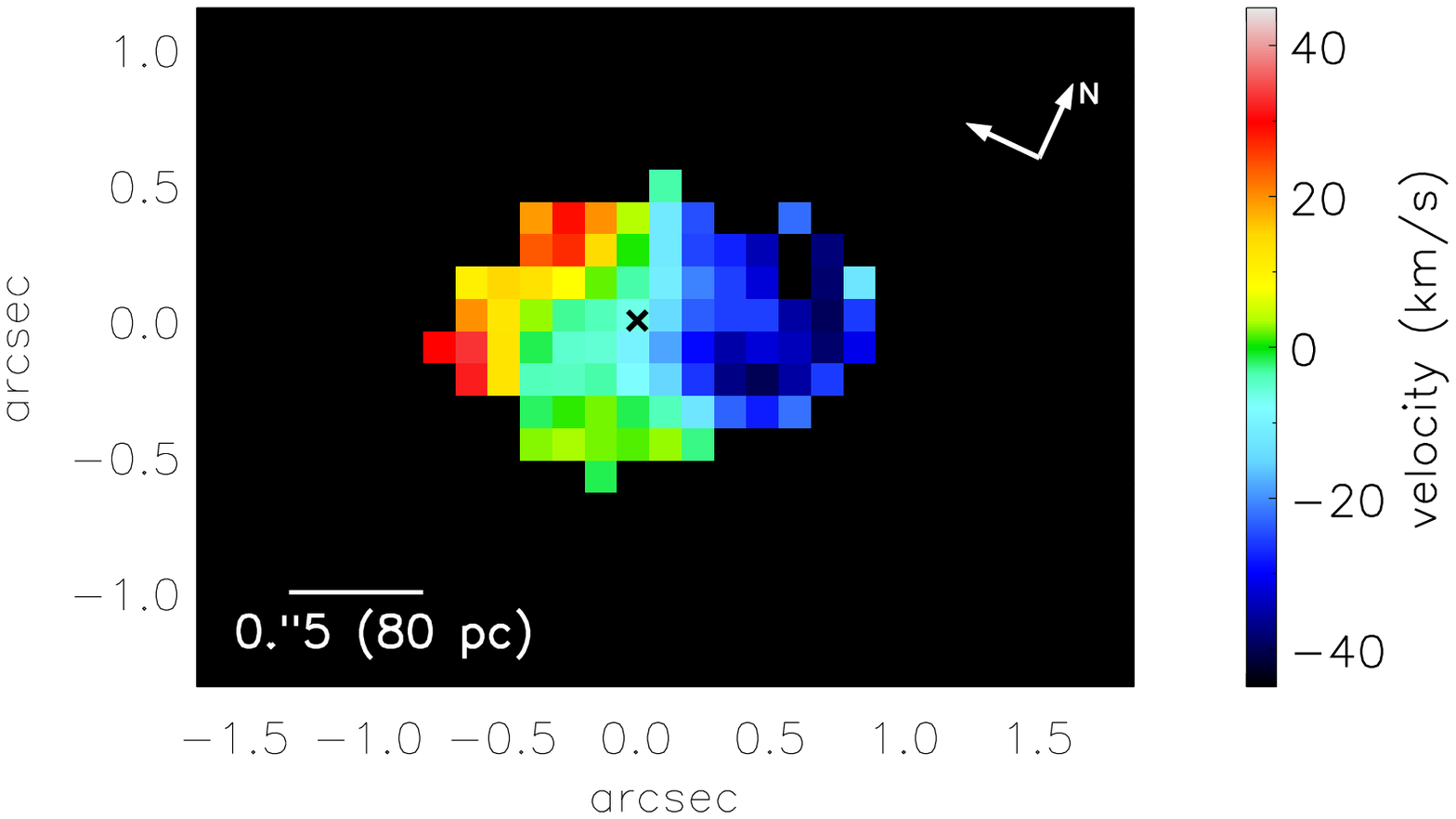,width=0.33\linewidth,clip=}
\epsfig{file=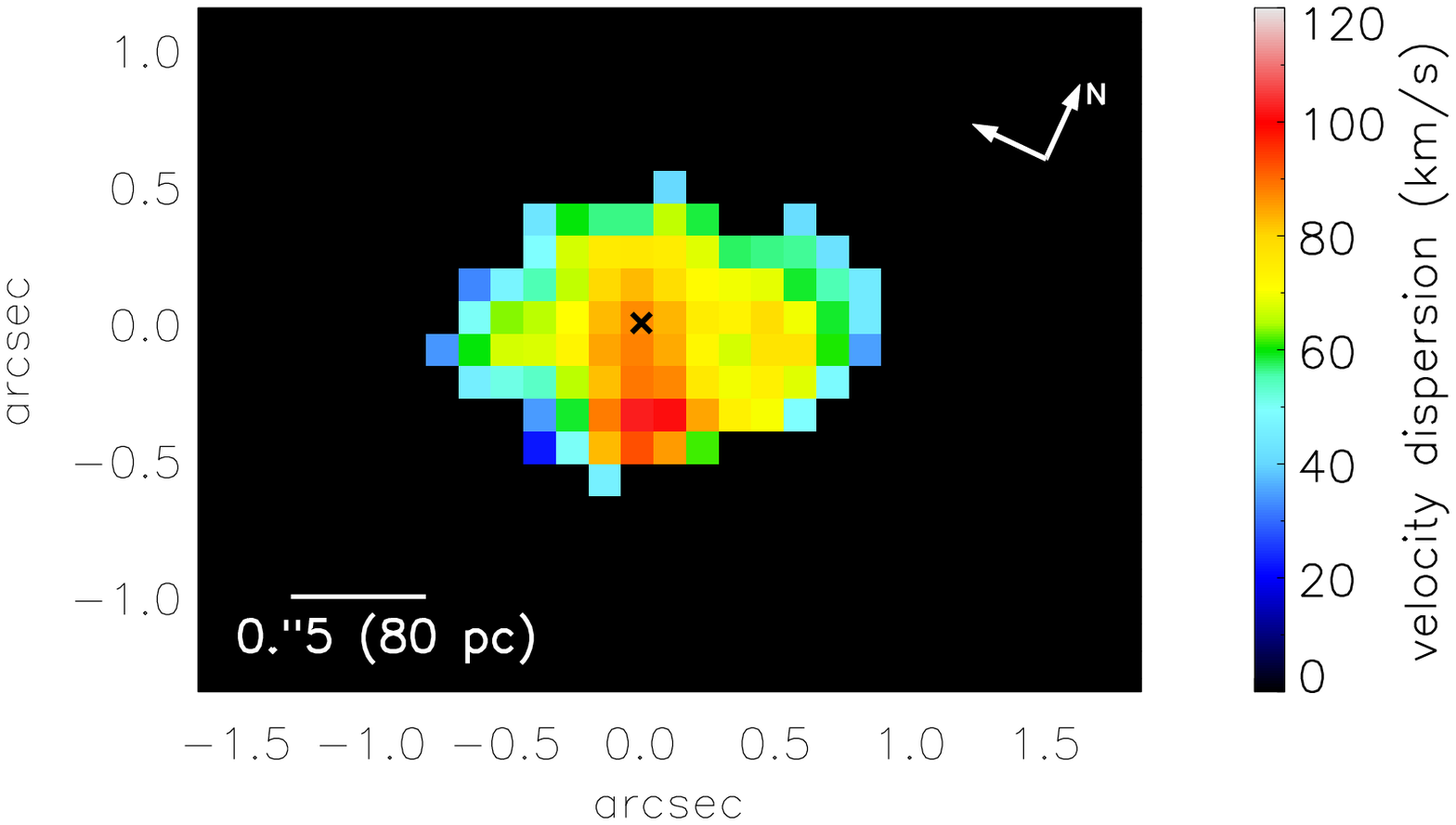,width=0.33\linewidth,clip=} 
\epsfig{file=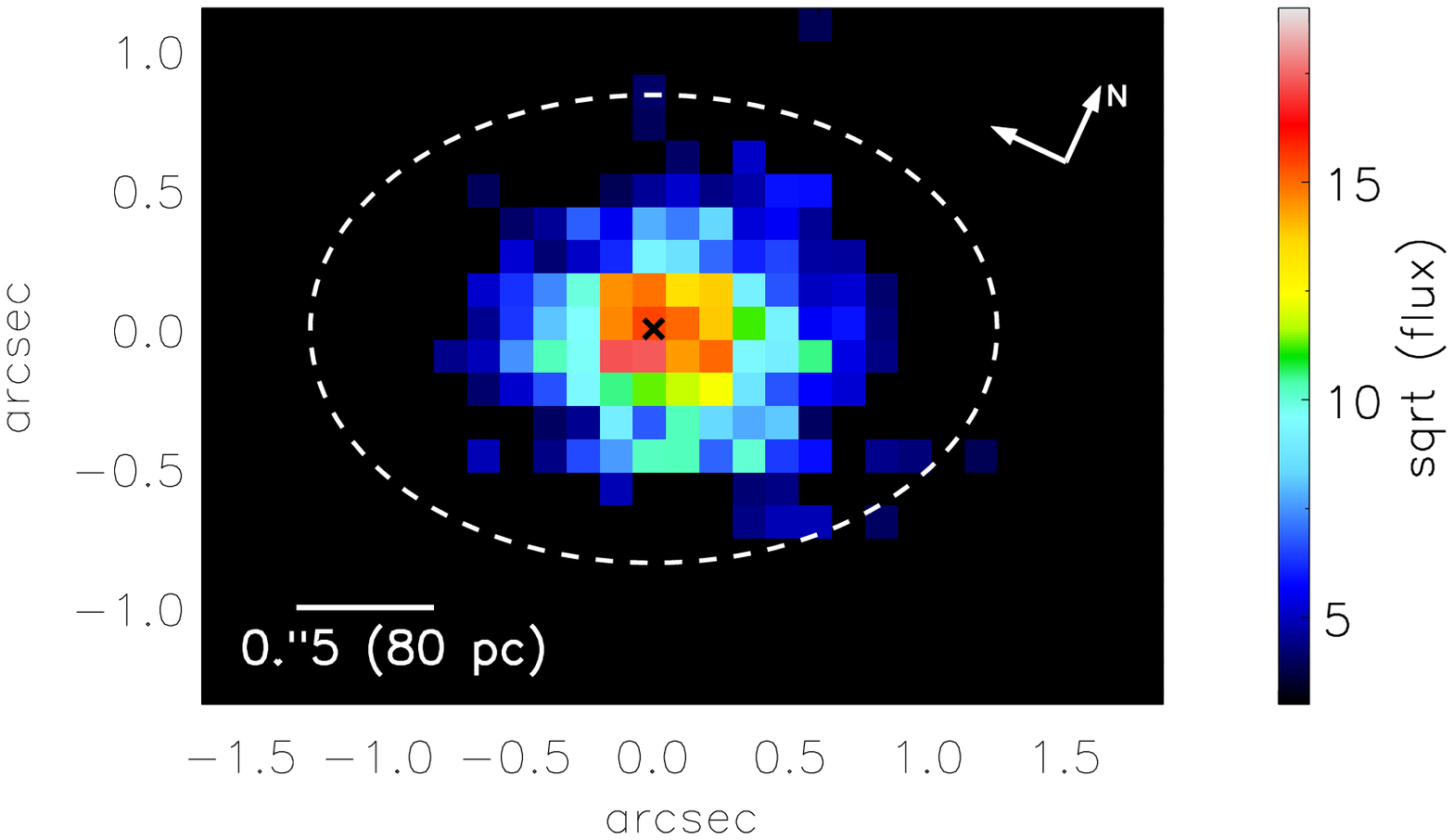,width=0.33\linewidth,clip=} \\[-0.6cm]
\epsfig{file=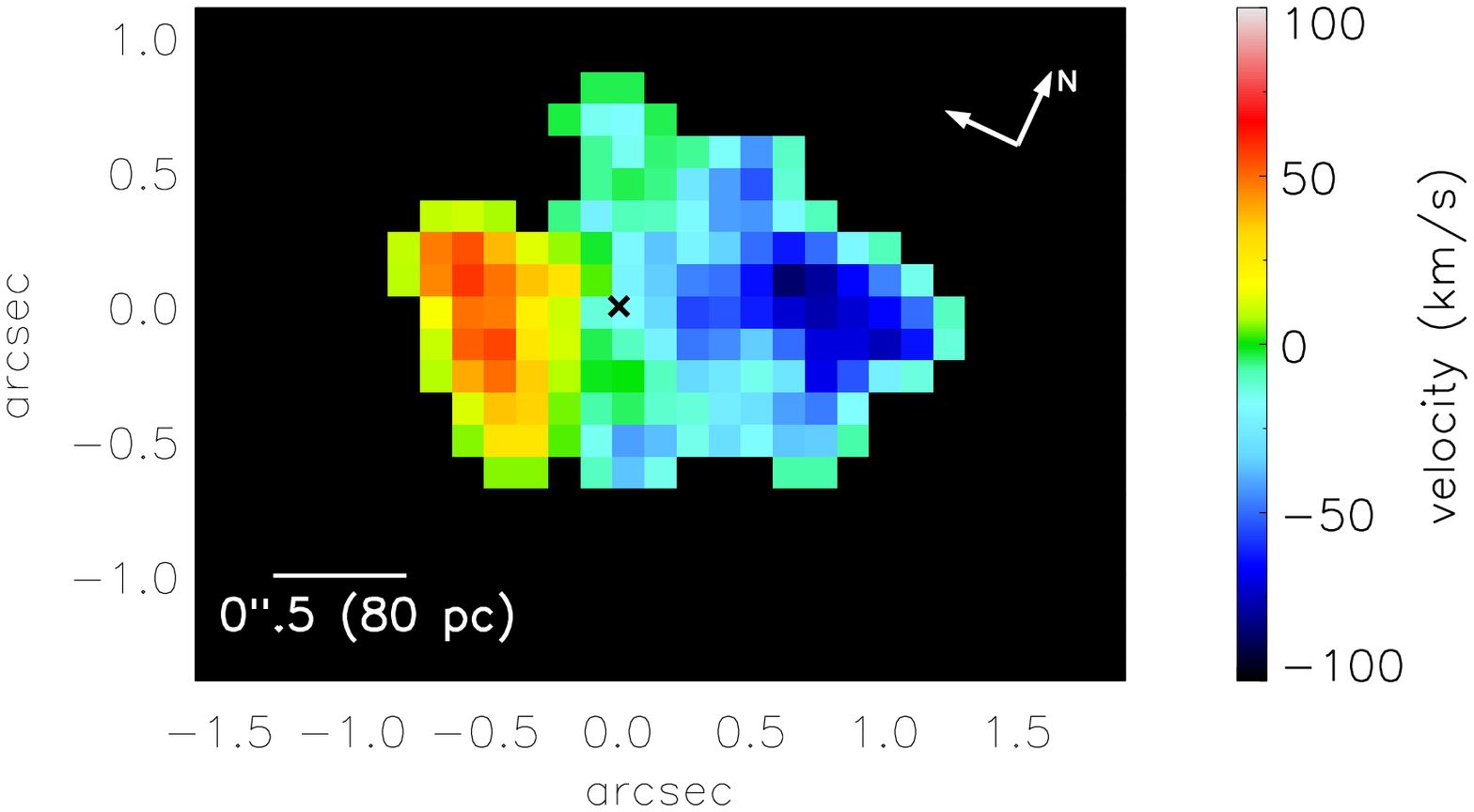,width=0.33\linewidth,clip=}
\epsfig{file=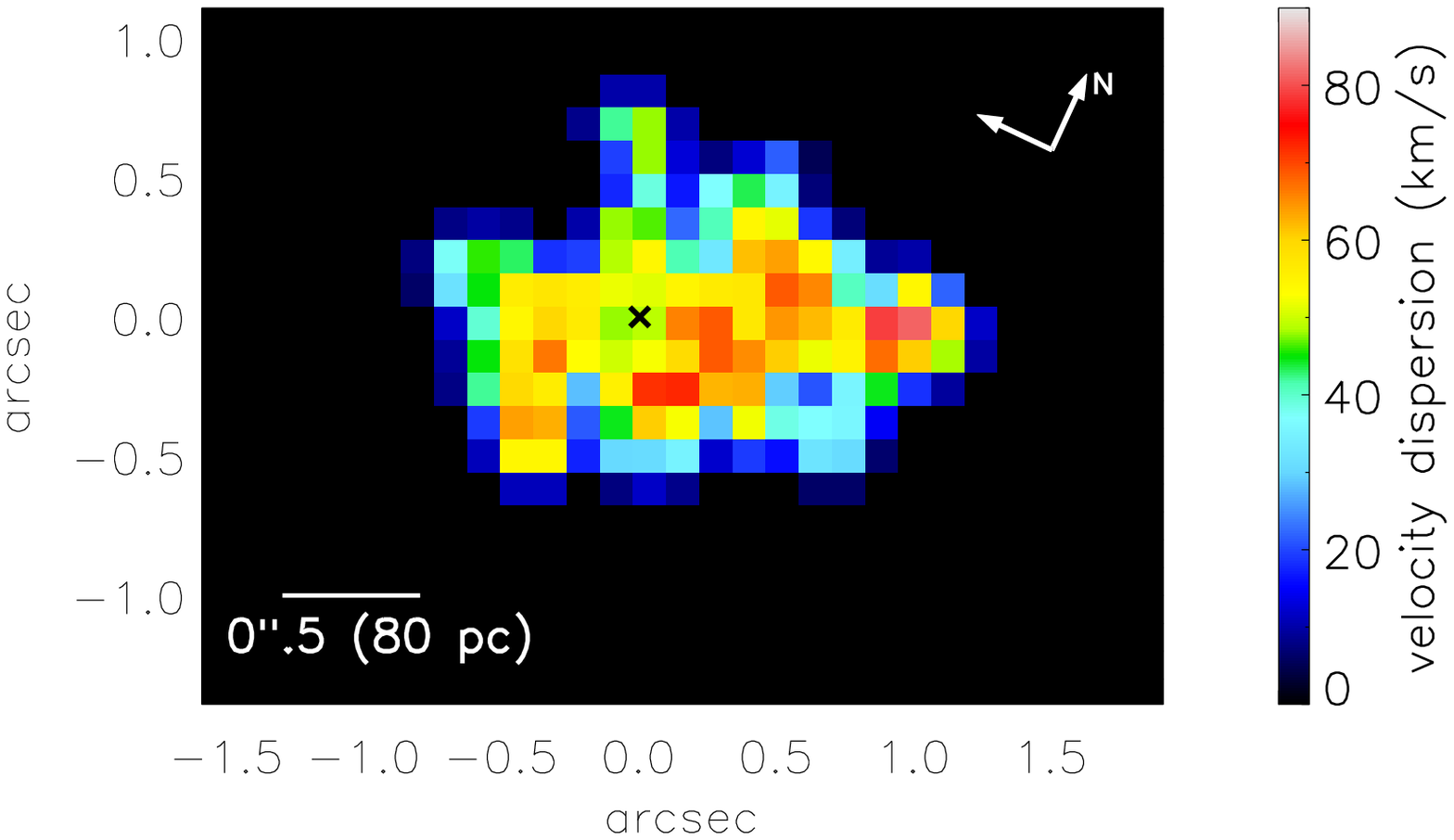,width=0.33\linewidth,clip=}
\epsfig{file=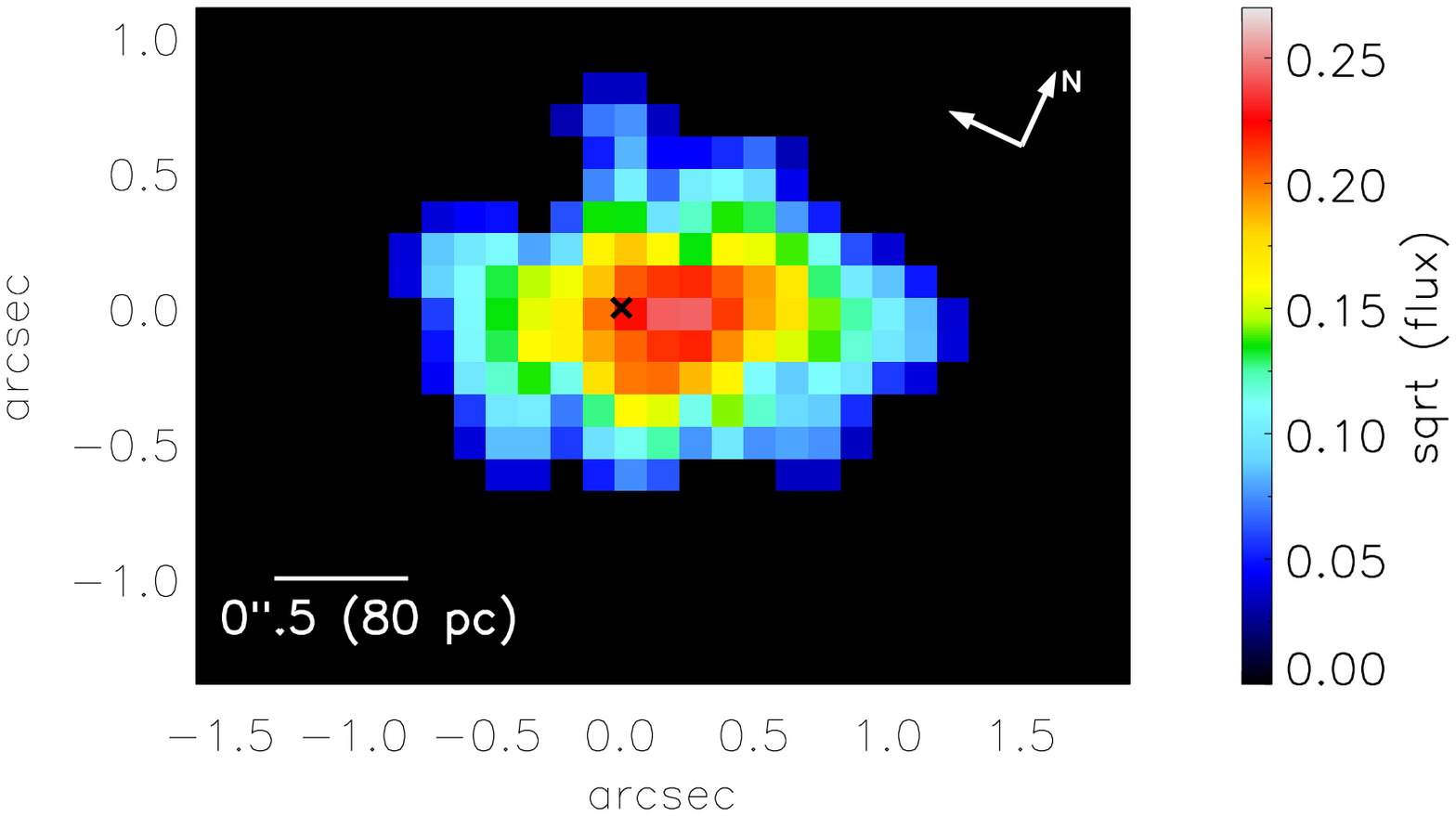,width=0.33\linewidth,clip=}\\[-0.6cm]
\epsfig{file=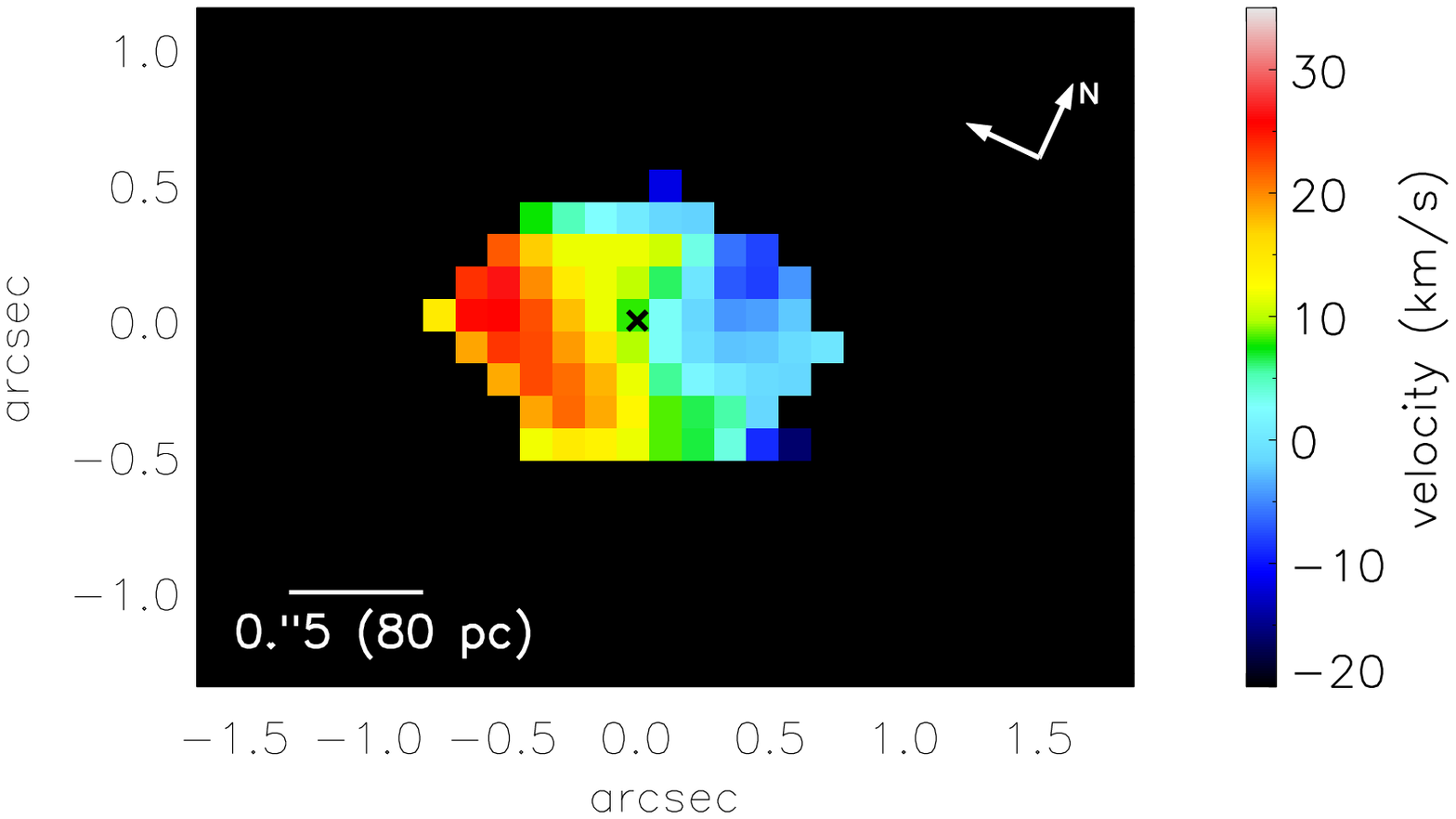,width=0.33\linewidth,clip=}
\epsfig{file=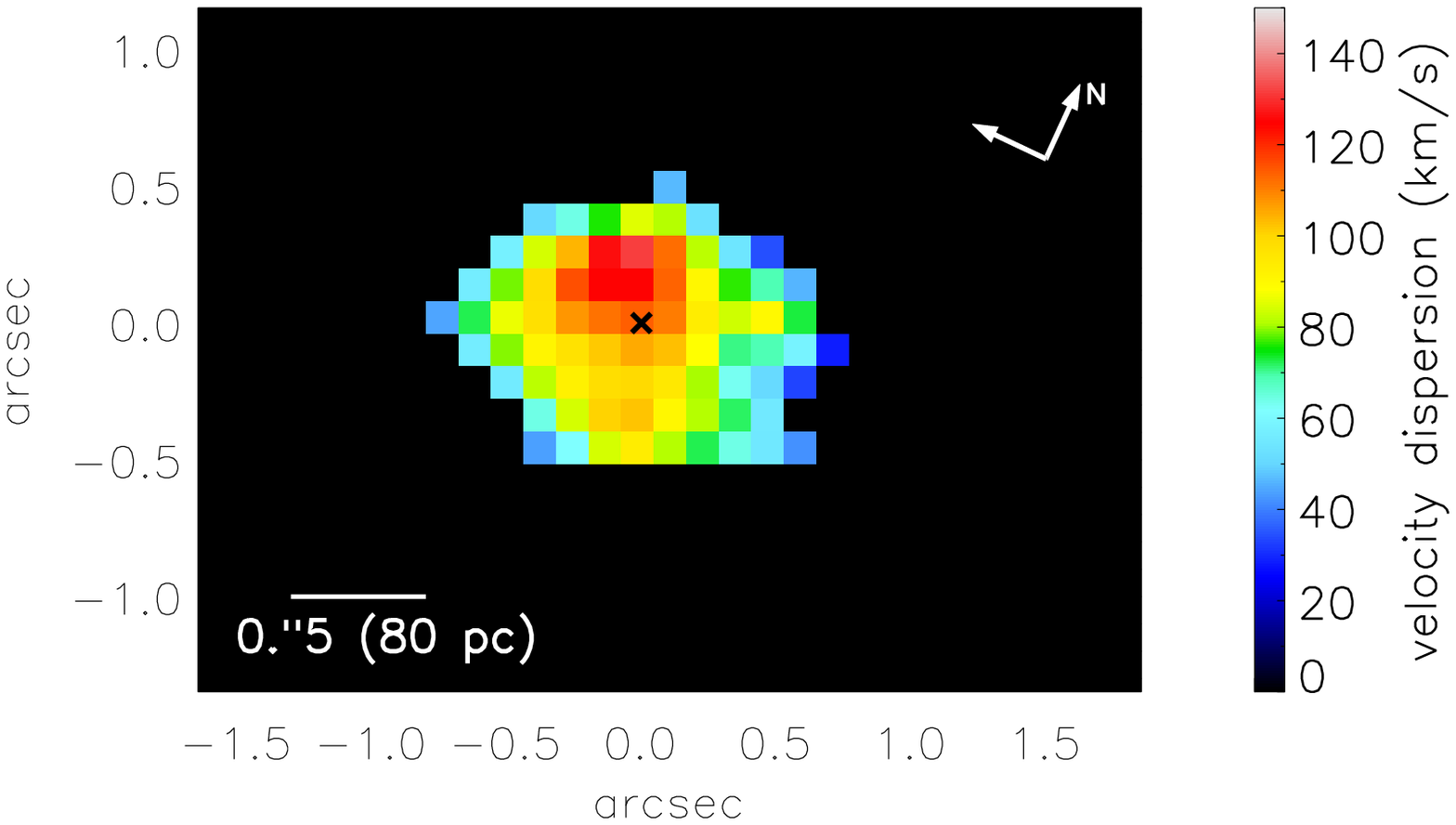,width=0.33\linewidth,clip=} 
\epsfig{file=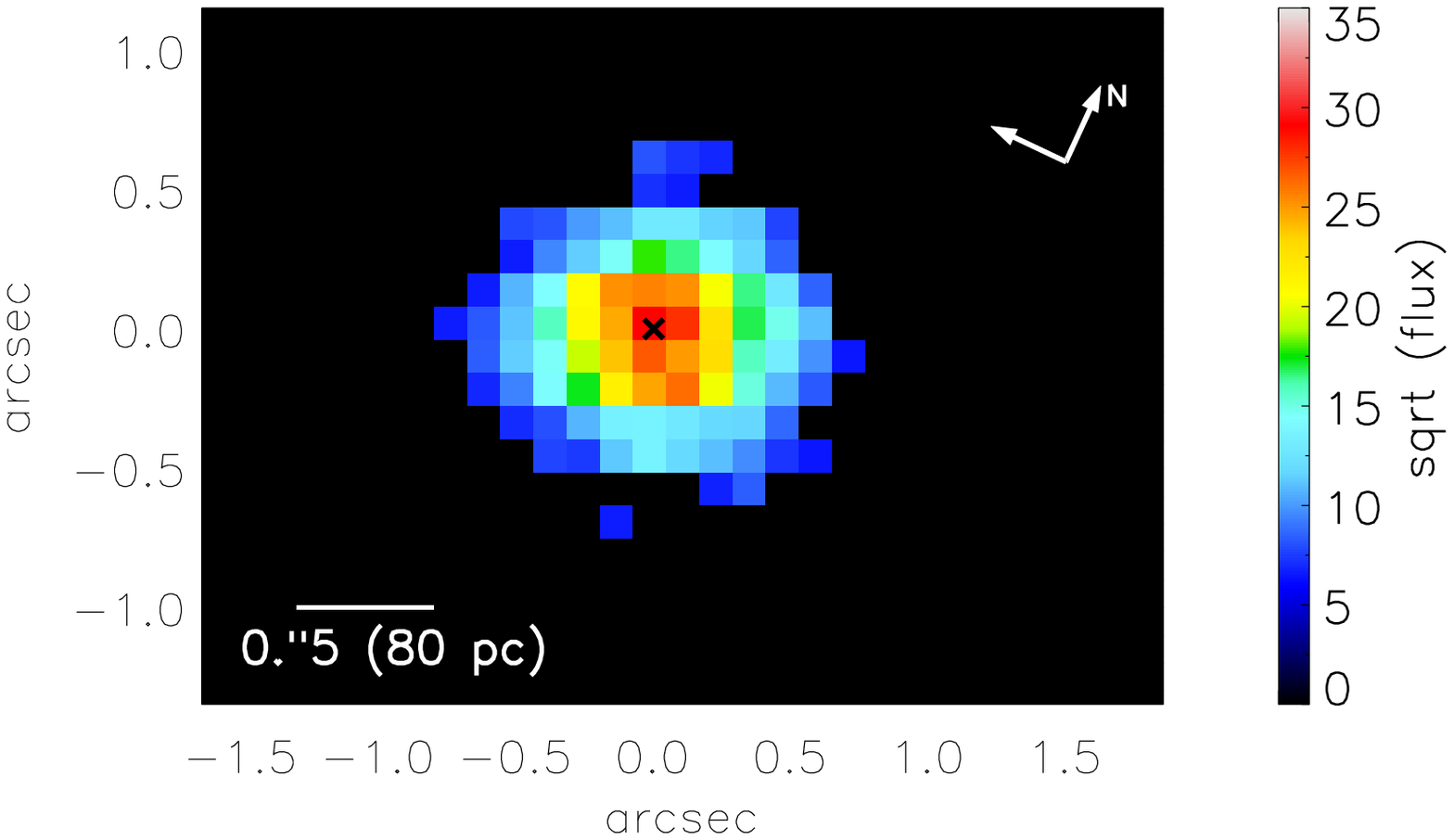,width=0.33\linewidth,clip=} \\[-0.2cm]
\epsfig{file=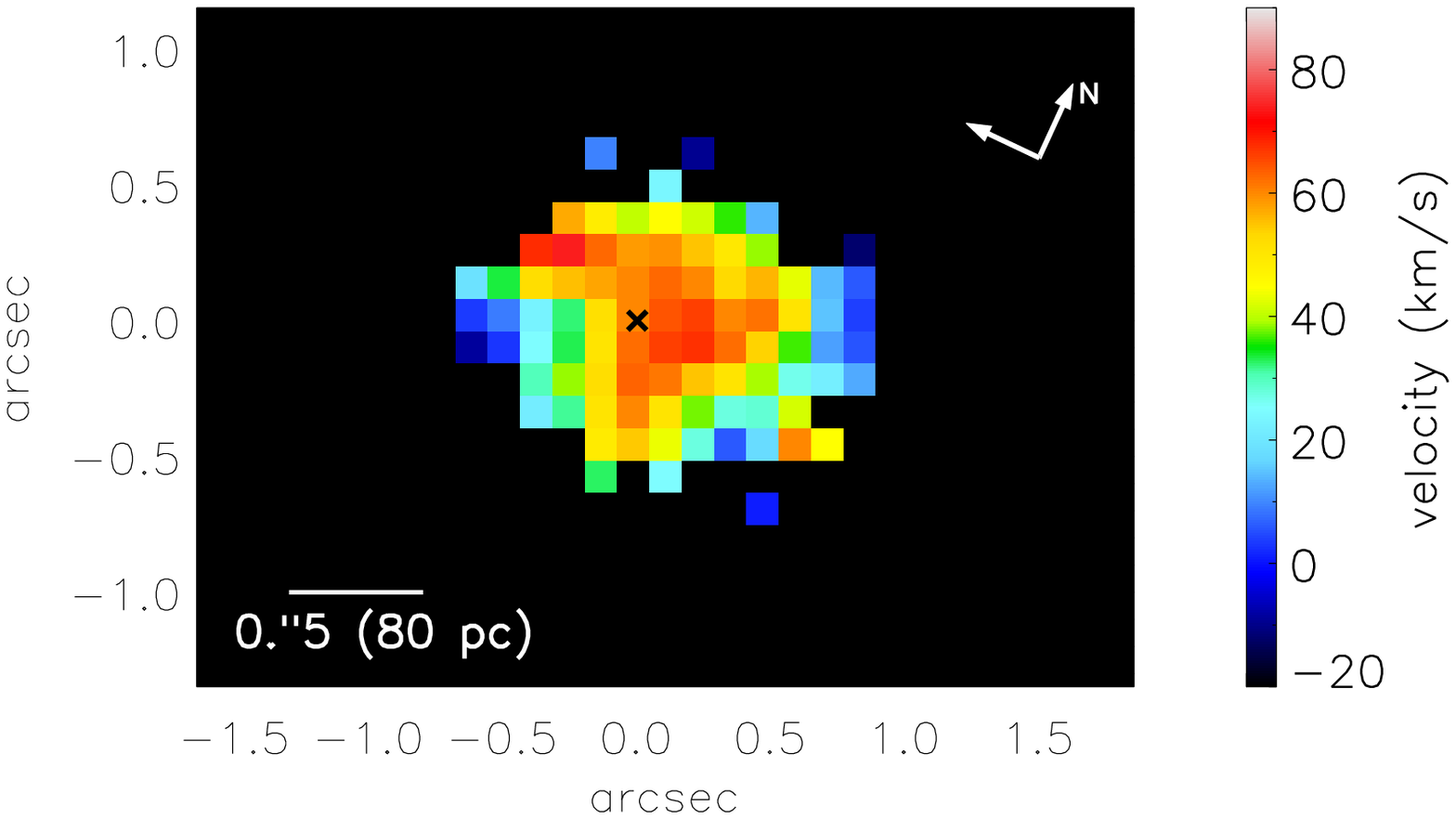,width=0.33\linewidth,clip=}
\epsfig{file=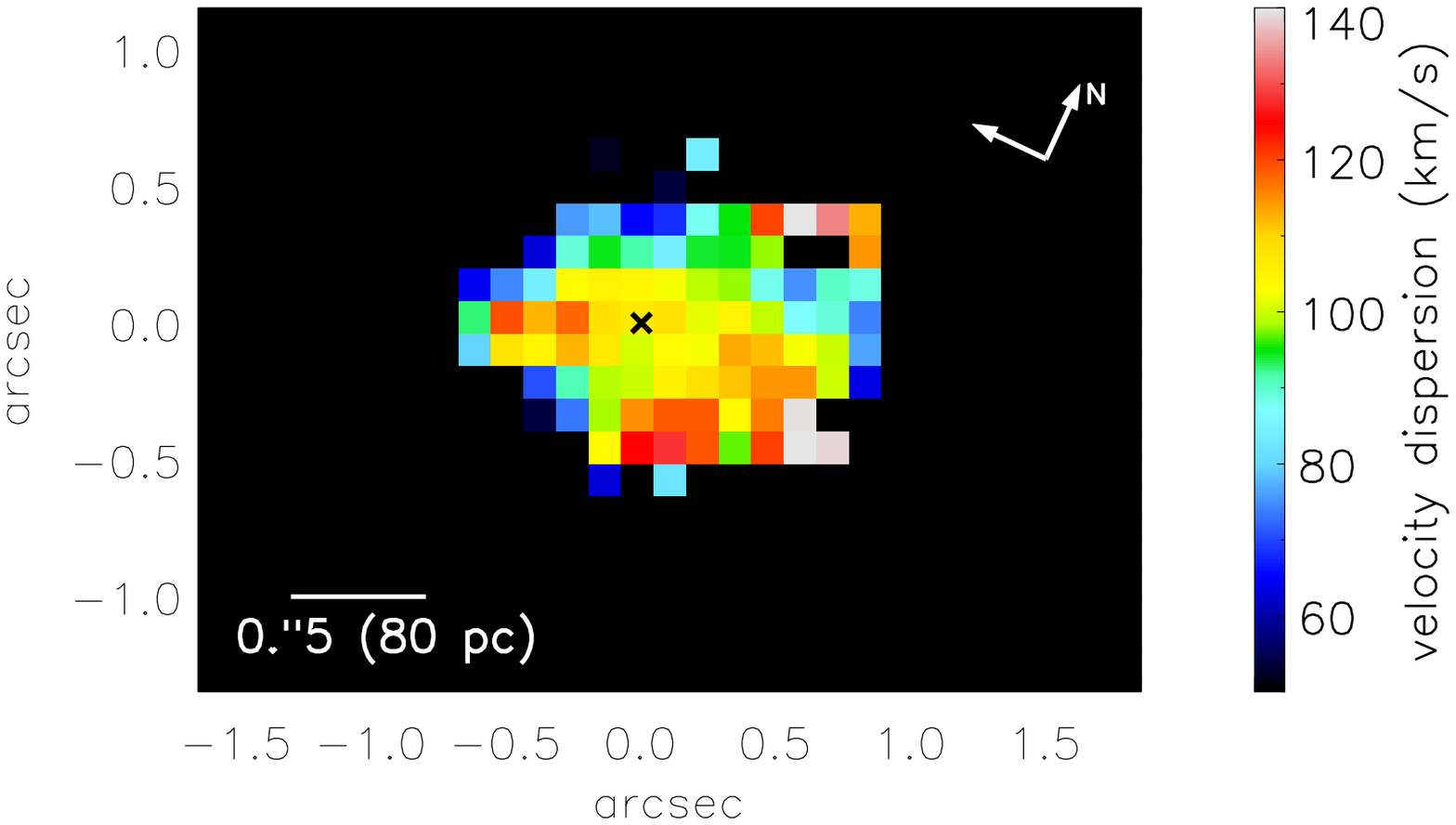,width=0.33\linewidth,clip=}
\epsfig{file=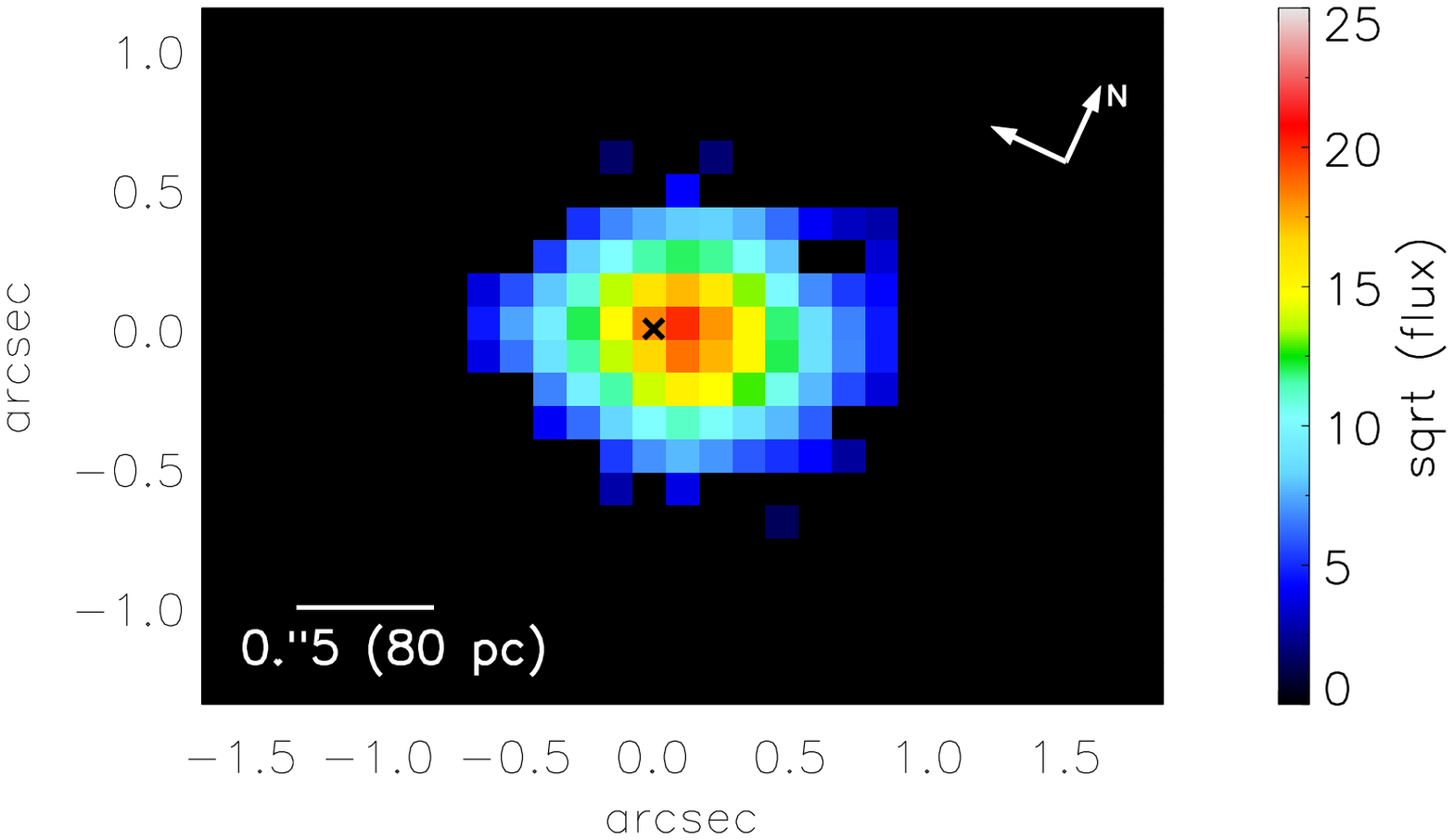,width=0.33\linewidth,clip=}
\caption {Velocity, velocity dispersion and flux maps of the ionised gas emission lines in our spectra. Top to bottom: Narrow Br$_{\gamma}$, [Fe II], [Si VI] and [Ca VIII]. The velocities were corrected for the systemic velocity of the galaxy and instrumental broadening. Minimum S/N is 3.0. The cross indicates the AGN position. The white ellipse in the top panel indicates the size of the stellar KDC.}
\label{gas_emission}
\end{figure*}
Each spatial pixel was then fit using pPXF in a similar procedure to what was done for the integrated spectrum in Fig.~\ref{central_H}. The maps were first binned to a minimum S/N $\sim$ 5 between the depth of the CO stellar absorption lines and the noise in a line free region of the spectra using the Voronoi binning technique described in \cite{cappellari&copin03}. The stellar features in the K-band are weaker and the map has to be more significantly binned to achieve the target S/N, furthermore, the inner regions show a high noise level due to the presence of the AGN emission, and therefore were masked out. The outer regions of both maps where the flux is low (less than 1 per cent of the peak flux) were also masked out. The maps of line-of-sight velocity and velocity dispersion for the H and K-bands are shown in Fig.~\ref{vel_field}. The kinematically distinct core, characterised by a change in the rotation direction of the velocity field, is clearly observed in the centre of both the velocity field maps. Despite the lower S/N in the K-band, the K-band velocity and velocity dispersion maps (Fig.~\ref{vel_field}, bottom panels) are consistent with the ones for the H-band considering that the velocities measured in the K-band have larger error bars (see below). Although for the final results we only fit for the first two moments of the velocity distribution, we tested a determination of the parameter $h3$, which is associated with asymmetric deviations from a gaussian shape for the line of sight velocity distribution. The $h3$ map showed an anti-correlation with the velocity structure of the distinct core.

Using Monte Carlo simulations where the wavelength ranges for the fit and initial parameters are changed randomly, we estimate the errors in the H-band velocity map to be typically 6 km/s except in the central regions close to the black hole where it reaches 13 km/s. The errors in the velocity dispersion are $\sim$7 km/s and 15 km/s in the centre. In the K-band map we estimate the error in the velocity and velocity dispersion to be $<$ 10 km/s out to r $< 2{''}$. The outer regions (r $> 2{''}$) and the central masked region have errors of $\sim$ 20 - 25 km/s in both velocity and velocity dispersion.

The systemic velocity and position angle (PA) are determined using the method outlined in \cite{krajnovic06}. The kinematic PA for the distinct core is determined from the H-band velocity map limited to the region of r $<$ 1${''}$.1. We find PA = 118 $\pm$ 22 degrees (3-$\sigma$ error) measured from North to East. The region outside the distinct core is modelled setting r $>$ 1${''}$.2 and r $<$ 3.${''}$5 and we find PA = 112 $\pm$ 8 degrees (3-$\sigma$ error). The similarity between these angles combined with the observation of the shift in the velocity field indicate that the distinct core is misaligned from the main body of the galaxy with an orientation consistent with counter-rotation (180 degrees shift in the velocity orientation with respect to the main body of the galaxy). 
\subsection{Ionised gas}
\subsubsection{Hydrogen Br$_{\gamma}$ emission}
\begin{figure*}
\centering
\epsfig{file=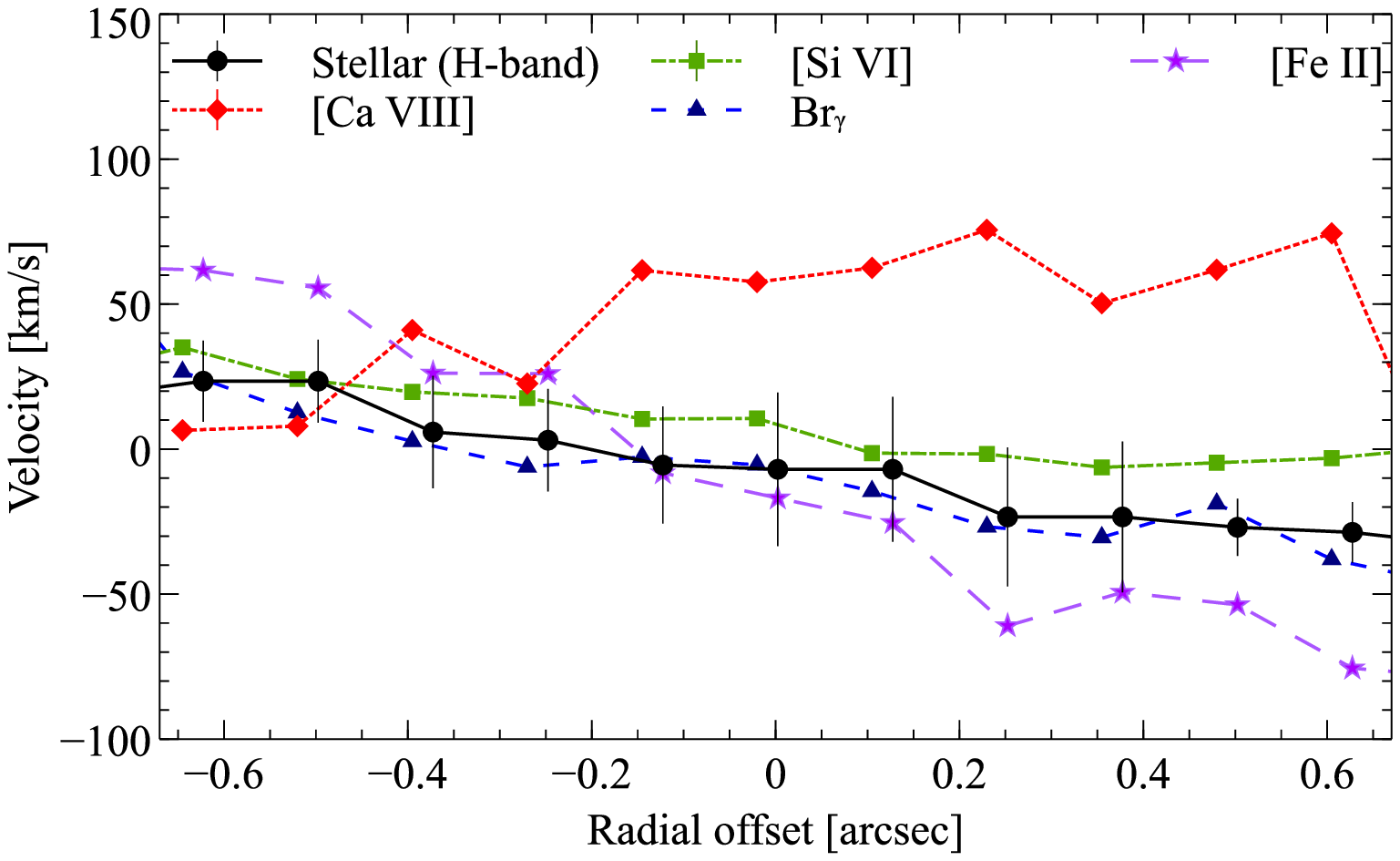,width=0.5\linewidth,clip=}\hspace{-0.1cm}
\epsfig{file=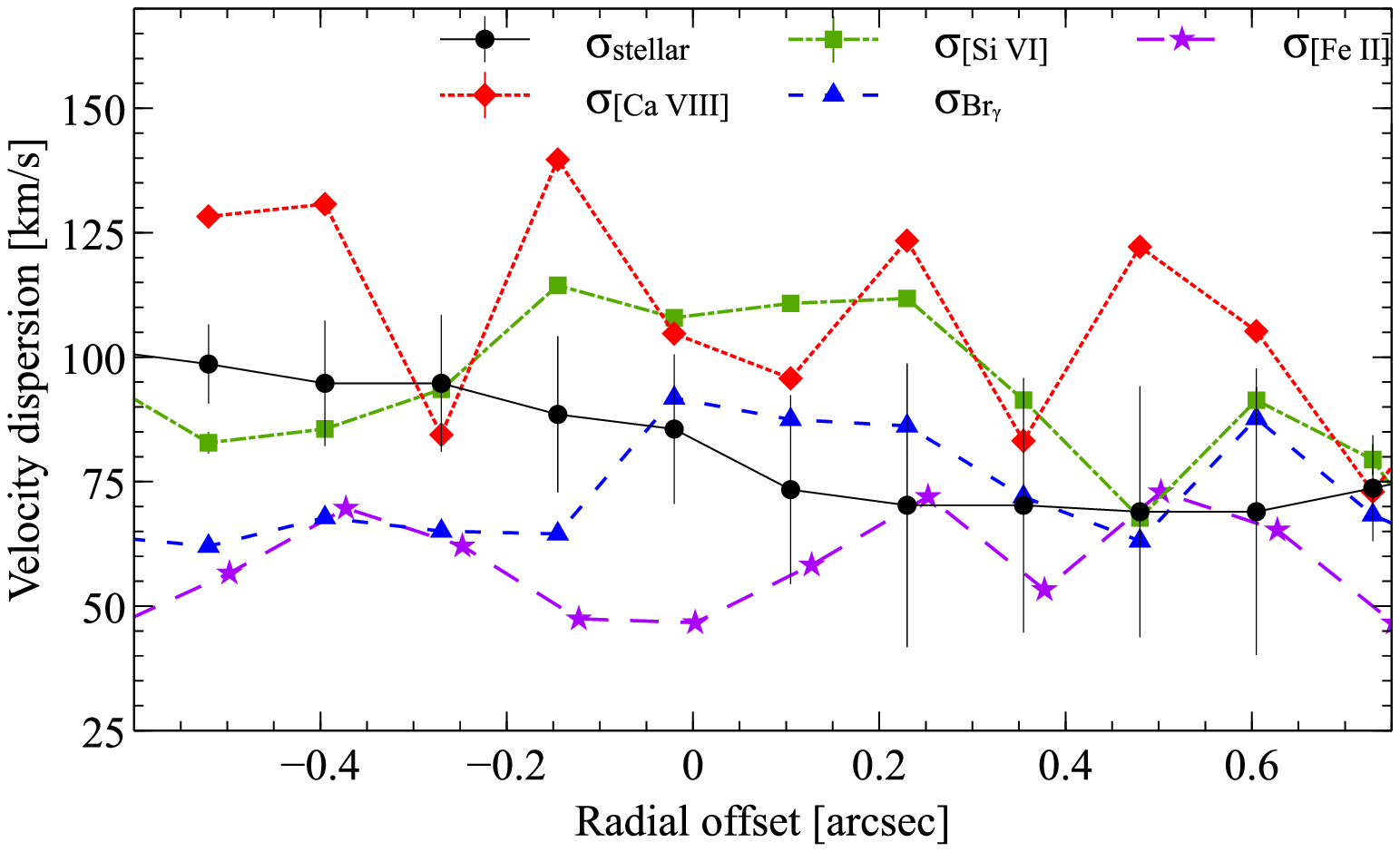,width=0.5\linewidth,clip=}
\caption{Horizontal cuts along the major axis of the galaxy to measure the ionised gas and the H-band measured stellar dynamics as a function of radius from the AGN position. Left: Velocity. Right: Velocity dispersion.}
\label{horizontal}
\end{figure*}

Broad hydrogen Br$_{\gamma}$ 2.166 $\micron$ emission is detected in the region close to the AGN and it is spatially unresolved by our data. The Br$_{\gamma}$ emission also shows a narrow component which can be separated from the underlying broad component. Fig.~\ref{brackett} shows the integrated emission in a region of 0${''}$.625 x 0${''}$.625 centred at the nucleus. This line shows a blueshift of $\sim$ 100-150 km/s with respect to the systemic velocity of the galaxy, as found in \cite{reynolds97}. The broad Brackett emission was used to determine the instrumental PSF and the centre of the PSF is taken as the spatial position of the AGN. Although the narrow emission may have a contribution from star formation, it is not detected significantly outside the PSF region. The velocity, dispersion and flux maps are shown in Fig.~\ref{gas_emission} (top row). Fig.~\ref{horizontal} shows the stellar and gas emission velocity profiles along the major axis of the galaxy, averaged in a pseudo-slit of 3 pixel ($\sim$0${''}$.4) width. The kinematics of Br$_{\gamma}$ resemble a rotation pattern and is very similar to the one of the stars in terms of the velocity gradient and low dispersion.

The Br$_{\gamma}$ narrow line flux can be used to estimate the star formation rate or to determine upper limits for it in galaxies where stars provide the only excitation mechanism for Br$_{\gamma}$ (e.g. \citealt{panuzzo03}, \citealt{valencia-s12}, \citealt{smajic14}). In the case of MCG--6-30-15, the spatial distribution of Br$_{\gamma}$ when compared with the stellar distribution profile leads us to believe that this emission is mainly associated with the AGN, although some smaller contribution from star formation is also possible. As noted by \cite{smajic14}, assuming that all of the central Br$_{\gamma}$ emission is due to star formation is an unlikely scenario for a Seyfert galaxy.
We can however use the equivalent width of the narrow Br$_{\gamma}$ line to set upper limits on the star formation. In the integrated central region of 0${''}.625 \times 0{''}.625$, the width of the narrow line is $\sim$ 15 \AA\thinspace. This is a low value (similar values have been found for other Seyfert galaxies - \citealt{davies07}), which indicates that there is little star formation ongoing in this galaxy. If star formation is occurring it is possibly in the form of short bursts, in line with what is observed in the ultraviolet spectra \citep{bonatto00}.
\subsubsection{[Fe II] emission}
\label{sec:feii}
The forbidden [Fe II] emission line at 1.644 $\micron$ was first detected in the H-band observations in \cite{raimundo13}, extending out to a radius of r $< 0{''}.8$. With the new data in the present work covering a wider field-of-view and at a higher S/N ratio we were able to study the distribution of ionised gas further out in the galaxy (Fig.~\ref{gas_emission} - second row). For the analysis of the ionised gas emission we use the code LINEFIT (appendix B of \citealt{davies11}) to fit the emission line at each spatial position of our field-of-view. LINEFIT uses a sky emission line observed close to the wavelength of interest as a template for the emission, and then convolves it with a gaussian kernel to match the observed line profile. The instrumental broadening is therefore automatically accounted for when fitting the lines. Individual pixels with S/N $<$ 3 (between the peak of the line and the \textsc{rms} noise of the continuum) were masked out of the image.
We observe that the [Fe II] line flux does not peak at the position of the nucleus but at a position north-west of the nucleus, as in the first time it was detected \citep{raimundo13}. The velocity map indicates an overall bulk rotational velocity direction similar to the one of the stellar counter-rotating core, albeit with a higher rotational velocity [-70, 60] km s$^{-1}$. We can now distinguish further structure in the velocity dispersion map. 
\begin{figure*}
\centering
\epsfig{file=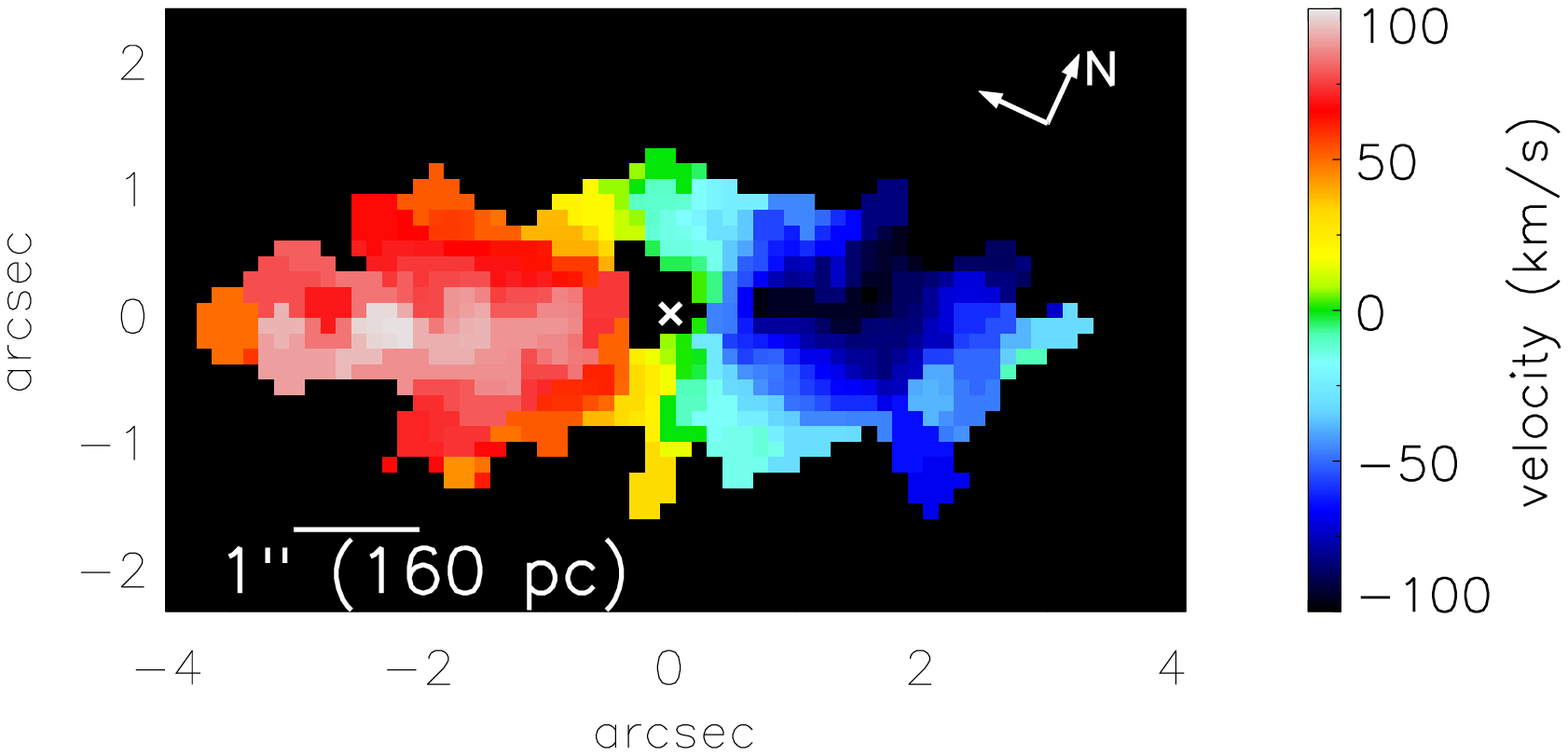,width=0.4\linewidth,height=4.0cm,clip=}\hspace{-0.15cm}
\epsfig{file=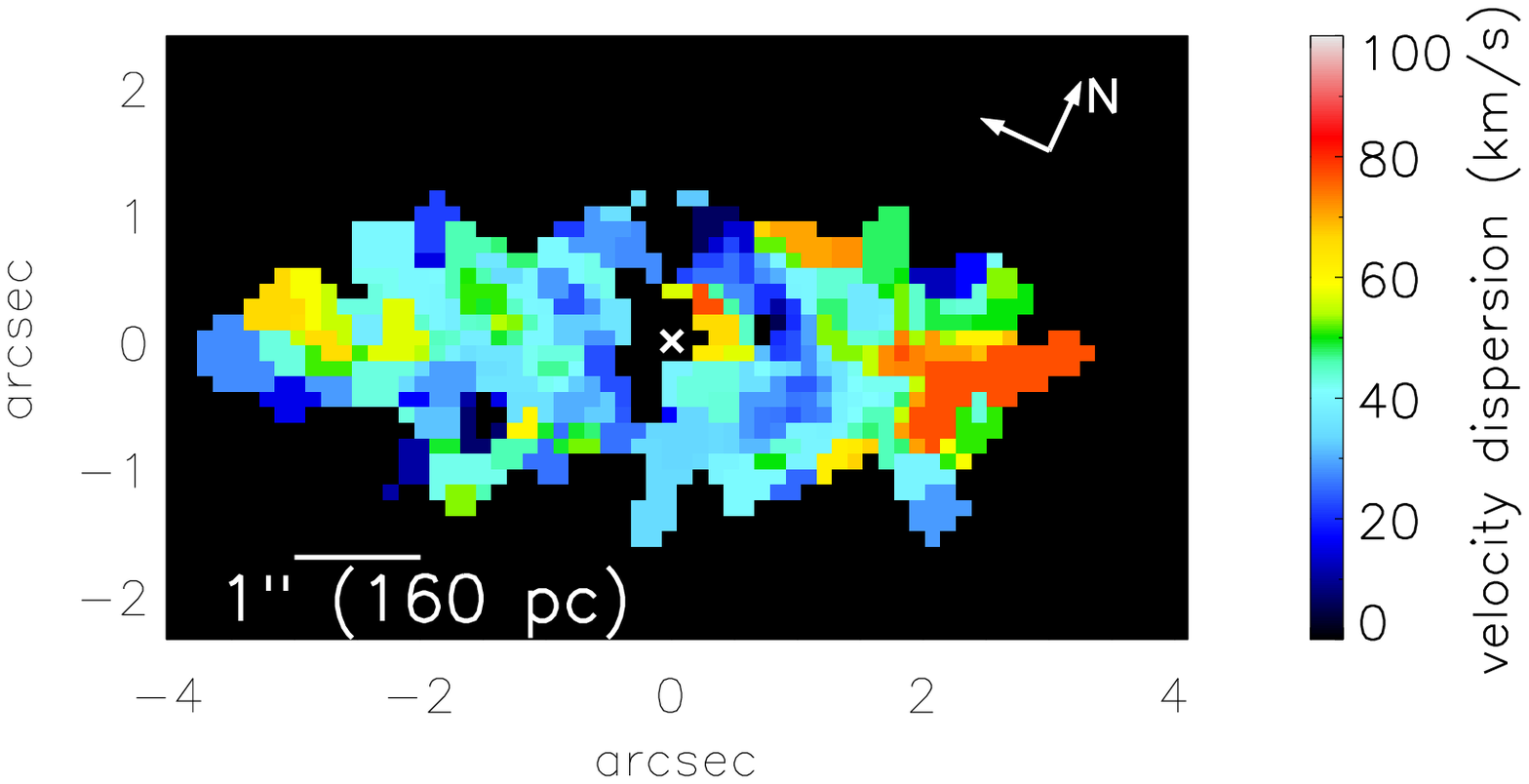,width=0.4\linewidth,height=4.0cm,clip=}\hspace{-0.15cm}\\
\epsfig{file=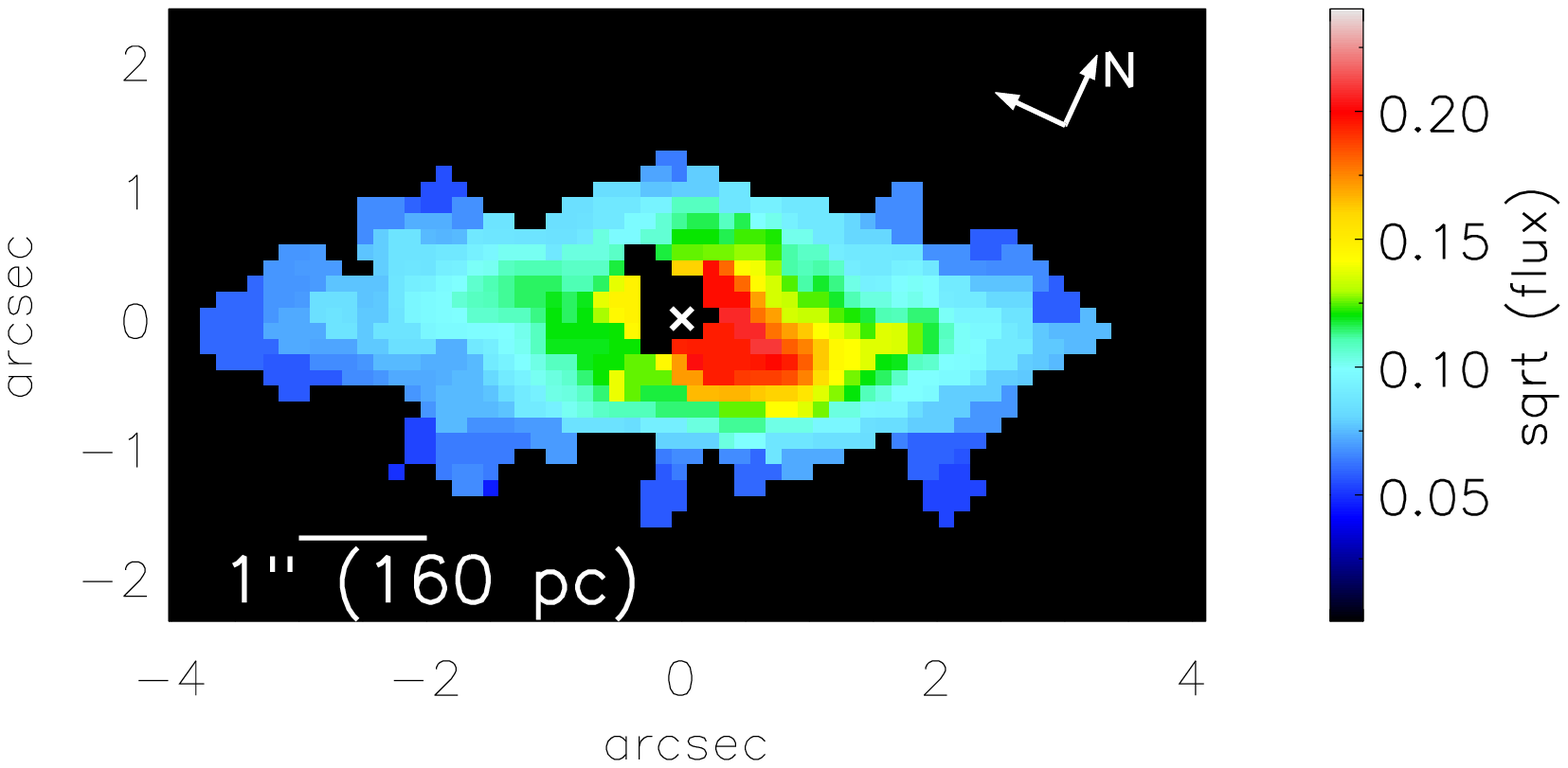,width=0.4\linewidth,height=4.0cm,clip=}\hspace{-0.15cm}
\epsfig{file=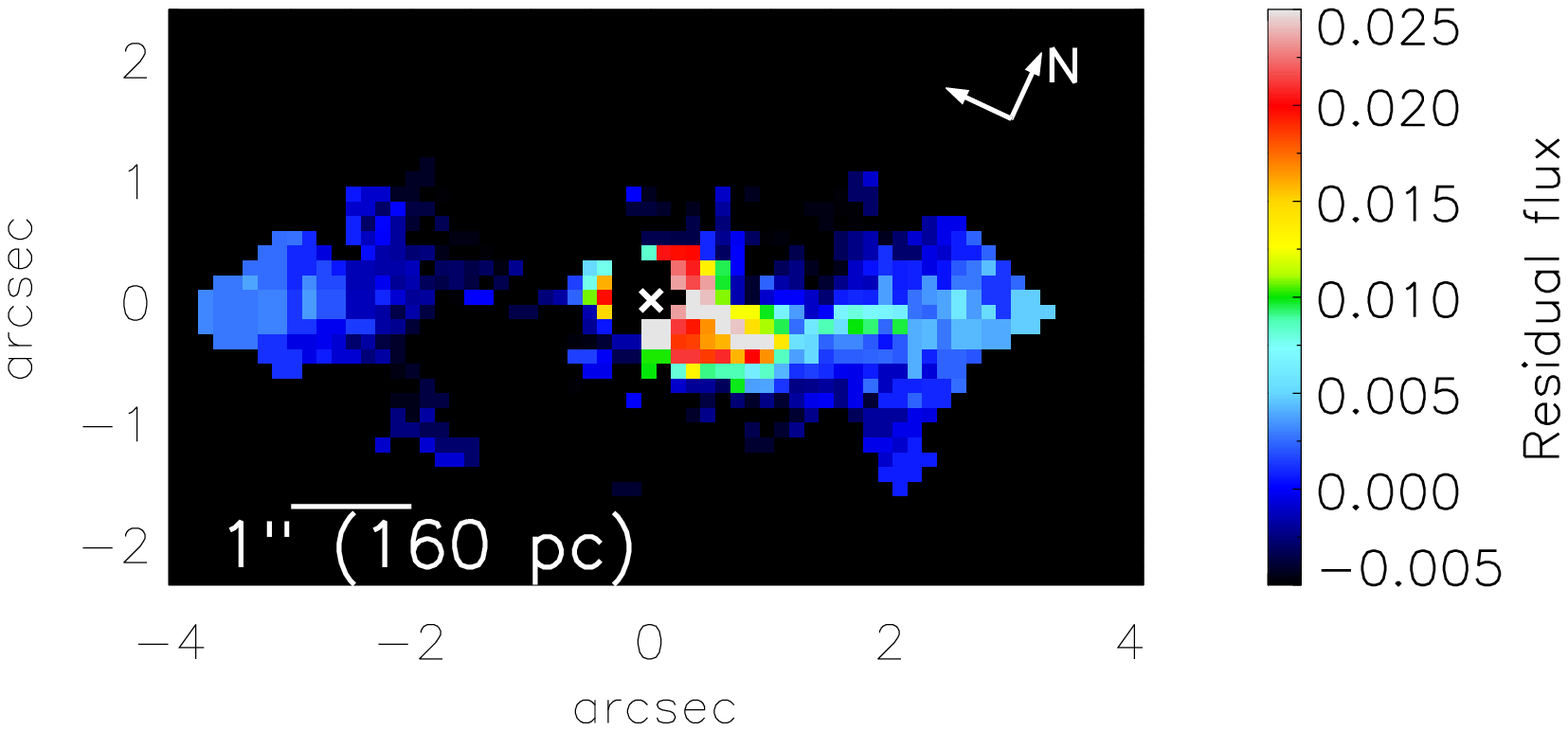,width=0.43\linewidth,height=4.15cm,clip=}\hspace{-0.15cm}
\caption {H$_{2}$ 2.12 $\micron$ distribution and dynamics. Top left: Velocity map. Top right: Velocity dispersion. Bottom left: Flux map. Bottom right: Flux residuals after subtracting elliptical isophotes. The H$_{2}$ rotation direction is the same as for the counter-rotating core. The spatial extent of the H$_{2}$ is larger than the counter-rotating core. The map was binned to a minimum S/N of 5. In the central regions the H$_{2}$ emission is not detected at a significant S/N level, therefore these central pixels were masked out of the image.}
\label{H2_vel}
\end{figure*}
The forbidden transition of [Fe II] is excited by electron collisions. To achieve the right conditions for this line emission, it is necessary to have a zone of partially ionised hydrogen where Fe$^{+}$ and e$^{-}$ coexist. 
In AGN in particular, [Fe II] emission can be generated by photoionisation by the central AGN or by shocks (caused by the interaction of radio jets with the medium, by nuclear mass outflow shocks with ambient clouds or by supernova-driven shocks) (\citealt{mouri00}, \citealt{rodriguez-ardila04}). 
Correlations have been found between the radio properties of Seyferts and the presence of [Fe II] (\citealt{forbes&ward93}, \citealt{riffel10}, \citealt{ramos-almeida06}) which indicates that shocks associated with radio jets may be the main excitation mechanism in some active galaxies. \cite{mundell09} analysis of radio data on MCG--6-30-15 find a possible elongation in the extended radio emission approximately perpendicular to the orientation of the [O III] emission which may indicate the presence of a jet-like or disc wind component. However, the [Fe II] emission has a similar orientation to the [O III] emission and therefore it is unlikely that shocks associated with a possible radio jet are the main excitation mechanism for [Fe II] emission in this galaxy.

In our previous work \citep{raimundo13} we hypothesised that the [Fe II] emission was caused mainly by supernova shocks from a previous star formation event and our new observations support this scenario. The spatial distribution of the [Fe II] coincides with
the region where we see the strongest S/N in the stellar absorption features and although we detect an AGN driven outflow in our data, the dynamics of [Fe II] resembles a rotation pattern and is distinct from the [Ca VIII] dynamics (Fig.~\ref{horizontal}). The following additional indications are present in the data.
First, we observe that the [Fe II] emission extends further out than the [Si VI] and Br$_{\gamma}$ emission (Fig.~\ref{gas_emission}). Second, the [Fe II] emission is not centred at the nucleus and seem to trace a similar distribution to what is seen for the molecular gas (traced by the H$_{2}$ emission in Fig.~\ref{H2_vel}), which would be expected if both these lines originate from regions of star formation. Third, the velocity dispersion map shows peaks in various regions of the [Fe II] emission which could be tracing the local dynamics and be associated with locations of major supernova events. The general [Fe II] velocity dispersion profile along the major axis, as can be seen in the right panel of Fig.~\ref{horizontal}, is lower than for the stars or the other ionised species, but closer to the dispersion profile of H$_{2}$, further indicating that these two lines may be tracing the same regions of star formation and supernova events. It would be expected that if [Fe II] and H$_{2}$ are being excited in the same regions, the ionised gas traced by [Fe II] would show more disturbed kinematics (higher velocity dispersion) than the molecular gas traced by H$_{2}$.

It is of course likely that AGN photoionisation also contributes to the [Fe II] emission, even if it is not the dominant mechanism \citep{rodriguez-ardila04}. The [Fe II] emission will then provide an upper limit for the number of supernova events.
With our new observations we can probe the [Fe II] emission within the full extent of the kinematically distinct core and improve the [Fe II] flux measurement in \cite{raimundo13}. Integrating for the regions with emission S/N $>$ 3, we obtain: F$_{\rm [Fe II]}$ = 3.0$\times 10^{-15}$ erg s$^{-1}$ cm$^{-2}$ and L$_{\rm [Fe II]}$ = 4$\times 10^{38}$ erg s$^{-1}$.

\subsubsection{Coronal line emission}
\begin{figure}
\centering
\epsfig{file=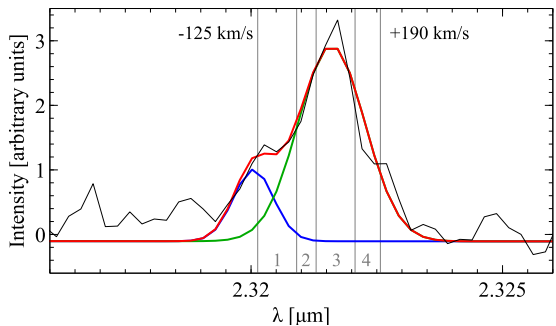,width=0.9\linewidth,clip=}
\caption{[Ca VIII] emission integrated in a 1${''}\times 1{''}$ region centred at the nucleus as a function of rest-frame wavelength. The figure shows the observed spectrum in black, the total line fit (red line) and the two gaussian components of the fit (blue and green). The vertical lines indicate the velocity bins used for the tomography analysis (-125 km/s, -25 km/s, +25 km/s, +125 km/s and +190 km/s). The numbers 1 to 4 indicate the corresponding panels of Fig.~\ref{caviii_tomography}.}
\label{caviii_integrated}
\end{figure}

\begin{figure*}
\centering
\hspace{-2.5cm}\epsfig{file=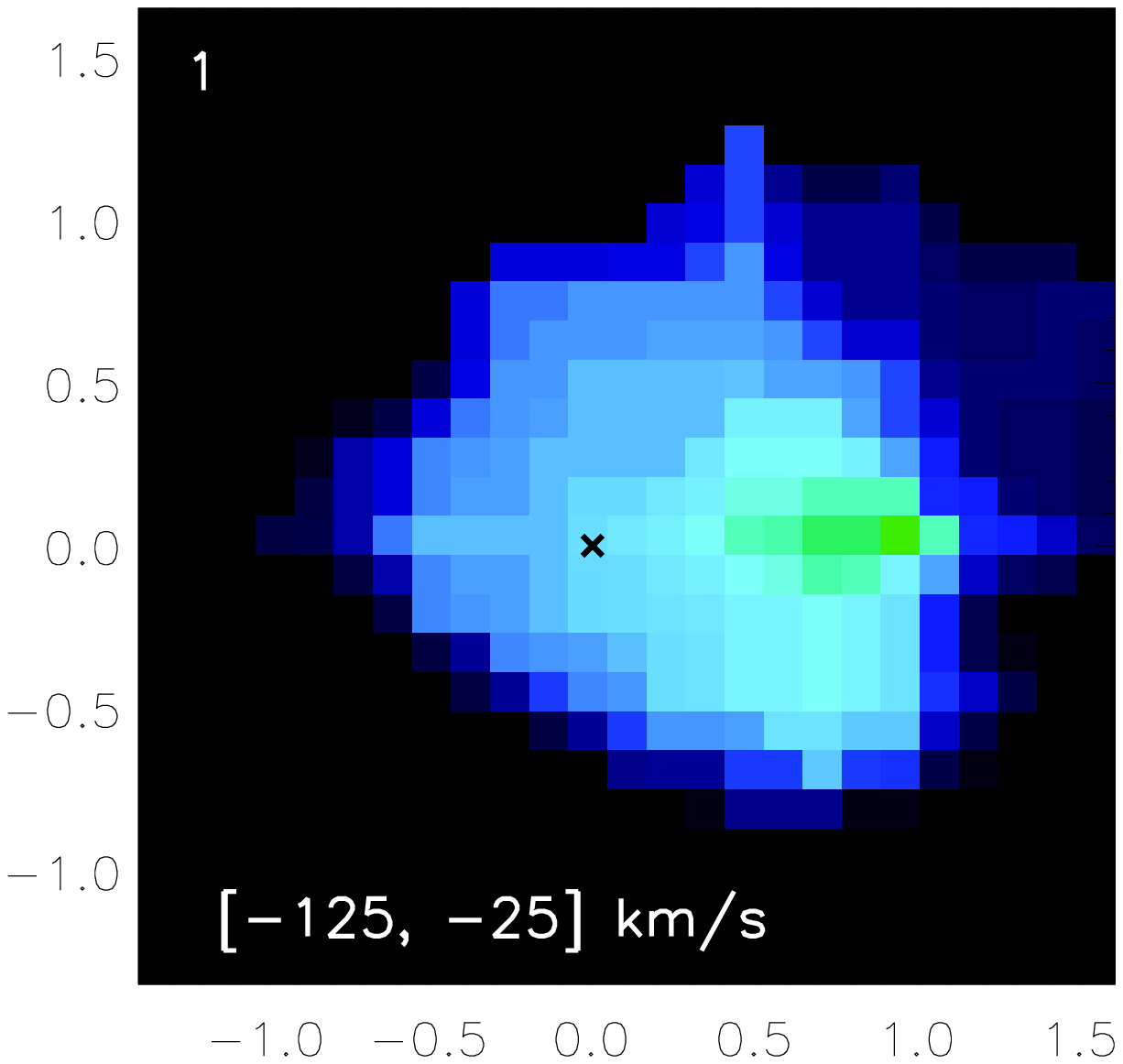,width=0.32\linewidth,clip=}\hspace{-1.6cm}
\epsfig{file=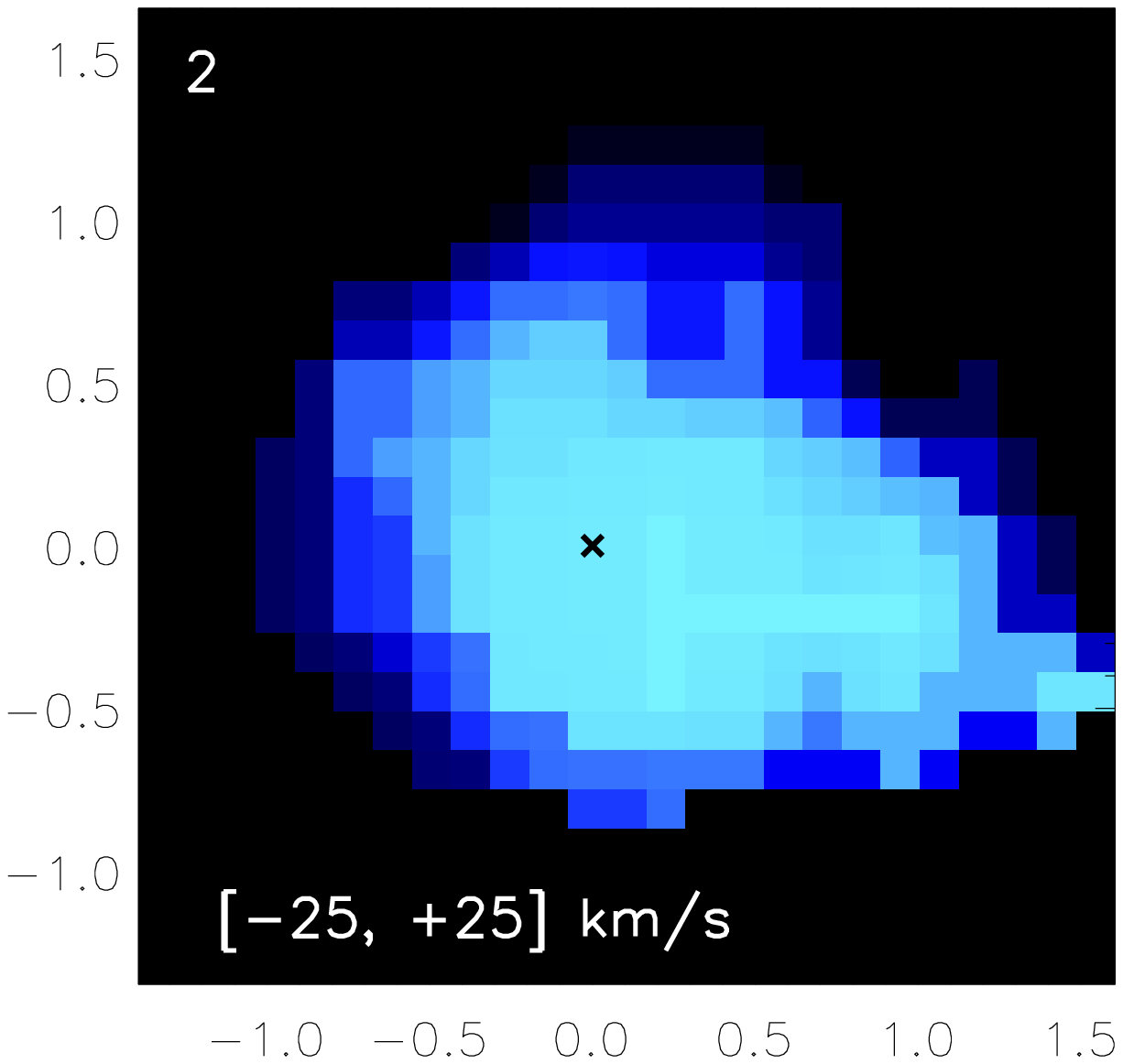,width=0.32\linewidth,clip=}\hspace{-1.6cm}
\epsfig{file=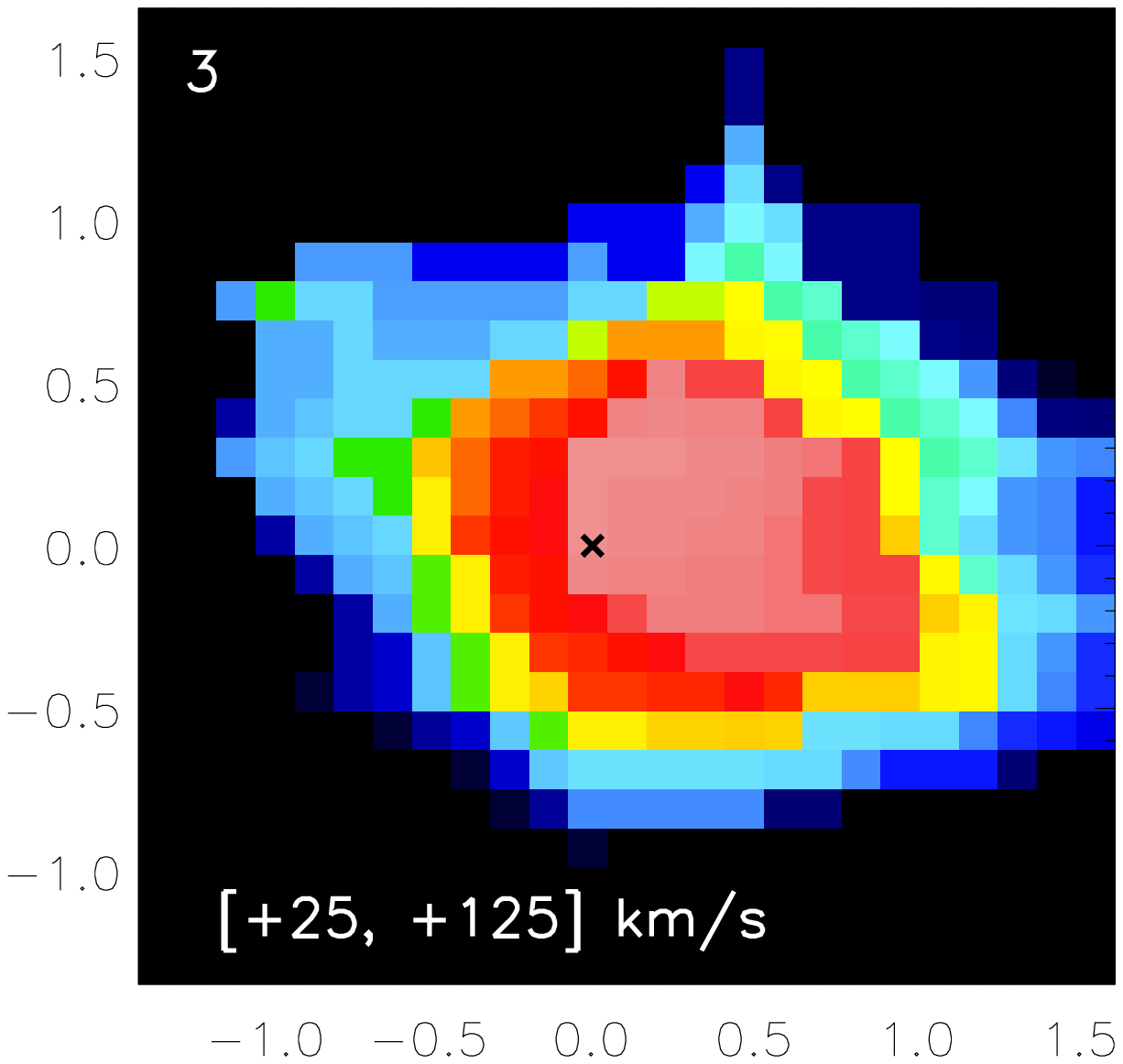,width=0.32\linewidth,clip=}\hspace{-1.6cm}
\epsfig{file=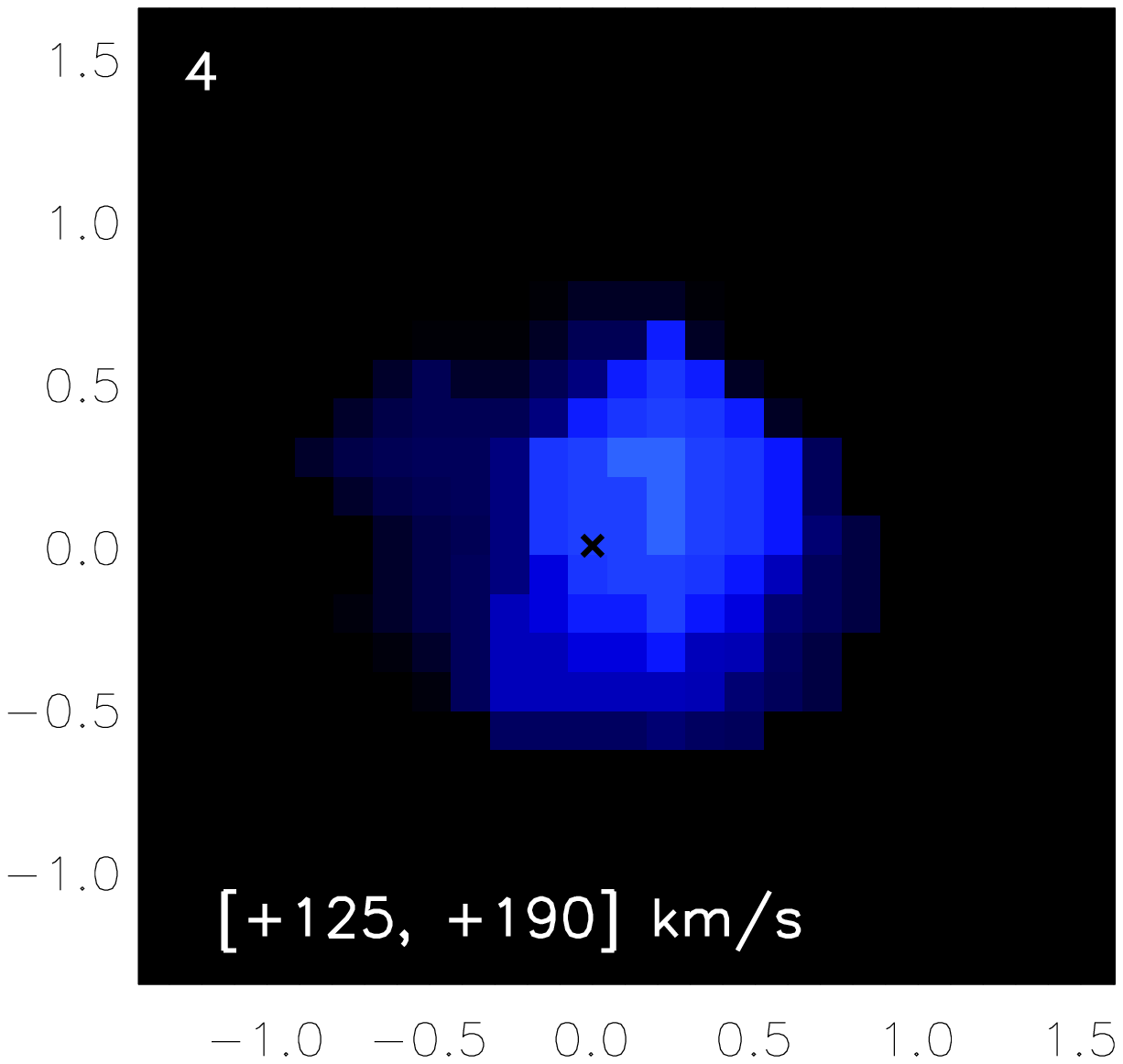,width=0.32\linewidth,clip=}\hspace{-1.6cm}
\caption {Velocity tomography of the [Ca VIII] line. Velocity intervals indicated at the bottom correspond to deviations from the systemic velocity of the galaxy and correspond to the regions (1 to 4) delimited by the vertical lines in Fig.~\ref{caviii_integrated}. The spatial scale is arcsec and the orientation is the same as in Fig.~\ref{gas_emission}. The first and third panels show the blueshifted and redshifted components of the outflow respectively. This suggests that we are seeing the approaching and receding sides of a one-sided ionisation cone.}
\label{caviii_tomography}
\end{figure*}

Two high ionisation coronal lines are detected in the K-band observations: [Si VI] 1.963 $\micron$ and [Ca VIII] 2.321 $\micron$. These lines are a good tracer of AGN activity due to their high ionisation potential (IP $>$ 100 eV). [Si VI] (IP = 167 eV) is mainly detected in the nucleus with a spatial extent of the order of the instrumental PSF. The velocity and velocity dispersion maps with a minimum S/N $\sim$ 3 are presented in Fig.~\ref{gas_emission} - third row. The regions with flux less than 5 per cent of that of the peak were masked out of the maps. [Ca VIII] has a lower ionisation potential (IP = 127 eV), and it is observed at a higher S/N than [Si VI]. Its emission is compact but resolved, slightly more extended than [Si VI] and asymmetric with its flux emission peak north-west of the nucleus. The shape of the [Ca VIII] emission line is complex, as can be seen in Fig.~\ref{caviii_integrated}. This plot shows the line profile integrated in a $1{''} \times 1{''}$ region centred at the nucleus. The emission line can be fitted using two gaussians, and that holds true for most of the pixels in the map with observed emission. The two lines are blueshifted and redshifted with respect to the systemic velocity of the galaxy and observed simultaneously at each spatial position, which is an indication that we are seeing the approaching and receding sides of a single ionisation cone. 
To better understand the properties of this emission we measured the flux at various velocity bins (`velocity tomography') and plot the result in Fig.~\ref{caviii_tomography}. We can see that the [Ca VIII] dynamics is dominated by an outflow. The first plot on the left maps the blueshifted component of the outflow at up to -125 km/s. The nuclear component is seen in the second plot from the left and the strong redshifted component is seen in third plot from the left, with velocities up to 125 km/s. In Fig.~\ref{gas_emission} we show the results of fitting the [Ca VIII] emission with a single gaussian line. Although the shape of the real emission is more complex, these maps reflect the dynamics of the dominating redshifted component outflowing at velocities $\sim$ 100 km s$^{-1}$.

The [Si VI] emission could also have a redshifted outflowing component, as the velocities north-west of the nucleus are slightly higher than expected based solely on rotation and the dispersion is higher than that observed in the narrow Br$_{\gamma}$ emission (Fig.~\ref{horizontal}). However, due to its higher ionising potential this outflowing component is not as strong as in [Ca VIII]. The velocity tomography for [Si VI] shows a very centrally concentrated emission, which does not extend beyond the PSF. The extent of the coronal line region for both these lines is shown in Table~\ref{coronal_extent}. We define the extent of the coronal line region at the position where the flux of the line drops to 5 per cent of its peak value.

When comparing the profiles of the ionised gas emission with the one of the stars, we see clearly that narrow Br$_{\gamma}$ has a profile similar to the stars. [Si VI] and [Ca VIII] show a redshift in relation to the stellar rotation. The redshift of the [Si VI] is small and it is close to the range of errors expected for the stellar velocity. [Ca VIII] has a clearly distinct velocity profile, dominated by the redshifted component of the outflow. Although also slightly distinct from the stellar, the Br$_{\gamma}$ and the [Si VI] profiles, the velocity gradient of [Fe II] is clearly different from [Ca VIII] which is another indication that the [Fe II] emission is not part of the AGN outflow but tracing star formation in the disc. 

\begin{table}
\begin{center}
\textsc{Table 1}
\vspace{0.1cm}

\textsc{Spatial extent of the coronal line emission region.}
\vspace{0.2cm}
\begin{tabular}{ | l | l | l | }
\hline
\hline

Emission line & IP & Extent (radial)\\
\hline
[Si VI] 1.963 $\micron$ & 167 eV & 0${''}$.7 ($\sim$ 110 pc) \\

[Ca VIII] 2.321 $\micron$ &  127 eV & 0${''}$.9 ($\sim$ 140 pc) \\

\hline
\end{tabular}
\end{center}
\caption{The radial extent is measured in radius from the nucleus.}
\label{coronal_extent}
\end{table}
The study of \cite{rodriguez-ardila06} on the size of the coronal line region in MCG--6-30-15 shows that the [Fe VII] line with an ionisation potential of IP = 100 eV close to the one of [Ca VIII] (IP = 127 eV) shows an extent of $\lesssim$ 140 pc (after converting to our cosmological parameters) similar to the spatial extent we observe in our study (Table~\ref{coronal_extent}). \cite{rodriguez-ardila06} also find that the broad component of [Fe VII] is blueshifted and has a FWHM $\sim$ 670 km/s which has been interpreted as signature of an outflowing wind.
Blueshifted outflows have also been detected in MCG--6-30-15 from X-ray studies of warm-absorbers (\citealt{sato03}, \citealt{blustin05}, \citealt{chiang11}). These outflows present blueshifted velocities of v$_{\rm out}$ = -1900 km s$^{-1}$ and v$_{\rm out}$ = -150 km s$^{-1}$ at distances of 0.036 pc and 5.7 pc from the nucleus, respectively, although it is not clear if the two outflows are part of the same accelerating wind \citep{blustin05}. The outflows at small scales may be associated with the [Ca VIII] outflow at scales of r $<$ 140 pc that we observe in this work, however it is difficult to establish a direct connection between such a large range of physical scales. Assuming that the faster outflow at $r \sim 0.036$ pc would lose velocity as it propagates, its velocity would have to decrease as $v_{\rm out} \propto r^{-0.4}$ to reproduce the outflow we observe in [Ca VIII] at $r \sim 140$ pc.
\subsubsection{Optical emission lines}
\begin{figure*}
\centering
[OIII]  \hspace{7cm} H$_{\beta}$\\
\vspace{-0.5cm}
\epsfig{file=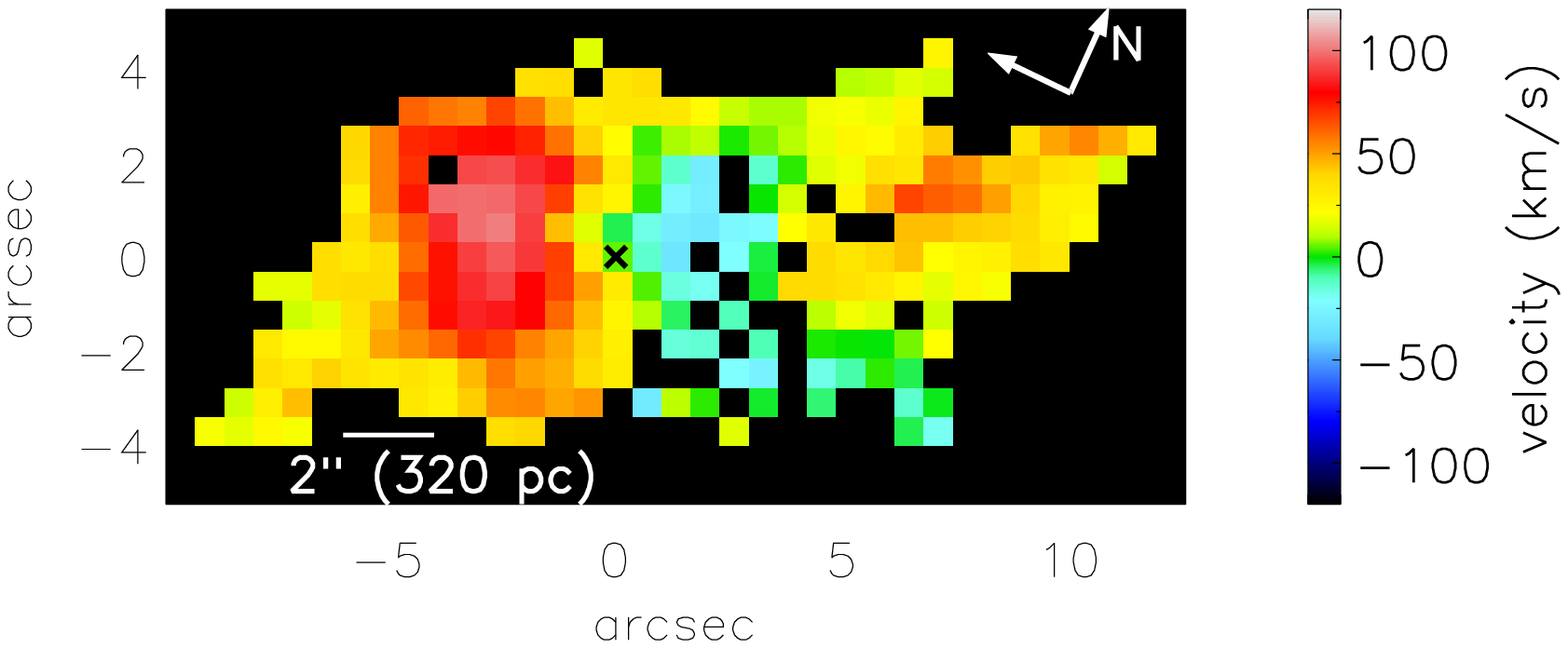,width=0.45\linewidth,clip=}\hspace{-0.2cm}
\epsfig{file=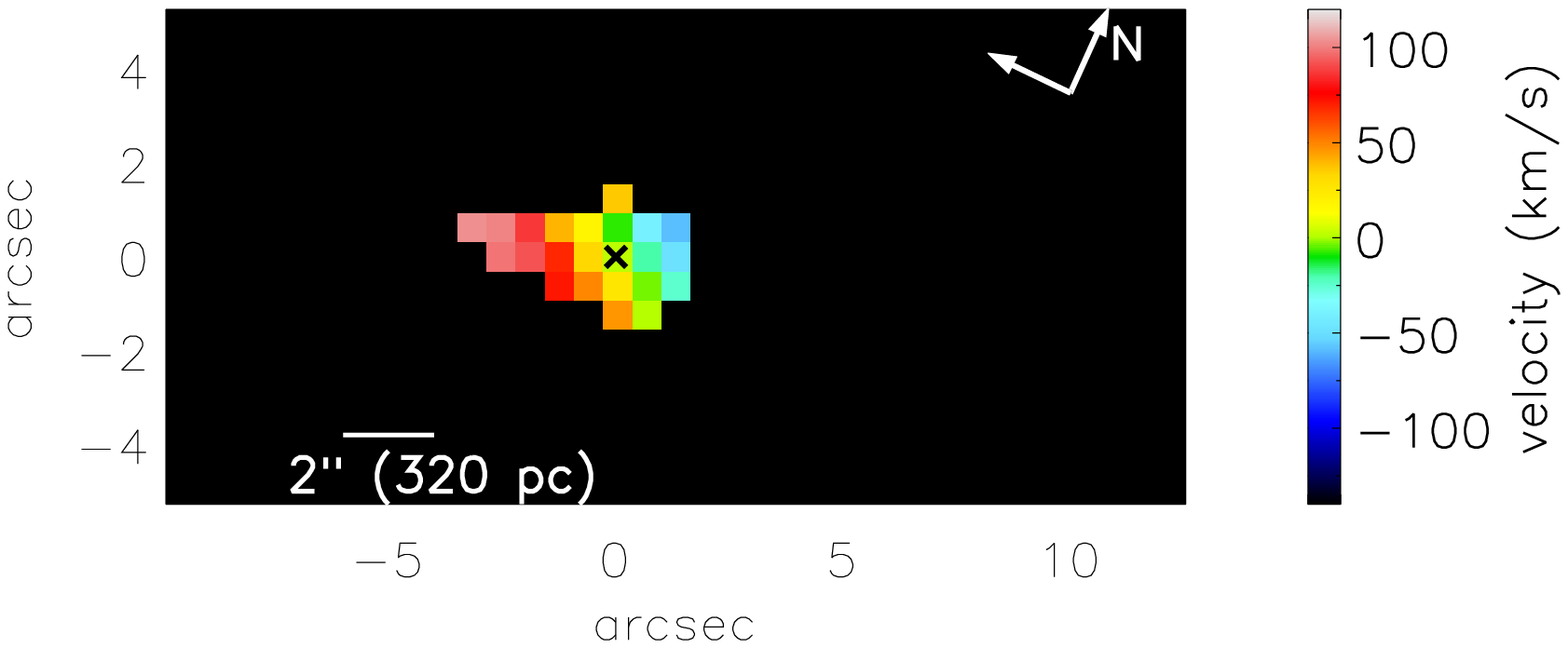,width=0.45\linewidth,clip=}\hspace{-1.8cm}\\
\vspace{-1.0cm}
\epsfig{file=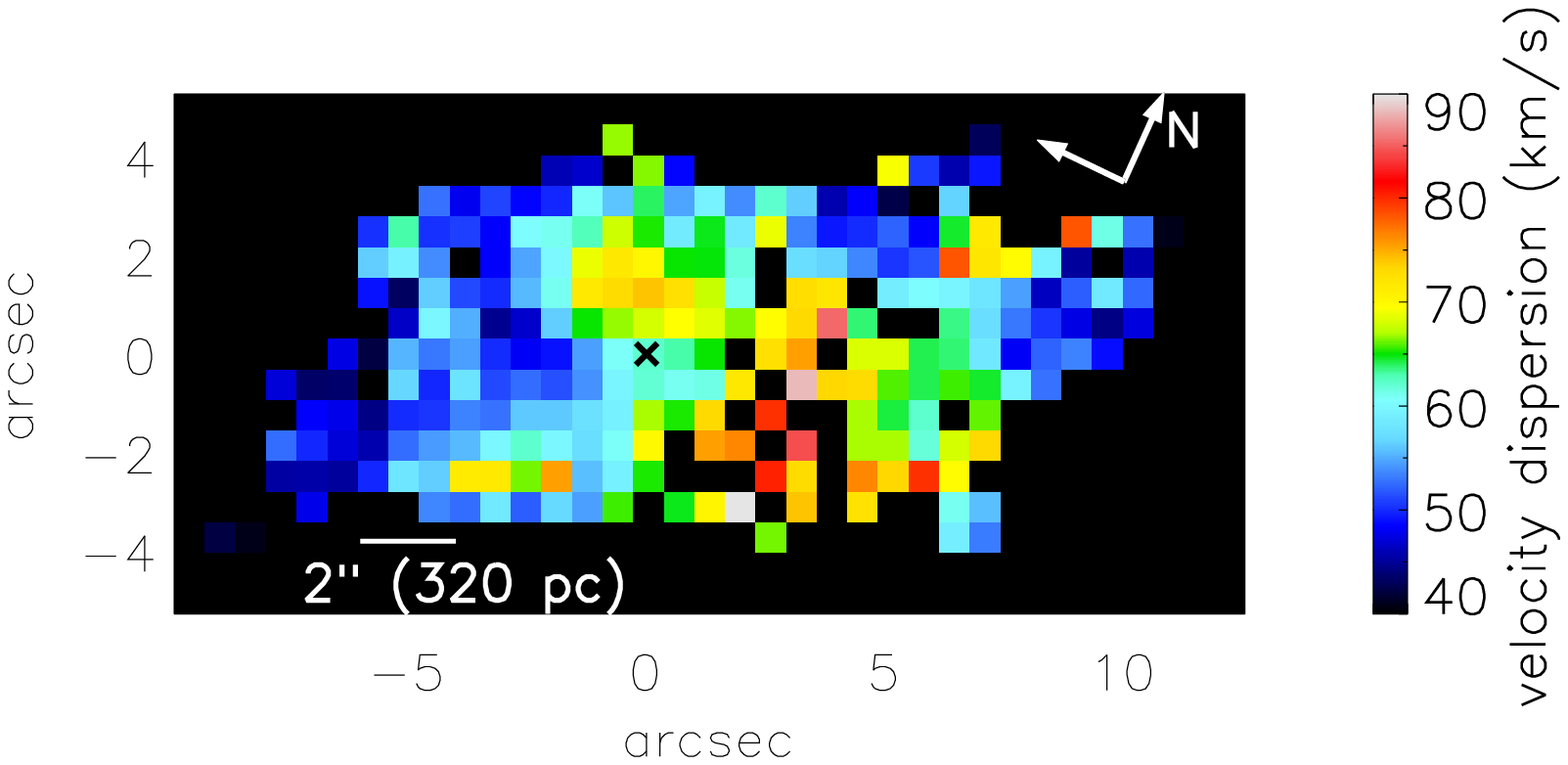,width=0.45\linewidth,clip=}\hspace{-0.2cm}
\epsfig{file=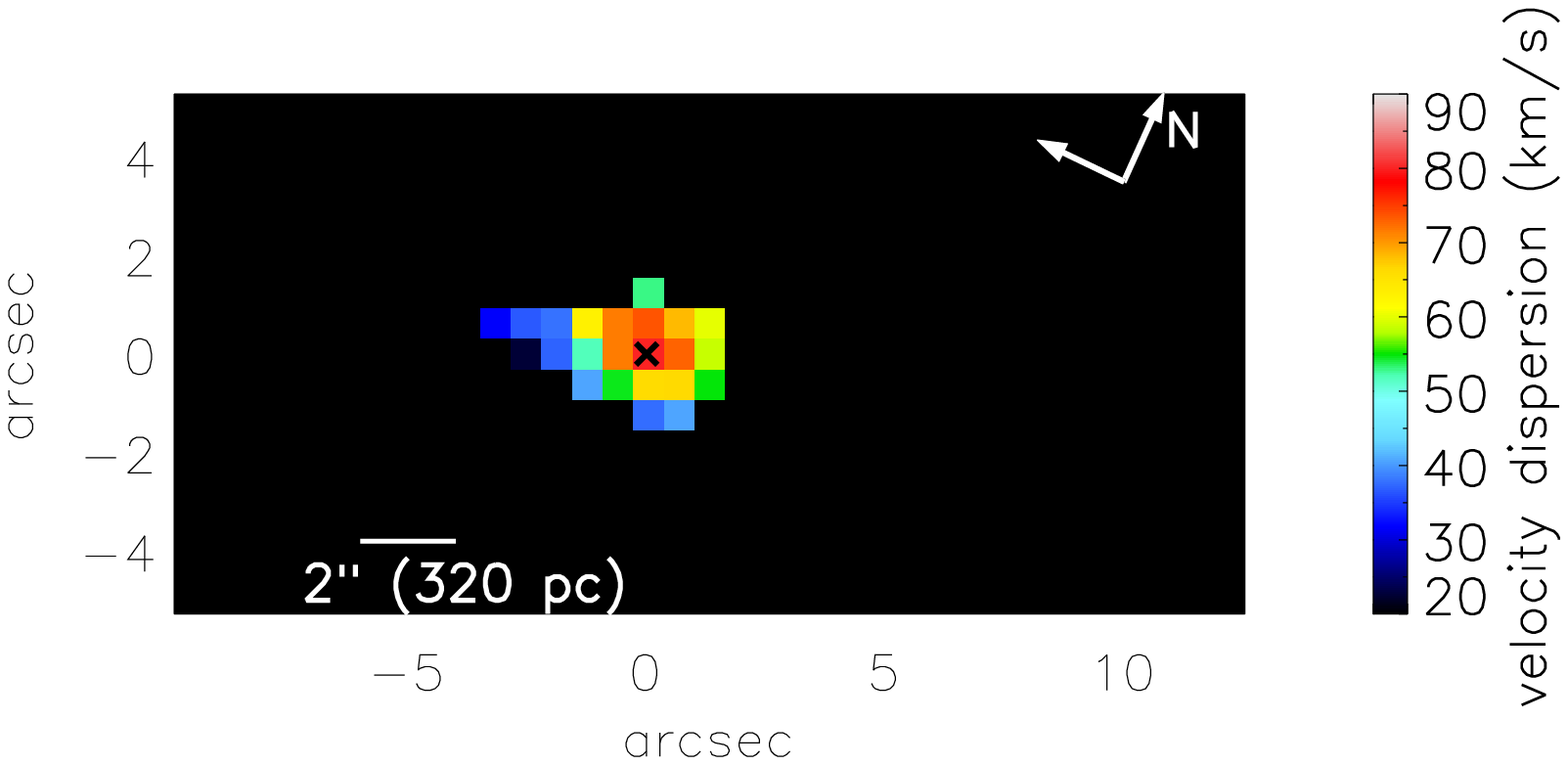,width=0.45\linewidth,clip=}\hspace{-1.8cm}\\
\vspace{-1.0cm}
\epsfig{file=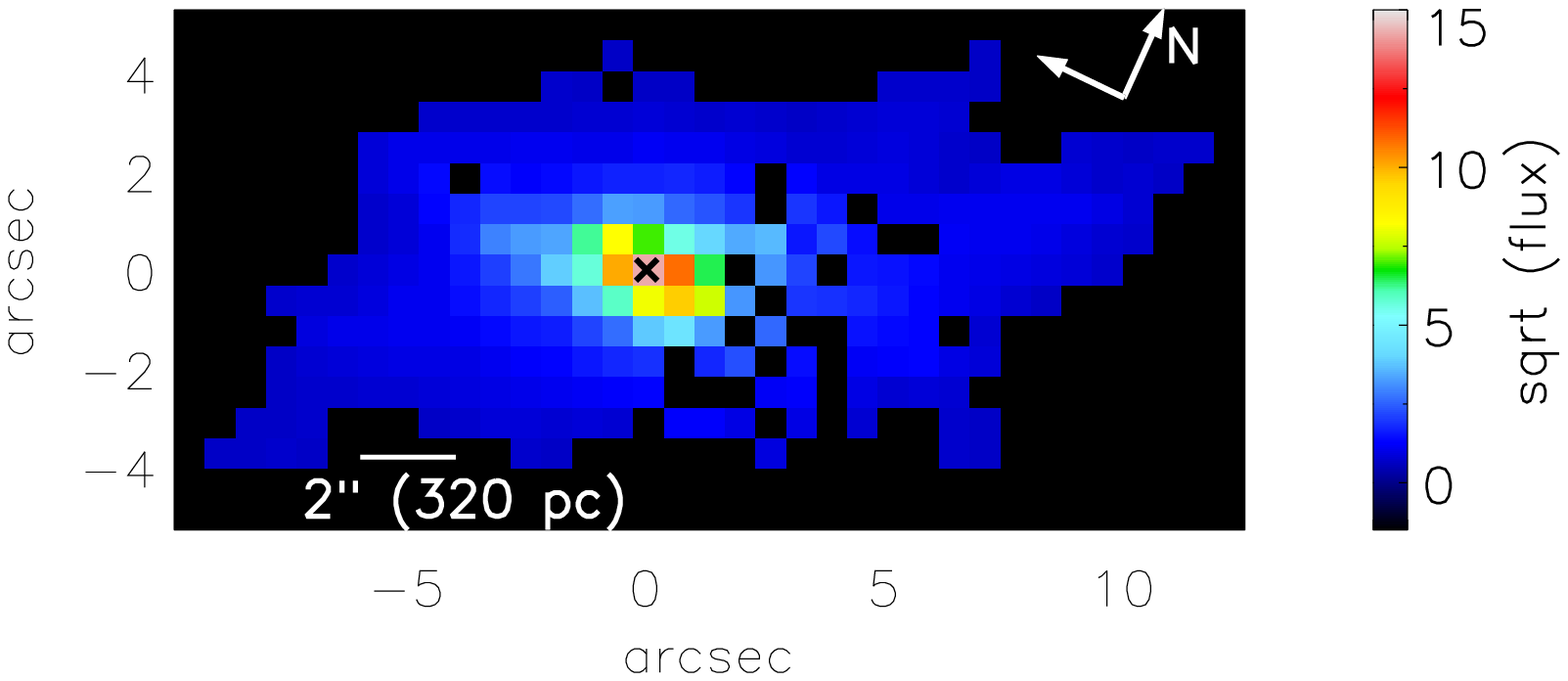,width=0.45\linewidth,clip=}\hspace{-0.2cm}
\epsfig{file=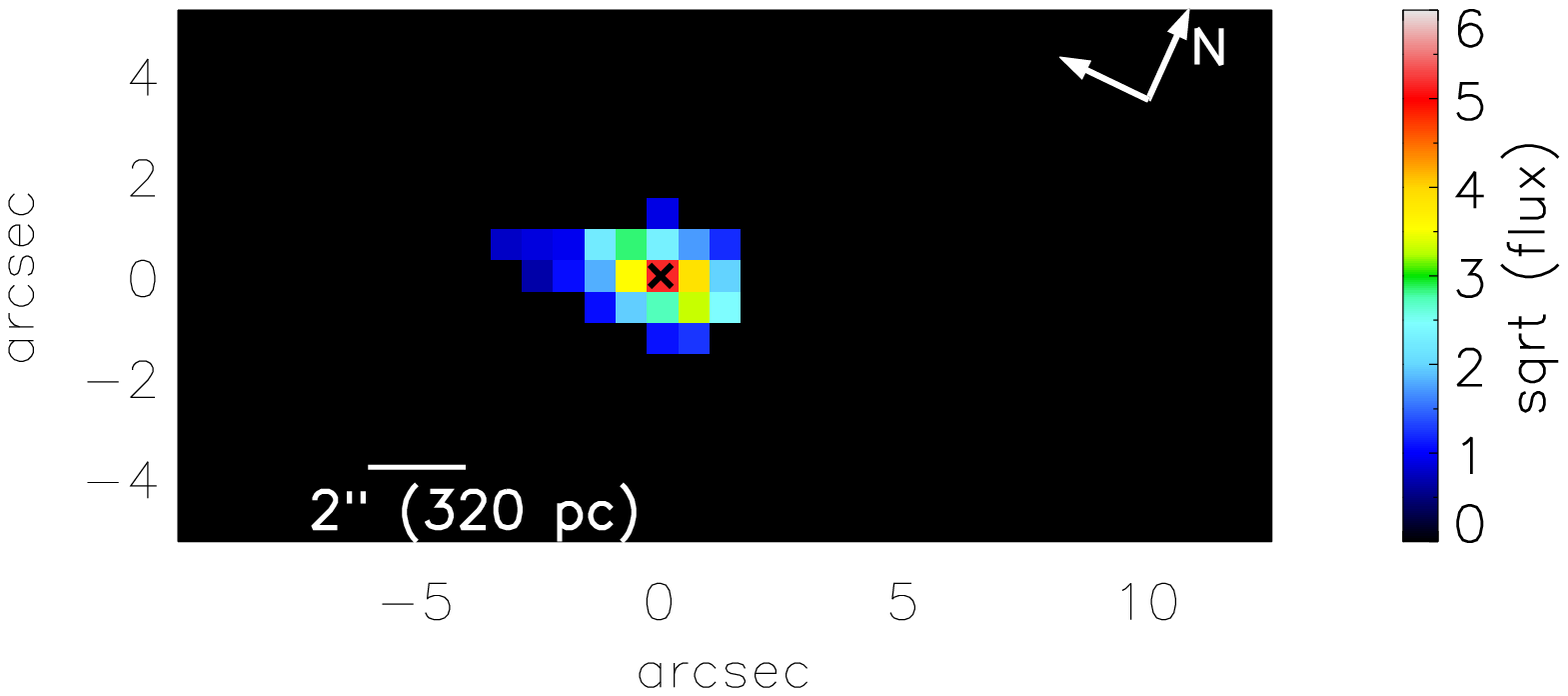,width=0.45\linewidth,clip=}
\vspace{-0.5cm}
\caption{Ionised gas optical line emission maps. Left column shows the [O III] line emission maps and the right column shows the narrow H$_{\beta}$ emission. Top row: Velocity maps. Centre: Dispersion maps corrected for instrumental broadening. Bottom: Flux maps. }
\label{optical_lines}
\end{figure*}

In the VIMOS data cube we detect two strong emission lines: [O III] 5007 \AA\ and H$_{\beta}$ 4861 \AA\ (with a nuclear broad component and a spatially extended narrow component). To study the gas dynamics we first use the stellar absorption lines to determine the galaxy's systemic velocity. We integrate the data cube spatially in the region of the field-of-view with highest contribution from the galaxy to determine an integrated spectrum. We then use pPXF and a set of stellar optical templates \citep{vazdekis99} to determine the systemic velocity. This velocity is then use as a reference for the study of the gas dynamics. To analyse the H$_{\beta}$ narrow component we subtract the broad H$_{\beta}$ component from the spectra. As noted previously by \cite{reynolds97}, the broad H$_{\beta}$ line can be fitted by two components, with the strongest flux being in a component blueshifted by $\sim$ 150 - 200 km/s with respect to the systemic velocity of the galaxy. In Fig.~\ref{optical_lines} we show the line emission maps for the [O III] and the narrow H$_{\beta}$. Pixels with S/N $<$ 5 between the peak of the line and the noise in a line-free region of the continuum were masked out. The H$_{\beta}$ emission extends out to r $\sim 400$ pc, significantly more extended than the coronal lines detected in the near-infrared. Its velocity map shows a rotation pattern counter-rotating with respect to the main body of the galaxy. 

The [OIII] emission is the most extended gas emission we detect. The line is detected out to r $<10{''} \sim 1.5$ kpc above a S/N of 5. The dynamics in the region to the south-east of the nucleus match what we observe in the molecular gas (in terms of the rotational velocity  expected and counter-rotation). To the north-west of the nucleus the velocity dispersion increases and the velocity does not match the rotational pattern. We are likely observing the large scale counterpart of the gas outflow observed in [Ca VIII]. In this region to the north-west of the nucleus we observe velocities of 50 - 60 km/s at $r \sim 1$ kpc which roughly matches the decrease in velocity expected if $v_{\rm out} \propto r^{-0.4}$.

\subsection{Molecular gas}
\begin{figure*}
\centering
\epsfig{file=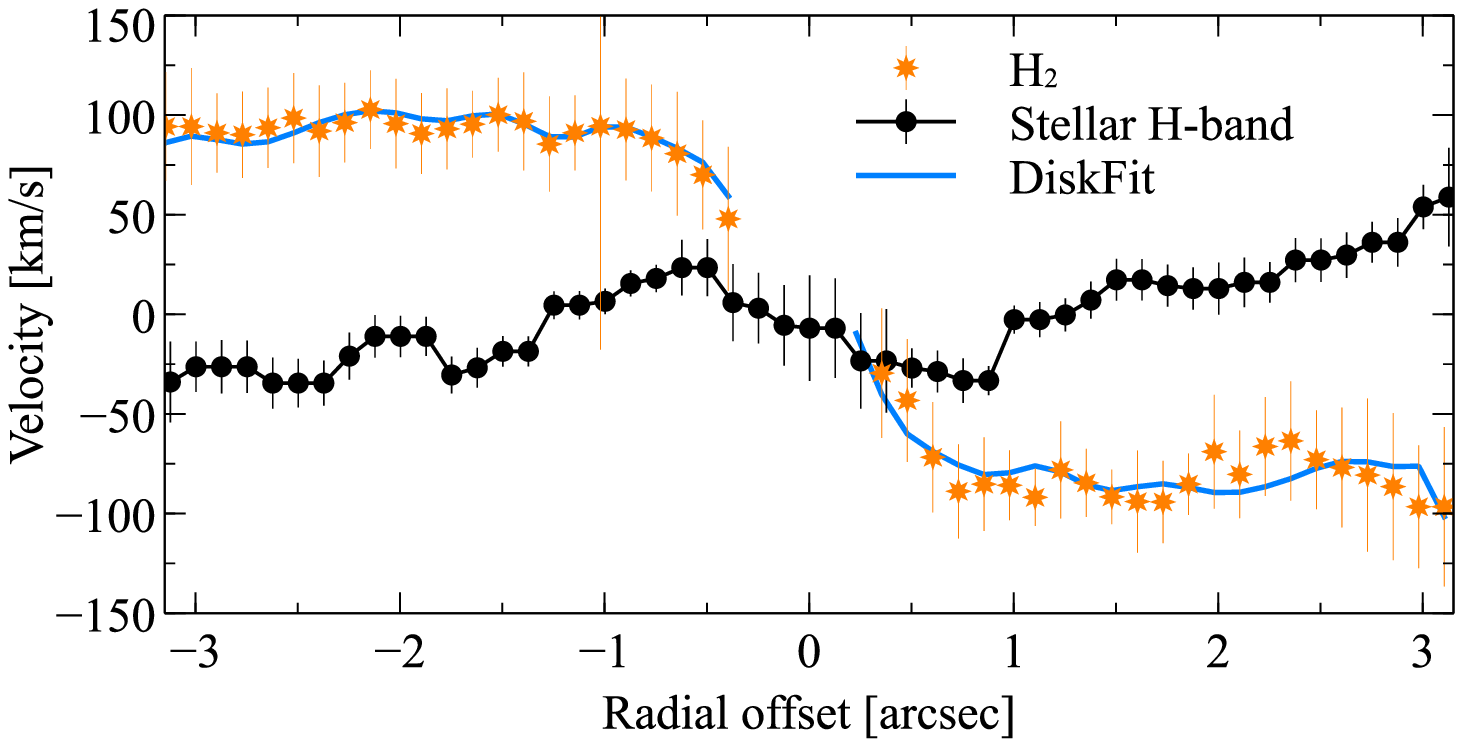,width=0.49\linewidth,clip=}
\epsfig{file=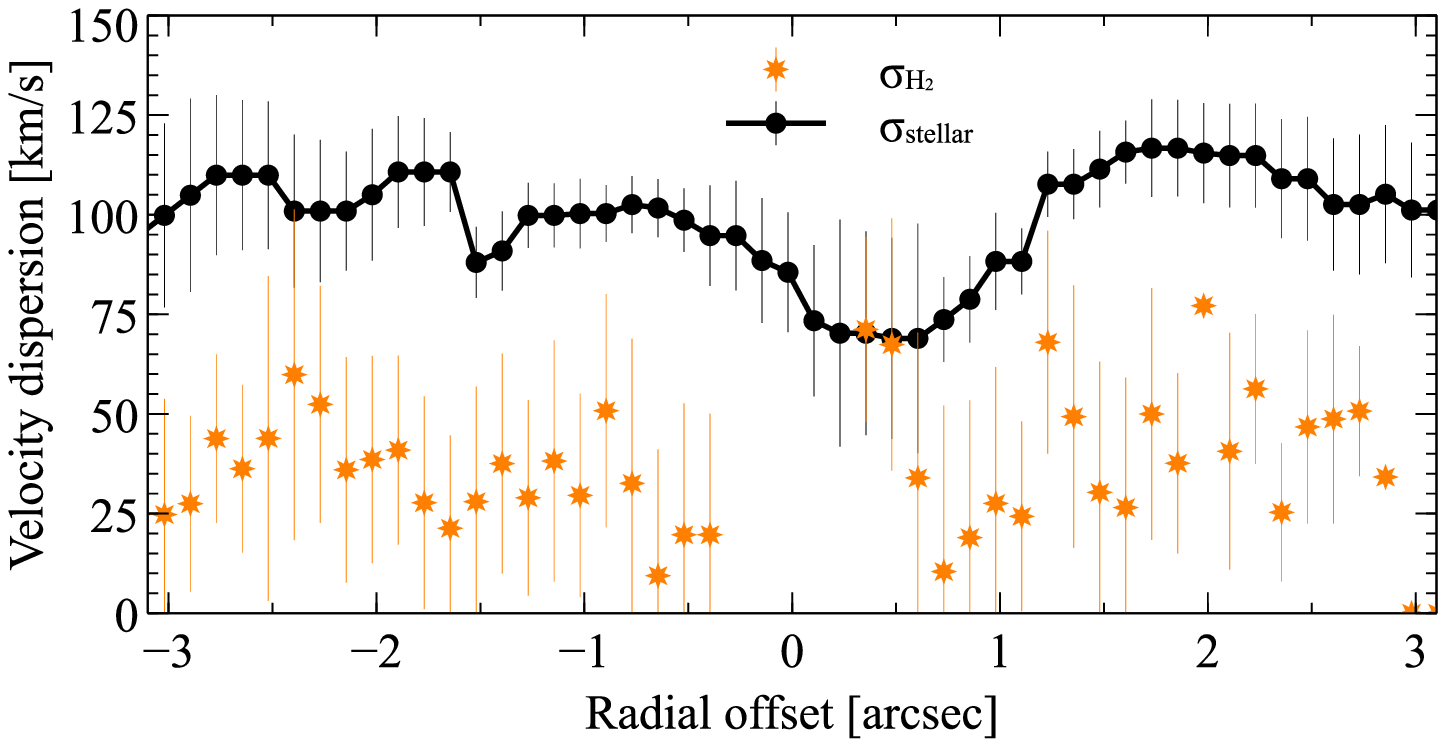,width=0.49\linewidth,clip=}
\caption{Horizontal cuts along PA $= 122.5$ deg for the molecular gas emission. The x-axis indicates the radial distance from the AGN position. Left: Velocity. Right: Velocity dispersion. The orange stars represent the H$_{2}$ properties while the blue solid line is the best fit velocity from the DiskFit routine.}
\label{horizontal_h2}
\end{figure*}

\begin{figure}
\centering
\includegraphics[width=0.9\columnwidth]{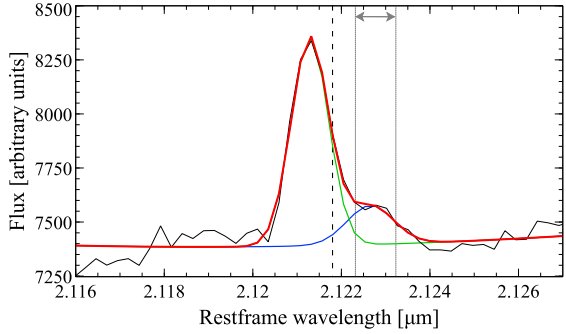}
\caption {Emission line profile of H$_{2}$ at the peak in line flux, 0${''}$.8 SW of the nucleus, integrated in a $0{''}.7 \times 0{''}.7$ region. The vertical dashed line indicates the line wavelength corresponding to the systemic velocity of the galaxy. The vertical dotted lines limit the region used in the analysis in Fig.~\ref{h2_redbump}.}
\label{h2_2lines}
\end{figure}

\begin{figure}
\centering
\includegraphics[width=0.9\columnwidth]{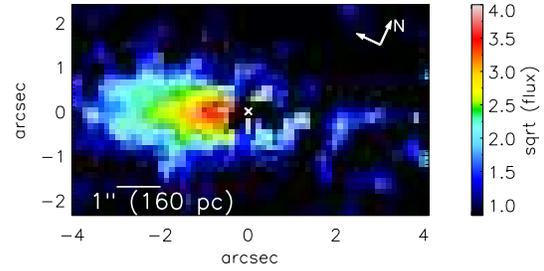}
\caption {Flux map of the H$_{2}$ line emission between velocities +70 km/s and +200 km/s (indicated by the vertical dotted lines in Fig.~\ref{h2_2lines}). In addition to the expected disc rotation component south-east of the nucleus, a weaker component is observed south-west of the nucleus and extending towards the west, apparently matching the distribution of flux residuals in the bottom right panel of Fig.~\ref{H2_vel}.}
\label{h2_redbump}
\end{figure}

The molecular gas is traced by the 1-0 S(1) H$_{2}$ distribution at 2.1218 $\micron$ which extends further out than the kinematically distinct core. However the direction of rotation for the gas is the same as for the distinct core - the molecular gas is also counter-rotating with respect to the main body of the galaxy, as can be seen in Fig.~\ref{H2_vel}. The emission line map was binned to a S/N of 5 and the pixels with S/N lower than 3 were masked out of the map. The regions with low line flux, less than 5 per cent of the peak value were also masked out. The velocity dispersion of the gas is low, typically $\sim$ 25 - 50 km s$^{-1}$. The molecular gas distribution is more disc-like than the stellar distribution, the axis ratio is $\sim$ 0.5 in comparison with the axis ratio of $\sim$ 0.6 from the stellar continuum. The H$_{2}$ rotational velocity is significantly higher than the H$_{2}$ velocity dispersion which indicates that the gas is rotationally supported. A comparison between the velocity and velocity dispersion of the molecular gas and stars along the major axis of the galaxy is shown in Fig.~\ref{horizontal_h2}. We use the method of \cite{krajnovic06} to determine the kinematic PA. Excluding the central r $<$ 0$''$.5 we find PA $ = 122.5 \pm 3.5$ degrees. The PA of the gas is marginally consistent with both the distinct core and the outer regions of the galaxy, indicating that the molecular gas shares the same orientation as the stars, i.e. counter-rotating with respect to the main body of the galaxy. As can be seen in Fig.~\ref{H2_vel}, the flux distribution for H$_{2}$ does not peak at the nucleus but in a region south-west of the nucleus. To better explore this asymmetry we fit elliptical isophotes to the H$_{2}$ flux map centred at the AGN position and having the same PA and ellipticity as the large scale H$_{2}$ distribution. In the bottom right plot of Fig.~\ref{H2_vel} we show the flux residuals after removing these elliptical isophotes from the original H$_{2}$ flux map. In this figure it is possible to see more clearly that there is a strong H$_{2}$ emission component 0${''}.8$ SW of the nucleus which extends to the West at least out to 2${''}$. In this region of higher flux, we also observe a second weaker component in the line emission spectrum, redshifted from what is expected from the rotation pattern. In Fig.~\ref{h2_2lines} we show the line profile integrated in a $0''.7 \times 0''.7$ region 0$''$.8 SW of the nucleus showing these two line components. 
The velocity of the gas in the redshifted component is $\sim$+120 km/s in a region where the H$_{2}$ disk shows a velocity of -75 km/s. The velocity dispersion is $\sigma \sim$ 57 km s$^{-1}$ which is slightly but not significantly higher than what we observe in the bulk disk component. In Fig.~\ref{h2_redbump} we show a map of the flux integrated within the waveband where we observe the redshifted ${`}$bump' component in the H$_{2}$ line emission. This waveband corresponds to velocities between [+70, +200] km/s indicated by the vertical dotted lines in Fig.~\ref{h2_2lines}. The map has been smoothed using a $3 \times 3$ boxcar median. When trying to identify the redshifted component at V $\sim$ 120 km/s we are also sensitive to the disc rotation redshifted component to the south east of the nucleus. In the region where the disc emission is blueshifted (v$_{\rm rot} \sim$ -50 km/s) we can detect the red bump at velocities $\sim$ +120 km/s. However, it is not clear if this red ${`}$bump' is present to the south-east of the nucleus as the main disc component there is expected to have velocities between 50 km/s and 100 km/s. In Fig.~\ref{h2_redbump} we can identify two regions: to the south-east of the nucleus we observe the expected redshifted component of the H$_{2}$ disc rotation; to the south-west flux residuals within the velocity interval $[+70, +200]$ km/s are observed, which illustrates the regions where the red ${`}$bump' emission is observed. Along the filamentary structure which extends to the West out to $\sim 2{''}$ we also observe that the H$_{2}$ line shows an asymmetric shape with an extended red wing, which can also be seen as emission in the map of Fig.~\ref{h2_redbump}. However, the S/N in this region is lower and we cannot obtain robust measurements of the velocity or velocity dispersion of the redshifted component here.

\begin{figure*}
\centering
\epsfig{file=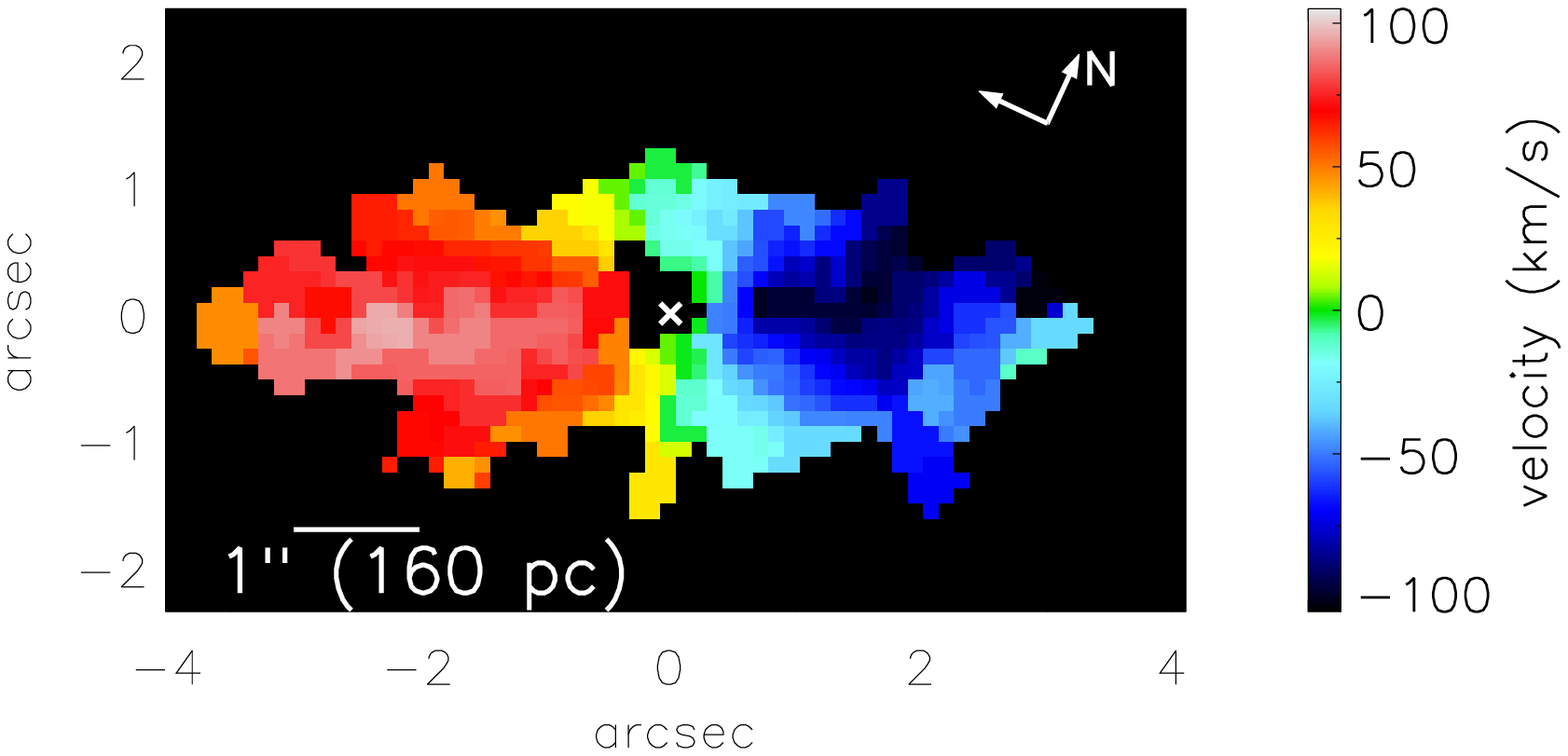,width=0.335\linewidth,height=4.0cm,clip=}\hspace{-0.15cm}
\epsfig{file=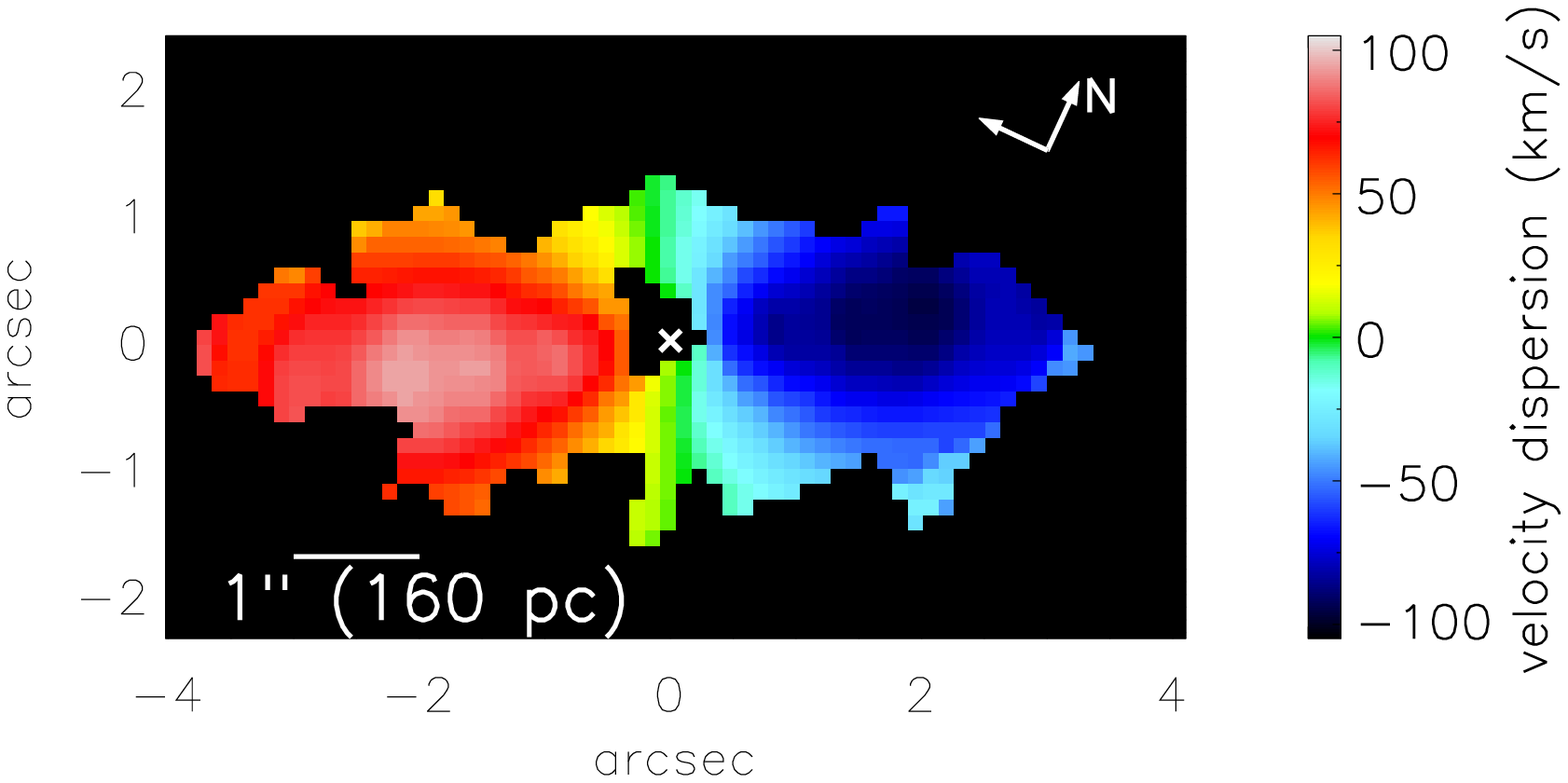,width=0.335\linewidth,height=4.0cm,clip=}\hspace{-0.15cm}
\epsfig{file=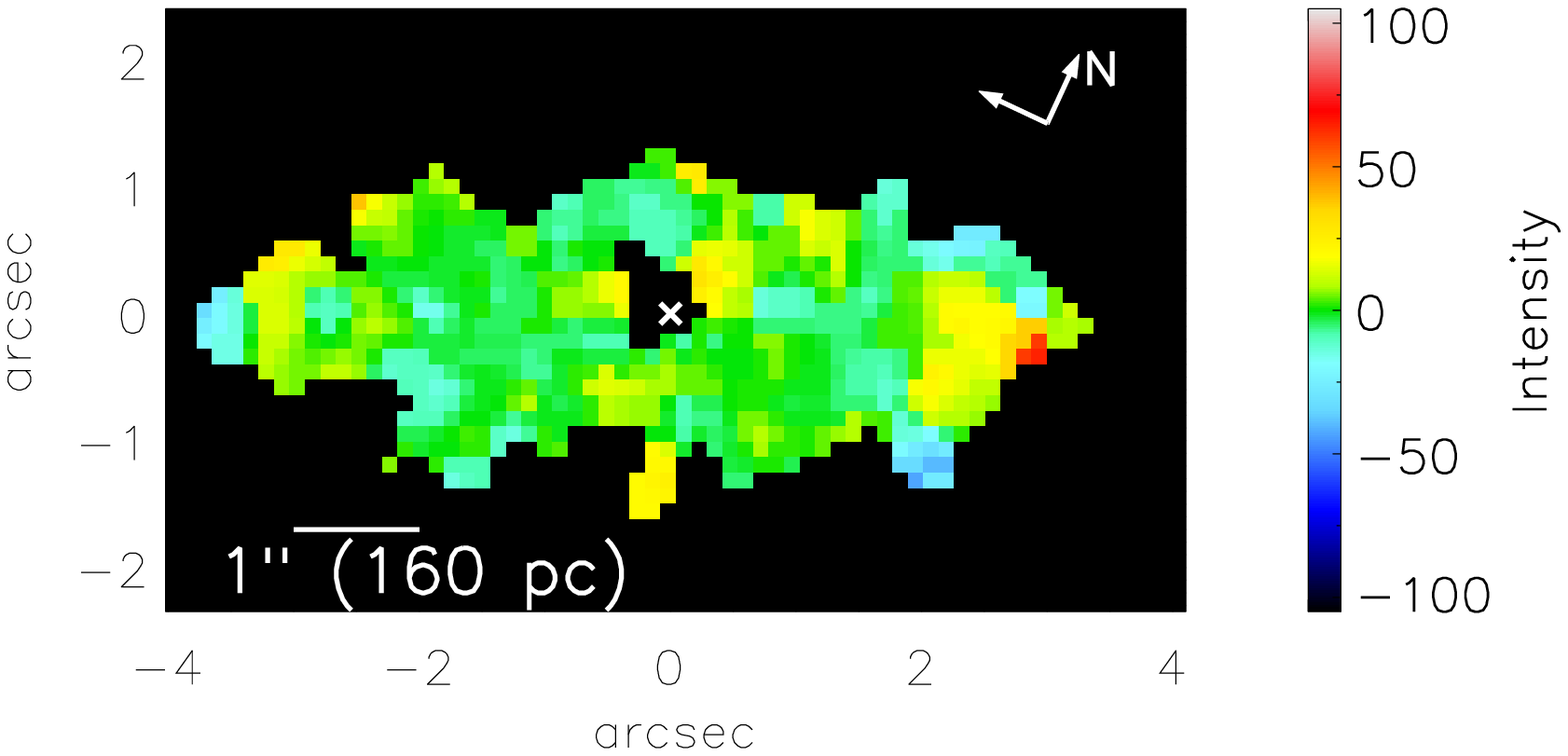,width=0.335\linewidth,height=4.0cm,clip=}
\caption {Velocity field, best fit Keplerian disc model from \textsc{diskfit} and velocity residuals for the H$_{2}$ emission. The maps were smoothed using a $3 \times 3$ pixel median.}
\label{h2_model}
\end{figure*}

The H$_{2}$ velocity map is modelled using the code \textsc{diskfit} version 1.2 (\citealt{spekkens&sellwood07}, \citealt{sellwood&sanchez10}, \citealt{sellwood&spekkens15}). We apply \textsc{diskfit} to our H$_{2}$ velocity field to fit the circular speed of the gas at each radius. The code minimises the $\chi^{2}$ difference between a model convolved with the PSF of the observations and the input data, which in our case is the 2D velocity field derived from the H$_{2}$ line fit. The code also takes into account the uncertainties in the velocity measurements. The input error map is obtained by doing Monte Carlo simulations on our line fit considering the typical noise observed in the spectra and varying the initial parameters for the fit. We fit a simple axisymmetric disk rotation model to the data, with the inclination, position of the centre, systemic velocity and position angle as free parameters. The data, best fit model and residuals are shown in Fig.~\ref{h2_model}. The systemic velocity used is the one calculated by \textsc{diskfit} (V$_{sys}$ = 2392 $\pm$ 1 km/s), which is lower by $\sim$18 km/s in relation to the systemic velocity determined from the stellar velocity field. The velocity field is well reproduced by a rotating disc with a PA $ = 122.4 \pm 1$ degrees consistent with what was found with the method of \cite{krajnovic06}, ellipticity of $(1 - b/a) = 0.51 \pm  0.03$ and a disc inclination of $i = 60.9 \pm 1.9$ degrees. The errors were obtained using a bootstrapping method. The velocity residuals have values lower than 25 km/s in general which shows that the H$_{2}$ dynamics can be relatively well represented by rotation in a disc. There are no significant trends in the velocity residuals which may indicate that due to its lower flux, the redshifted component observed in the SW region does not affect significantly the velocity measurements. The largest velocity residuals from \textsc{diskfit} are observed $\sim$3\arcsec\ south-west of the nucleus in a region that coincides with the residuals in flux after subtracting isophotes and also with a region of higher velocity dispersion. The flux and velocity residual and the higher velocity dispersion in this region could be caused by an outflow superimposed on rotation or by gas flow in the disc departing from circular rotation as observed in NGC 3227 \citep{davies14}.
In Fig.~\ref{h2_contours} we compare the spatial distribution of the H$_{2}$ flux with the position of the dust lane just south of the nucleus. We use the HST archival image of MCG--6-30-15 taken in the WFC3/UVIS1 medium-band filter f547M with peak wavelength at 5475 \AA. The two panels on the left show the flux contours and the two panels on the right show the flux residual contours (after subtracting elliptical isophotes) superimposed on the HST image. The H$_{2}$ emission seems to originate from a large scale region which includes the dust lane. The contours of the flux residuals seem to indicate that this flux component is coming from the dust lane. This can be an explanation for the redshifted H$_{2}$ component we observe in this region which also appears in the region of the flux residuals as can be seen in Fig.~\ref{h2_redbump}. We are likely in the presence of two components, one from a rotating disc close to the AGN and another originating in the dust lane and therefore presenting a different velocity. If the dust lane is between us and the centre of the galaxy, with our line of sight to the nucleus passing just above the dust lane (as it seems to be based on the obscuration of the optical light and from the X-ray study of the warm absorber in this galaxy - \citealt{ballantyne03}), we would expect to observe a blueshifted H$_{2}$ component with respect to the systemic velocity of the galaxy. The fact that we observe a redshifted velocity indicates that the gas in the dust lane is moving towards the centre of the galaxy.
Although the H$_{2}$ flux residuals appear to be spatially coincident with the dust lane, this may be a coincidence. If molecular clouds in the disc of the galaxy are being excited by ultraviolet fluorescence from the AGN, then the redshifted `bump' could be associated with the receding side of the ionisation cone that is associated with the [Ca VIII] emission as well. Unfortunately the S/N in the H$_{2}$ line does not allow us to rule out one of these scenarios. Better S/N observations of H$_{2}$ could trace the presence (or absence) of emission in the dust lane to the south-east of the nucleus. High resolution observations in the submillimeter could also trace colder molecular gas and determine if its distribution matches the dust lane.

\subsubsection{Molecular gas excitation and molecular gas mass}
The main processes responsible for H$_{2}$ emission are collisions with warm gas (T$\sim$ 2000 K) which has been heated by shocks, either from supernova or AGN driven outflows, or X-ray irradiation (i.e. thermal processes), or by UV fluorescence, where excitation by a UV photon produces a radiative cascade through various rotational and vibrational states of H$_{2}$ (i.e. non-thermal processes) (e.g \citealt{mazzalay13}). Each of these mechanisms will produce a different emission spectrum and in theory it is possible to determine which one dominates.
The only other transition of H$_{2}$ we detect is 1-0 S(3) 1.9576 $\micron$. Its spatial distribution follows the one of 1-0 S(1) but at a lower S/N level and it is only significantly detected in the region of highest H$_{2}$ flux south-west of the nucleus. The line ratio  1-0 S(3)/1-0 S(1) appears to be high ($\gtrsim$ 1) south-west of the nucleus which would suggest that excitation by thermal processes may dominate (e.g. \citealt{rodriguez-ardila04}). However, the 1-0 S(3) line is in a spectral region affected by telluric absorption and therefore we cannot robustly determine the line ratios.
For the reason given in the previous sentence, we will use the 1-0 S(1) H$_{2}$ line for our diagnostics and will refer to it as H$_{2}$ in the following discussion.
Another option is to compare the line ratios of H$_{2}$/Br$_{\gamma}$. In the entire region where Br$_{\gamma}$ is detected (r $<$ 0${''}$.8) H$_{2}$/Br$_{\gamma} < 1$, but Br$_{\gamma}$ is not significantly detected outside the PSF. H$_{2}$ is typically more intense relative to Br$_{\gamma}$ in Seyfert 2 galaxies than in starburst galaxies which may be an indication that AGN make an important contribution to the excitation mechanisms of molecular gas \citep{riffel06}. In our source the ratio is lower than one in the centre, but higher than one for the outer regions. This could indicate that in the central regions non-thermal processes contribute to the excitation of H$_{2}$ while in the outer regions we see an increased contribution from shock excitation \citep{valencia-s12,mazzalay13}.
\begin{table*}
\begin{center}
\textsc{Table 2}
\vspace{0.1cm}

\textsc{Molecular gas mass estimates}\\

\vspace{0.2cm}
\begin{tabular}{ | l | c | c | c |}
\hline
\hline

Spatial region & Flux [W/m$^{2}$] & Luminosity [erg/s] & Total molecular gas mass [M$_{\odot}$]\\
\hline
r $<$ 3${''}$.5 & 1.4$\times 10^{-18}$ &  2$\times 10^{38}$ &  5.8$\times$10$^{7}$\\
r $<$ 1${''}$.25 (KDC) & 8.4$\times 10^{-19}$ & 1$\times 10^{38}$ &  3.4$\times$10$^{7}$ \\
Redshifted emission & 4.2$\times 10^{-20}$ & 5.5$\times 10^{36}$ & 1.7$\times$10$^{6}$ \\

\hline
\end{tabular}
\end{center}
\caption{Estimates for H$_{2}$ flux, H$_{2}$ luminosity and total molecular gas mass for various spatial regions in the galaxy. The spatial region is defined by the radius of the circular aperture where the flux is integrated. For the redshifted emission the region of integration is 0${''}$.7 x 0${''}$.7.}
\label{h2_mass_table}
\end{table*}

The H$_{2}$ line flux was integrated to determine the luminosity and estimate the total molecular gas mass. The redshifted emission line component is also integrated to determine its luminosity. 
Determining the total gas mass requires first the conversion of H$_{2}$ luminosity to warm H$_{2}$ gas mass. To then determine the total gas mass it is necessary to know the ratio of cold gas mass to warm gas mass.
Converting the 1-0 S(1) H$_{2}$ luminosity to total gas mass is not trivial. There are several estimates for a conversion between these two values from studies which compare the H$_{2}$ luminosity with total gas mass derived from sub-mm observations of CO for various samples of galaxies (e.g. \citealt{dale05}, \citealt{muller-sanchez06}, \citealt{mazzalay13}).

Here we use the conversion from \cite{mazzalay13}, who determine the conversion factor by studying the inner regions ($r < 300$ pc) of six nearby galaxies (consisting of low-luminosity AGN, Seyferts and one quiescent galaxy). They find that a factor of $\beta$ = 1174 M$_{\odot}$/L$_{\odot}$ provides a good correlation between the independent estimates of molecular gas mass (from CO observations and from H$_{2}$ observations):
\begin{eqnarray}
\frac{M_{gas}}{M_{\odot}}\sim 1174 \times \frac{L_{1-0 S(1)}}{L_{\odot}},
\end{eqnarray}
\noindent Where M$_{gas}$ is the total molecular gas mass. We do the calculations in the masked flux map so that we are confident that we are measuring the flux in H$_{2}$. Therefore the values we obtain are a lower estimate for the total flux. Extinction will also lower the H$_{2}$ flux we measure with respect to its intrinsic value. Pa$_{\alpha}$ is not within the wavelength range of our observations and therefore we are not able to derive accurate line ratios from the hydrogen recombination lines. However the effect of extinction will also make our measured value of H$_{2}$ flux to be a lower estimate for the total flux. The estimates are summarised in Table~\ref{h2_mass_table}.
\subsection{Stellar population}
\label{sec:stellar_pop}
\begin{figure}
\centering
\epsfig{file=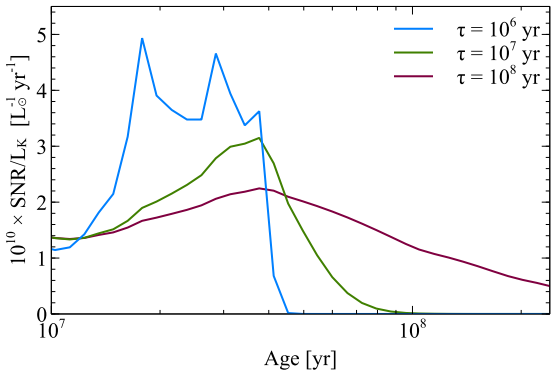,width=0.9\linewidth,clip=}
\caption{\textsc{stars} theoretical profiles of the SNR/L$_{K}$ ratio as a function of stellar age for three different star formation decay timescales ($\tau$).}
\label{stars}
\end{figure}

The distribution and kinematics of [Fe II] indicate dynamics distinct from the ionised species of [Si VI] and [Ca VIII] which trace outflowing gas (Fig.~\ref{horizontal}). As described in Section \ref{sec:feii} we argue that supernova shocks are the main excitation mechanism for [Fe II]. As the [Fe II] emission is restricted to the region of the distinct core, with an underlying counter-rotating velocity component, it is likely that these supernova events are a consequence of the formation of the stellar counter-rotating core. The supernova rate we are observing results from the stellar evolution of the counter-rotating population of stars that were formed from the initial inflow of gas. Using the measured [Fe II] flux and stellar evolution models we can derive constraints on the age of the stellar population. The [Fe II] flux we measure is higher than our previous estimate \citep{raimundo13}, mainly due to the larger field-of-view, which allow us to determine more accurately the extent of the counter-rotating core, and due to the higher S/N in the [Fe II] line emission in our new data. Using Eq. 2 in \cite{raimundo13} taken from \cite{rosenberg12}, and a theoretical ratio of [Fe II] $\lambda$ 1.64 \micron /[Fe II] $\lambda$ 1.26 \micron\thinspace= 0.7646 \citep{nussbaumer&storey88} we can convert the spatially integrated [Fe II] luminosity into a supernova rate estimate:
\begin{figure*}
\centering
\includegraphics[width=1.05\columnwidth]{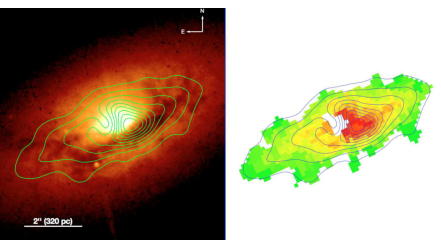}\hspace{-0.15cm}
\includegraphics[width=1.05\columnwidth]{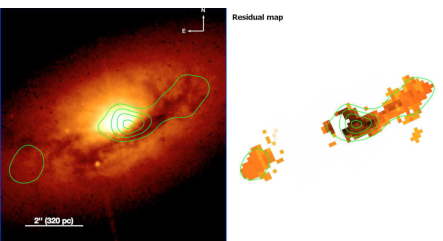}
\caption {H$_{2}$ flux contours overlaid in the HST WFC3/UVIS1 f547M archival image of MCG--6-30-15. Left pair of panels: Flux contours. Right pair of panels: Flux residual contours.}
\label{h2_contours}
\end{figure*}

\begin{eqnarray}
\log \frac{\rm SNR}{[\rm yr^{-1}]} = (0.89 \pm 0.2) \times \log \frac{L_{\rm [Fe II] 1.64\thinspace\micron}}{0.7646\thinspace[\rm erg\thinspace s^{-1}]} - 36.19 \pm 0.9
\end{eqnarray}
Which gives SNR = 1.9$\times$10$^{-2}$ yr$^{-1}$ in the region of the distinct core. To determine the age of the stars which would produce the supernova rate we currently observe, we used the evolutionary synthesis code \textsc{stars} \citep{sternberg98,thornley00,sternberg03}, which models the evolution of individual stellar clusters. We evolved a stellar population using \textsc{stars} to determine the predicted SNR and K-band luminosity for a specific stellar age, similar to the analysis in \cite{raimundo13}. We assume two possible values of 1 Myr and 10 Myr as the characteristic exponential decay timescales of the star formation ($\tau$) and a Salpeter initial mass function for the stellar evolution modelling. We calculate the luminosity in the K-band from the absolute magnitude using the relation in \cite{davies07}: $M_{K} = -0.33 -2.5\thinspace {\rm log\thinspace} L_{K}$, where $L_{K}$ is the total luminosity in the K-band in units of bolometric solar luminosity ($L_{\odot}$). We obtain $L_{K} = 5.1 \times 10^{7}\thinspace L_{\odot}$ in the region of the distinct core. To use \textsc{stars} as a diagnostic on the stellar age, we compare results normalised to our K-band luminosity as in \cite{davies07}, i.e. we compare the SNR/L$_{K}$ ratio predicted by \textsc{stars} and the one we observe. In Fig.~\ref{stars} we show the predicted evolution of the SNR/L$_{K}$ as a function of stellar age. Considering the higher SNR obtained from our new data and the K-band luminosity within the distinct core (r $< 1{''}.25$), we estimate 10$^{10} \times$ SNR/L$_{K}\sim 3.7$ which indicates a stellar age of around 40 Myr and a mass-to-light ratio of M/L$_{\rm K}$ =  0.7 M$_{\odot}/L_{\odot}$ for a characteristic star formation timescale $\tau = 10^{6}$ yr, which agree within the uncertainties with our previous estimate of 65 Myr \citep{raimundo13}. Assuming that all the [Fe II] flux is due to supernova events provides an upper limit estimate for the SNR. If the AGN contributes to the excitation of [Fe II] then the values for the stellar age that we estimate are lower limits. What we can conclude from this analysis is that there is strong indication of at least one recent starburst in the nucleus of MCG--6-30-15, with an age younger than 100 Myr, most likely close to 50 Myr. 

From the \textsc{stars} modelling we can determine the current mass in stars for our SNR and $L_{K}$. Within the KDC we calculate that the mass in stars is M$_{*} \sim 10^{6}$ M$_{\odot}$ (considering $\tau = 10^{6}$ yr or $\tau = 10^{7}$ yr), while the molecular gas mass estimated from the previous section is: M$_{\rm mol\thinspace KDC}$ = 3.4 $\times$ 10$^{7}$\thinspace M$_{\odot}$ which is significantly larger. This indicates that the starburst which formed the KDC did not exhaust the total gas mass that was accreted to the centre of the galaxy. It is possible that another starburst will occur in the near future from the molecular gas reservoir we observe.
\subsection{Dynamical modelling}
\begin{figure}
\centering
\epsfig{file=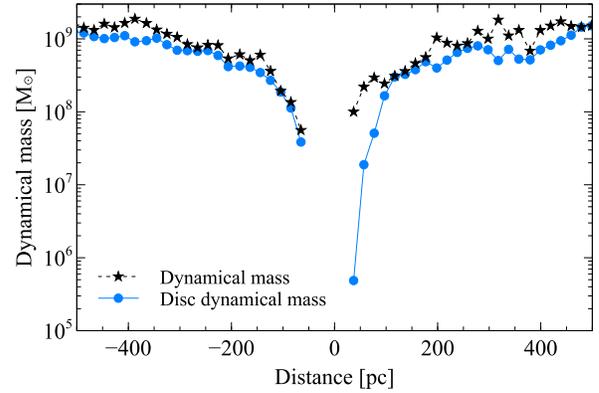,width=1.0\linewidth,clip=}
\caption{Dynamical mass (total enclosed mass) as a function of distance from the black hole based on the H$_{2}$ kinematics for two types of distributions: first considering that the velocity dispersion makes a significant impact in the dynamics (black points) and second considering that the distribution is approximately a disc (blue points).}
\label{dynamical_mass}
\end{figure}

One of our goals is to understand the dynamics in the central regions of MCG--6-30-15. In our previous work we observed that the second velocity moment of the line-of-sight velocity distribution was not very sensitive to the presence of a black hole with mass lower than $\sim\thinspace10^{7}$\thinspace M$_{\odot}$ \citep{raimundo13}. Here we explore various approaches to better constrain the black hole mass, based on the stellar and gas dynamics. The ionised gas dynamics show that this gas component is affected by the AGN outflow and possibly supernovae shocks as well and therefore is not a good tracer of the underlying gravitational potential. The molecular gas however rotates in a disc like structure, and although there is an area of the map that shows disturbed dynamics (and possible inflow), most of the H$_{2}$ dynamics appears to be dominated by the central gravitational potential. Since the H$_{2}$ emission is consistent with rotation in a thin disc, its velocity distribution can be used as a first estimate of the enclosed mass: 

\begin{eqnarray}
M_{\rm dyn}(r)=\frac{(V_{\rm obs}/{\rm sin}(i))^{2} + 3\sigma^{2}}{G}\thinspace r,
\end{eqnarray}
Where V$_{\rm obs}$ is the observed rotational velocity, $i$ is the disc inclination (with $i$ = 0 degrees being face on and $i$ = 90 degrees being edge on), $\sigma$ is the velocity dispersion of the gas and $r$ the distance from the centre of the galaxy (which is defined as the AGN position). If we assume that the gas is rotationally dominated and the velocity dispersion contribution to the dynamics is negligible (in Fig.~\ref{horizontal_h2} we see that V/$\sigma \sim$ 3 for an inclination of 61 degrees) then:

\begin{eqnarray}
M_{\rm dyn}(r)=\frac{V_{\rm obs}^{2}}{G\thinspace {\rm sin}(i)^{2}}\thinspace r,
\end{eqnarray}
An estimate of the total (stars + gas + dark matter) enclosed dynamical mass at various radii is obtained from the H$_{2}$ kinematical modelling (Fig. \ref{dynamical_mass}). If we consider the innermost radius where we can get an accurate measurement of the H$_{2}$ dynamics, and use the average velocity and velocity dispersion measured at both sides of the nucleus along the H$_{2}$ PA, and the inclination measured from {\sc diskfit} we obtain an upper limit for the enclosed mass within the PSF of our data, r $< 0.{''}5$ (r $\lesssim$ 80 pc) and therefore for the black hole mass of: 
M$_{\rm enc} < 1.8 \times 10^{8}$\thinspace M$_{\odot}$ or if we consider the velocity dispersion negligible: M$_{\rm enc} < 7.8 \times 10^{7}$\thinspace M$_{\odot}$.
\subsubsection{Jeans anisotropic model}
To model the overall potential of the galaxy we use the Jeans Anisotropic Models (JAM) method \citep{cappellari08}. The general goal in dynamical modelling is to determine the underlying gravitational potential, or the mass, from the position and velocity of a sample of stars. The Jeans equations reduce the problem of determining the distribution function (position and velocity) of stars by focusing on determining only the velocity moments of the distribution function.  Axisymmetry is assumed to reduce the complexity of the problem: in cylindrical coordinates the symmetry is assumed in relation to the coordinate $\phi$. Furthermore a `semi-isotropy' condition is often assumed, where $\overline{v_{R}^{2}} =  \overline{v_{Z}^{2}}$. In the anisotropic Jeans solutions (\citealt{cappellari08}), the assumptions are not as restrictive, and a constant anisotropy factor is included to account for possible anisotropies: $b = \overline{v_{R}^{2}}/ \overline{v_{Z}^{2}}$.

\begin{figure*}
\centering
\epsfig{file=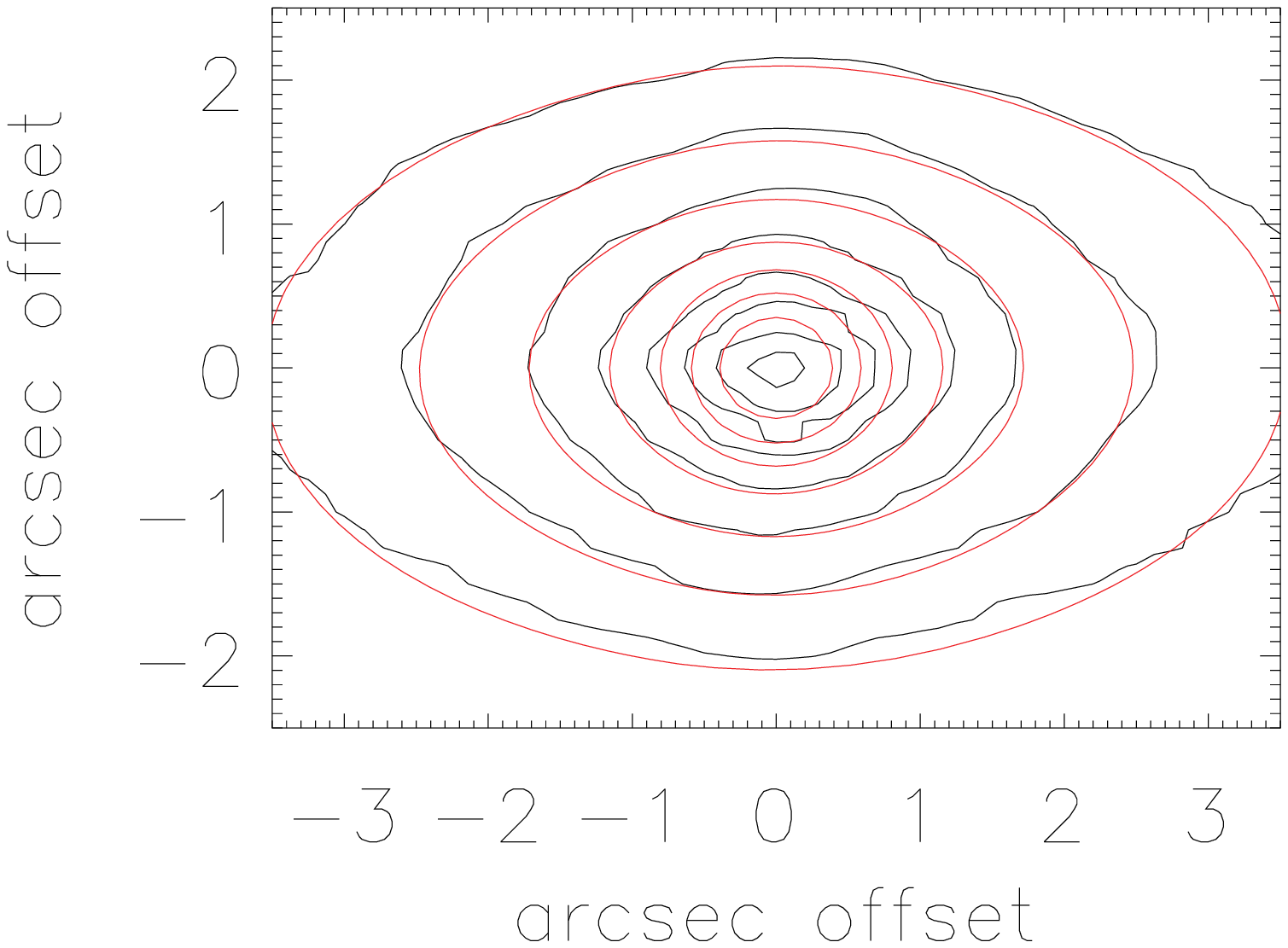,width=0.45\linewidth,clip=}
\epsfig{file=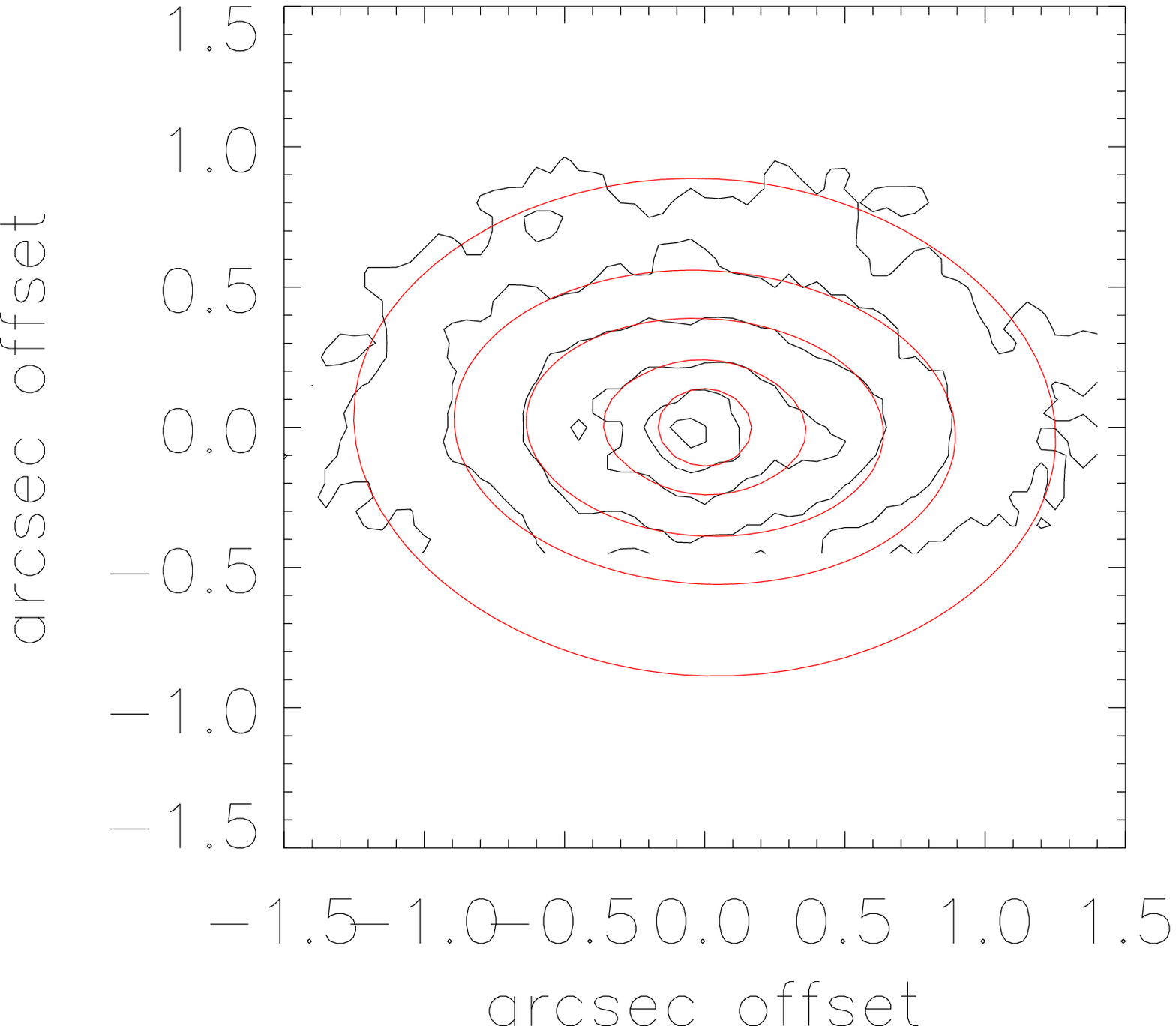,width=0.4\linewidth,clip=}
\caption{MGE best-fit parameterisation (red contours) for the H-band stellar continuum (black contours). Left: Larger field-of-view. Right: Fit to the nuclear regions using the higher resolution data of \citealt{raimundo13}.}
\label{mge_stellar_cont}
\end{figure*}

As input for the dynamical model, we first start by modelling the galaxy surface brightness from the stellar emission. The surface brightness is more accurately modelled in the H-band, the signal to noise ratio is higher in the CO absorption bands detected in this waveband and therefore the stellar continuum is better isolated. We use the Multi-Gaussian Expansion (MGE) method of \cite{emsellem94} and the MGE fitting software developed by \cite{cappellari02} to parametrise our surface brightness in terms of gaussians. To determine the MGE decomposition we used the H-band image with the hydrogen broad Br$_{\gamma}$ and the AGN continuum removed, from our method of AGN and stellar continuum decomposition based on the equivalent width of the CO absorption lines (see Section~\ref{sec:continuum} and Fig.~\ref{stellar_agn_cont}). The stellar continuum data cube is then integrated in wavelength to obtain the stellar emission at each pixel of the image. In the MGE gaussian decomposition, the minimum gaussian axial ratio determines the minimum inclination allowed in the subsequent dynamical model. We therefore explore the $\chi^{2}$ variation of the gaussian fit as the minimum axial ratio is changed and set a limit for the minimum ratio allowed in MGE, as suggested in \cite{cappellari02}. For the H-band the minimum projected axis ratio is set at 0.56. The best fit gaussians are plotted in Fig.~\ref{mge_stellar_cont} - left. To model the inner regions of the field-of-view we use our previous H-band data at resolution of $0{''}.1$. The MGE modelling for the nuclear region is shown in Fig.~\ref{mge_stellar_cont} - right.

The MGE parameters which characterise the Gaussians are converted to the units required by the \textsc{jam} code using the guidelines for the \textsc{mge fit sectors} package release by Michele Cappellari. The input for \textsc{jam} is then the stellar surface brightness based on the \textsc{mge} parameterisation (peak surface brightness, dispersion and observed axial ratio of each Gaussian),  the inclination, the distance, the observed kinematics and its associated uncertainties and (optionally) the black hole mass.

\begin{figure*}
\vspace{-1.0cm}
\centering
\epsfig{file=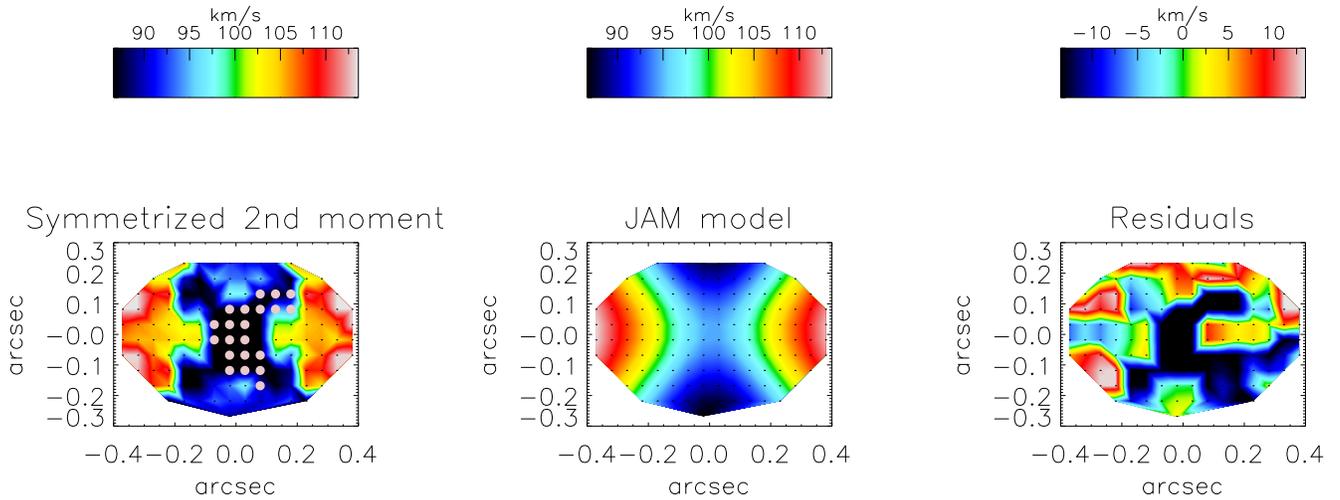,width=0.53\linewidth,angle=90,clip=}
\caption{JAM modelling. The second moment ($V_{\rm rms} = \sqrt{V^{2} + \sigma^{2}}$) of the stellar velocity distribution (left panel) is compared with the output from the model (middle panel). The residuals are shown on the right panel.}
\label{jam}.
\end{figure*}

\begin{figure}
\centering
\epsfig{file=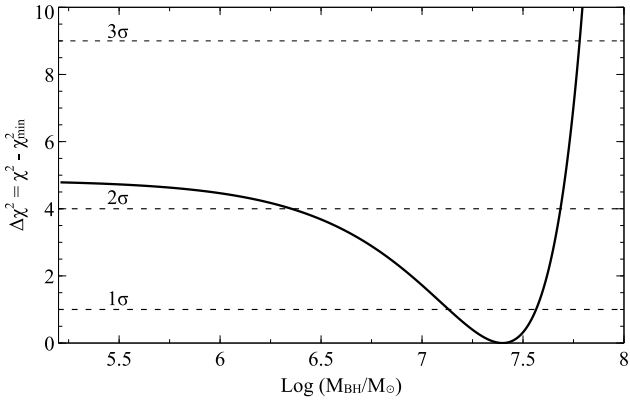,width=1.0\linewidth,clip=}
\caption{$\Delta\chi^{2}$ as a function of black hole mass for the JAM fit. Confidence levels for a $\chi^{2}$ distribution with one degree of freedom are indicated by the horizontal dashed lines.}
\label{delta_chi}
\end{figure}

To determine the first moment of the line of sight velocity distribution, the line of sight velocity, further constraints are needed because one needs to separate the contribution of ordered and random motions.
To compare the second moments, the problem has essentially three unknown parameters, the mass to luminosity ratio (M/L), the anisotropy $\beta = 1 - 1/b$ and the inclination $i$. M/L is weakly dependent on the anisotropy. 
However there is a degeneracy between the inclination and $\beta$. As discussed in \cite{cappellari08}, the inclination can be recovered if we assume certain observational constraints on $\beta$, as in their sample of fast rotators for example, or vice-versa. 
A dynamical signature of the presence of the black hole would be an increase of the stellar velocity dispersion in the centre of the galaxy. This would be indicated by a central increase in the observed second velocity moment (V$_{\rm RMS}$ = $\sqrt{V^{2} + \sigma^{2}}$). However, for the low black hole masses ($<$10$^{7}$ M$_{\odot}$) this increase is expected to be present at very small radii and it is not observed in our data. In \cite{raimundo13} we determined an upper limit for the black hole mass, which was a more robust constraint considering the information we had on the source. In these new data we are probing even larger radii (the H-band spatial resolution in this work is 0${''}$.6, higher than the 0${''}$.1 in \citealt{raimundo13}). However with the molecular gas information we have a new constraint on the galaxy inclination if we consider that the H$_{2}$ gas is in circular orbits, and can fix the inclination to 61 degrees in our dynamical model.
We used the H-band data at resolution of 0${''}$.1 from \cite{raimundo13} as input for the new JAM modelling. The model takes in the maps for the stellar velocity and stellar velocity dispersion and the errors in these parameters determined by Monte Carlo simulations. We restricted our fit to the inner r $<$ 0${''}$.4 where the black hole potential will have the highest impact on the second moment of the line of sight velocity. The inclination is fixed at 61 degrees and the M/L, $\beta$ and M$_{\rm BH}$ are allowed to vary. The bins which show the highest errors in velocity and velocity dispersion due to the strong AGN continuum were masked out of the fit to ensure that they would not bias our result and are represented in the plot by the grey filled circles. The observed second velocity moment (V$_{\rm RMS}$ = $\sqrt{V^{2} + \sigma^{2}}$) is compared with the projected (and convolved with the instrumental PSF) V$_{\rm rms}$ from the JAM modelling and we use the least-squares minimisation algorithm MPFIT \citep{markwardt09} to find the best-fit set of parameters. The maps for the best fit result are shown in Fig.~\ref{jam}. The best-fit black hole mass is $2.5 \times 10^{7}$\thinspace M$_{\odot}$, however, a lower black hole mass or even a zero black hole mass are consistent within a 3$\sigma$ error. Although the absolute values for the model parameters may be affected by degeneracies, we can analyse their overall trend: the negative ($\beta < 0$) value obtained indicates a tangential anisotropy in the regions close to the black hole (observed in other galaxies with counter-rotating components as well - \citealt{cappellari07}), and the M/L ratio is higher than what was determined from the \textsc{stars} code modelling in Section~\ref{sec:stellar_pop}, which is consistent considering that there may be a contribution from dark matter and/or gas to the total mass. To evaluate the confidence limits for the black hole mass we determine the chi-square difference as a function of black hole mass. This method relies on estimating $\Delta\chi^{2} = \chi^{2} - \chi^{2}_{\rm min}$, which is a quantity that only depends on the number of `interesting' parameters and not on the total number of parameters or degrees of freedom used in the initial fit \citep{avni76}. In our case the only `interesting parameter' is the black hole mass and for each fixed value of M$_{BH}$ we minimise $\chi^{2}$ over the remaining parameters. The result is shown in Fig.~\ref{delta_chi} with the confidence limits $\Delta\chi^{2} = 1$  (1$\sigma$), $\Delta\chi^{2} = 4$ (2$\sigma$) and $\Delta\chi^{2} = 9$ (3$\sigma$) considering a $\chi^{2}$ distribution with one degree of freedom. Our best fit black hole mass assuming a fixed inclination of 61 degrees is: M$_{\rm BH} = 2.5_{-2.3}^{+2.3}\times$ 10$^{7}$ M$_{\odot}$ with the confidence limits referring to a 2$\sigma$ (95.4 per cent) confidence level. As can be seen from Fig.~\ref{delta_chi} our data is not very sensitive to lower black hole masses and it is essentially consistent within 3$\sigma$ with values down to zero black hole mass. Therefore the strongest constrain from our modelling is a 3$\sigma$ upper limit for the black hole mass of M$_{\rm BH} < 6 \times$10$^{7}\thinspace$M$_{\odot}$. \footnote{Our estimates are consistent with the black hole mass measurement of (1.6 $\pm$ 0.4) $\times 10^6$ M$_{\odot}$ \citep{bentz16} from a recent reverberation mapping campaign that was published after the submission of this paper.}
\subsection{The kinematically distinct core}
To determine the size of the distinct core we use the stellar velocity map and define as the radius of the KDC the spatial region where the superposition of the two velocity components has a local minimum (as in \citealt{mcdermid06}). We determine a size of r $\sim$ 1.$''$25 (200 pc) for the KDC in MCG--6-30-15. The H$_{2}$ distribution extends to larger radii (r $<$ 600 pc) than the counter-rotating stellar core (r $<$ 200 pc). If the stars of the counter-rotating core formed in-situ and from this gas, they are expected to trace the highest gas density which may explain why they are confined to a smaller diameter than the gas. The counter-rotating system could have a larger diameter but not be observable in the infrared - we are limited to observing the regions where the luminosity of the counter-rotating stars dominates over the stars co-rotating in the main body of the galaxy.

In each position of the field-of-view, the velocity distribution we observe is a superposition of the stars in the KDC and the stars in the main body of the galaxy, weighted by their relative contributions to the total light at that particular position. We explored the possibility of decomposing the spectra into two line-of-sight velocity components at every pixel. We modified pPXF to include two different sets of templates, one set to model the stars in the KDC and another set to model the stars in the main body of the galaxy, similar to what has been done in \cite{coccato11} and \cite{johnston13}. Due to the small difference in velocities between the two components, in particular in the central 2${''}$, it is not possible to obtain an unambiguous decomposition for the full field-of-view without further constraints. We also notice a significant dependence of the final result on the initial parameters chosen and therefore do not attempt to fit different stellar populations or determine the line-of-sight distribution for the two components independently. Instead we follow a different approach to determine the relative contribution of each of the components to the total light, which will give us an idea of how much the KDC contributes to the light outside its apparent size. We assume that if the counter-rotating stars formed from the molecular gas, their dynamics should be similar, therefore the line-of-sight velocity can be approximated by a superposition of a component with the dynamics of the H$_{2}$ disc and another component with dynamics similar to the main body of the galaxy (which we consider to be the stellar dynamics at r $>$ 2${''}$). We build two different templates: one which is a convolution of K-type stellar templates with a gaussian kernel with the velocity and velocity dispersion of the H$_{2}$ gas profile, and another one which is a convolution of K-type stellar templates with a gaussian kernel with velocity and velocity dispersion of the stars between $2{''} < r <$ 3${''}$. The two templates are used as input to pPXF. The velocity and velocity dispersion of the KDC template are allowed to vary within the measurement errors while the parameter space for the velocity and velocity dispersion of the main galaxy template is broader to take into account the fact that even at $r > 2 {''}$ the line of sight distribution may still be affected by the KDC component. We find that the galaxy spectra can be well reproduced by a combination of these two templates. For radius $r \gtrsim 1{''}$ the contribution of the KDC component to the light is $\sim 20 - 30$ per cent. Within $r < 1{''}$ the relative contributions of the templates is harder to determine but we observe that the KDC component dominates and is responsible for $\sim 60 - 80$ per cent of the light. We also obtain higher absolute velocities for the main galaxy component, V $ \sim$ 50 km/s at r $\sim 2{''}$ along the major axis compared with $V \sim$ 30 km/s when fitting the line-of-sight velocity with just one component. This indicates that the rotation curve for the main body of the galaxy has a higher absolute velocity than what we measure in Fig.~\ref{horizontal_h2} for the combined KDC + main galaxy line-of-sight velocity distribution, which is expected since the counter-rotation of the KDC will cause a shift in the centroid of the absorption line profiles.
\subsubsection{Formation of the KDC}
We use the observations and results of this work to constrain the formation scenarios for the KDC. To fully constrain the origin and age of the distinct core an important diagnostic would be to determine the age and metallicities of the stellar populations in the counter-rotating core and in the main body of the galaxy. The scenario where gas accretion and subsequent in-situ star formation occurs implies that the stellar age of the counter-rotating component is always younger than the main body of the galaxy. In the case of minor mergers as the origin for the stars and gas in the distinct core, the stellar population can be younger or older than the population in the main body of the galaxy. High spatial resolution optical observations that could model the stellar populations would allow us to set better constraints on the formation scenario for the KDC. However with the information we currently have, we can exclude some of the formation mechanisms and derive conclusions about the recent past of MCG--6-30-15 and its AGN activity.

We observe four key features: 1) The KDC is relatively small (diameter of less than 1 kpc); 2) The molecular gas is also counter-rotating in relation to the main body of the galaxy; 3) There is evidence, from our [Fe II] and supernova rate discussion and from UV observations (\citealt{bonatto00}), that there are traces of recent star formation (age $< 100$ Myr) in this galaxy; 4) A dust lane is observed which points towards a past dynamical interaction.

There are various hypothesis for the formation of KDC that we can exclude as unlikely in the case of our galaxy: a) the size of the KDC is too small to have been formed during a major merger; b) the counter-rotating orientation and the presence of molecular gas indicate that dissipation was important in the formation of the KDC, if a merger occurred it most likely was a minor merger and involved gas; and c) the scenarios for an internal origin for the KDC are extremely unlikely in our galaxy, the lack of a stellar bar and the presence of a large quantity of counter-rotating ionised and molecular gas in addition to a counter-rotating stellar core indicate that the origin of the KDC is not internal. 
Our observations are in line with the scenario of the KDC being formed by recent accretion of gas into the nuclei and subsequent in-situ star formation, as proposed by \cite{mcdermid06} for the formation of small KDCs.
It is unclear if the KDC was formed by the accretion of a small satellite galaxy or external gas accretion, induced by tidal interactions with one of its neighbours for example, which drove gas to the nuclei and subsequently formed stars. In any of these cases we can conclude that gas with an external origin was accreted into the nucleus.

There are several spiral galaxies (e.g. NGC 5719 
\citealt{coccato11}, NGC 4138 - 
 \citealt{pizzella14}), and S0 galaxies: (e.g. NGC 3593,
 NGC 4550 - 
 \citealt{coccato13} and NGC 4191 - 
 \citealt{coccato15}), that show counter-rotating ionised gas in addition to counter-rotating stars, although these counter-rotating stars are typically on the form of large-scale discs and not small cores. For these galaxies the gas has an external origin, and for S0 galaxies in particular the formation of the counter-rotating stars involve accretion of external gas into an already existing gas-free disc and subsequent in-situ star formation (e.g. NGC 3593 - 
 \citealt{bertola96}, a galaxy which also shows counter-rotating molecular gas - \citealt{garcia-burillo00}).
\subsubsection{The environment of MCG--6-30-15}

The environment of MCG--6-30-15 is important to determine the origin of the gas whose remnants we observe in the form of the H$_{2}$ gas distribution and that possibly formed the stellar counter-rotating core. There is a strong indication that MCG--6-30-15 is in a small group based on the distance to its 4th nearest neighbour, which is only 4.1 Mpc (from the 2MASS redshift survey), much smaller than what is found for isolated galaxies (15 - 30 Mpc). It is likely that these neighbours are the source of the recently accreted gas. H I images could determine the gas distribution in the outskirts and neighbouring regions of the galaxy but they are not available in the literature.
Tentative evidence for dynamical interactions come from the \cite{bonatto00} identification criteria which includes MCG--6-30-15 in the set of nearby disturbed galaxies and from the presence of a dust lane in MCG--6-30-15 which is in line with a possible past interaction between this galaxy and one of its neighbours.
\subsubsection{Implications for AGN fuelling}
From a study of the distribution of gas in the inner kpc of S0 galaxies, \cite{davies14} find that the AGN in these galaxies are typically fuelled by external gas accretion. The observations and results for MCG--6-30-15 reinforce this scenario of AGN fuelling in S0 galaxies. In MCG--6-30-15 we observe a small stellar counter-rotating core and most importantly counter-rotating molecular gas which indicates that accretion of external gas occurred in this galaxy. The counter-rotating molecular gas has significantly replenished the AGN fuelling reservoir at scales of hundreds of parsecs, with a total mass of 3.4 $\times$ 10$^{7}$ M$_{\odot}$ in the central 400 pc. We also see that the molecular gas is present at radius r $\lesssim$ 50 - 100 pc which indicates that this external gas accretion event was able to drive gas to the central hundred parsecs. Considering these observations we conclude that external accretion has likely provided the gas needed to replenish the AGN fuelling reservoir in MCG--6-30-15. 

MCG--6-30-15 was famously the first galaxy to show a broad Fe K$_{\alpha}$ emission line in the X-ray spectrum (\citealt{tanaka95}) which allowed the black hole spin to be measured. The most recent measurements indicate a high black hole spin of a $>$ 0.98 (\citealt{brenneman&reynolds06}). Having a molecular gas reservoir counter-rotating in relation to the main body of the galaxy provides an interesting laboratory for tests on accretion physics. From theoretical models, the black hole spin increases due to continuous accretion of material with similar angular momentum. If the black hole has been accreting from the counter-rotating gas for the past 50 Myr or more, it could mean that the black hole is currently decreasing its spin. \cite{middleton16} find a misalignment between the inclination of the inner accretion disc and the host galaxy large scale stellar disc. They suggest that this misalignment could be related with the AGN in MCG--6-30-15 being fuelled by an external accretion event, which is supported by our findings.

In S0 galaxies such as MCG--6-30-15, KDCs are relatively rare ($< 10$ per cent \citealt{kuijken96}). Out of the tens of KDCs discovered and published in the literature, MCG--6- 30-15 has one of the highest AGN X-ray luminosities (L$_{\rm X(2-10)}$ keV = $4\times10^{42}$erg s$^{-1}$). Most of the remaining galaxies are quiescent and the ones with active nuclei have lower X-ray luminosities typical of low-ionisation nuclear emission regions (LINERs). As small KDCs can only be detected with high spatial resolution observations, we may be severely underestimating the number of galaxies with KDCs. Integral field spectroscopy observations are one of the best techniques to resolve the small KDC, however galaxies which host luminous AGN are often not considered as targets for integral field spectroscopy studies due to AGN contamination to the emission from the galaxy.  We are therefore likely biased to low luminosity AGN or quiescent galaxies. MCG--6-30-15 is therefore a unique and important case study since it has both a bright AGN and a small KDC.

Our current best theory to explain the formation of the distinct core is that earlier than 50 - 100 Myr ago MCG--6-30-15 accreted external gas from a neighbour or had a minor merger with a small gas-rich satellite galaxy. Out of this interaction, gas with a distinct angular momentum was driven to the centre of the galaxy, settling on a disc counter-rotating with respect to the galaxy's main body rotation. From this gas a younger stellar population formed which generated the stellar kinematically distinct core we observe. Possibly this gas formed or replenished the AGN fuelling reservoir at scales of hundreds of parsecs and we currently observe a remnant of the initial gas. A redshifted component in the H$_{2}$ velocity in our data, if proven to be associated with the dust lane would indicate that inflow from this disc of gas down to the centre of the galaxy may provide fuelling to the black hole in the near-future.
\section{Conclusions}
\label{sec:conclusions}
In this work we analysed the central 1 - 2 kpc of the active galaxy MCG--6-30-15 using new SINFONI observations in the H and K bands and VIMOS observations $\lambda$ [4200 - 6150] \AA. We mapped for the first time the molecular gas distribution traced by the H$_{2}$ emission and the extent of the counter-rotating core first detected in the H-band \citep{raimundo13}. Our main goal was to investigate if the formation of the stellar counter-rotating core was associated with inflow of gas into the centre of the galaxy. Our main conclusions can be summarised as follows:\\[0.2cm]
$\bullet$  \textbf{The molecular gas is in a counter-rotating disc:} The molecular gas shows counter-rotation with respect to the main body of the galaxy similar to what is observed in the counter-rotating stars but more extended (detected out to r $\sim$ 600 pc). If the stars of the counter-rotating core formed from this gas, the fact that they are confined to a smaller diameter could trace the regions of higher gas density. Alternatively the counter-rotating stellar core could indeed extend further out but we only observe the regions where the luminosity of the counter-rotating stars dominates over the main body of the galaxy. We determine that the contribution of the counter-rotating stars to the total light is $\sim$ 20 - 30 per cent for r $\gtrsim 1{''}$. 
The molecular gas dynamics are well modelled by a rotating disc with inclination $\sim$ 61 degrees assuming circular rotation and the PA of the molecular gas is similar to that of the stars in the distinct core. The counter-rotating orientation suggests that dissipative processes were important in the formation of the distinct core. An H$_{2}$ residual redshifted component is observed to the West of the nucleus, coinciding with the peak in the H$_{2}$ flux. This component has a velocity of 120 km/s and velocity dispersion similar to the bulk of the H$_{2}$ disc. The redshifted component seems to trace the dust lane which is observed south of MCG--6-30-15's and just below our line of sight towards the nucleus. According to this geometry the redshifted component would be associated with gas moving towards the nucleus of the galaxy. If this feature indicates counter-rotating gas inflow to the centre of the galaxy remains to be investigated with higher sensitivity observations.\\[0.2cm]
$\bullet$  \textbf{The ionised gas shows outflows:}  The various ionised gas tracers indicate a combination of outflows and counter-rotation when circular motions dominate. The narrow line Br$\gamma$ emission originates in the nuclear region and is very compact, while narrow H$_{\beta}$ is more extended (r $< 400$ pc). They both show counter-rotation. The [Ca VIII] dynamics is dominated by non-rotational motions, and the spectra shows two emission line components: a redshifted and a blueshifted one at each spatial position which trace the approaching and receding sides of an outflow cone. 
From the dynamical analysis we conclude that [Ca VIII] is tracing an outflow with velocity $\sim$ 100 km s$^{-1}$, where the redshifted component of the outflow dominates the flux we observe. The [O III] emission shows disturbed dynamics at $\sim$ kpc scales which indicates that the [Ca VIII] detected outflow may extend to larger radii. [Fe II] emission presents velocities above what is expected from the stellar rotation but also distinct from [Ca VIII]. We argue that [Fe II] is excited by supernova shocks and is tracing the local dynamics. From its flux we estimate that the star formation in the distinct core happened $\sim$ 50 Myr ago.\\[0.2cm]
$\bullet$  \textbf{The black hole mass is constrained dynamically:} From the dynamics of the H$_{2}$ emission we can determine the total enclosed mass in the inner 80 pc to be M$_{\rm enc} < 7.8 \times 10^{7}$M$_{\odot}$ which sets an upper limit for the M$_{\rm BH}$. Using JAM and fixing the inclination to 61 degrees as indicated from the molecular gas dynamics, we determine a best-fit black hole mass of M$_{\rm BH} = 2.5_{-2.3}^{+2.3}\times$ 10$^{7}$ M$_{\odot}$ with 2$\sigma$ (95.4 per cent) confidence limits. As our data is not very sensitive to lower black hole masses the strongest constrain from our modelling is a 3$\sigma$ upper limit for the black hole mass of M$_{\rm BH} < 6 \times$10$^{7}\thinspace$M$_{\odot}$.\\[0.2cm]
$\bullet$  \textbf{The counter-rotating core was formed by external gas accretion:} With the data presented here we improved on the size measurement of the counter rotating core first estimated in \cite{raimundo13} and determine that the core has a diameter of $\sim$ 400 pc extending out to a radius of r $\sim$ 1{$''$}.25. 
The results from this work point to a scenario where the KDC was formed by external gas accretion: The KDC is small, molecular gas is present and counter-rotating in relation to the main body of the galaxy, there are traces of recent star formation (age $<$ 100 Myr) in this galaxy and a dust lane is observed which points towards a past dynamical interaction.
Currently a molecular gas mass of M$_{\rm gas} = 3.4\times$ 10$^{7}$ M$_{\odot}$ is present in the central 400 pc of MCG--6-30-15. The fact that we observe counter-rotating molecular gas at radius r $\lesssim$ 50 - 100 pc indicates that this external gas accretion event was able to drive gas to the central hundred parsecs. Our observations agree with what was found by \cite{davies14}, that AGN in S0 galaxies are typically fuelled by external accretion.\\[0.2cm]
The results in this work suggest a scenario where the formation of the stellar kinematically distinct core is associated with gas inflow. External gas (from a galaxy neighbour or part of a minor merger with a gas rich satellite) was driven into the centre of the galaxy where it settled in the form of a counter-rotating disc in the equatorial plane of the galaxy. This event provided gas to replenish the AGN fuelling reservoir on scales of hundreds of parsecs. Kinematically distinct core formation when associated with gas can provide a tracer for the presence of fresh gas for AGN fuelling.
\section{Acknowledgements}
The authors would like to thank the referee, Cristina Ramos Almeida, for a thorough reading of the paper and for the constructive comments and suggestions which improved this work. The authors would also like to thank Michele Cappellari, Craig Markwardt, Jeremy Sanders and Kristine Spekkens for making their codes and/or software available to the community. The plots in this paper were produced with the plotting package \textsc{Veusz} by Jeremy Sanders. A.C.F acknowledges support from ERC Advanced Grant 340442. P.G. acknowledges support from STFC (grant reference ST/J003697/2). This research has made use of the NASA/IPAC Extragalactic Database (NED), which is operated by the Jet Propulsion Laboratory, California Institute of Technology, under contract with the National Aeronautics and Space Administration. In this work we used data obtained as part of the Two Micron All Sky Survey (2MASS), a joint project of the University of Massachusetts and the Infrared Processing and Analysis Center/California Institute of Technology, funded by the National Aeronautics and Space Administration and the National Science Foundation. This research has made use of NASA's Astrophysics Data System.

\bibliographystyle{mn2e}
\bibliography{AGN}
\end{document}